\documentclass[aps,pra,reprint,amsmath,amssymb,nofootinbib,longbibliography,floatfix]{revtex4-2}
\usepackage[T1]{fontenc}
\usepackage{lmodern}
\usepackage{graphicx,bm,xcolor,booktabs}
\usepackage{tikz,pgfplots}
\pgfplotsset{compat=1.18}
\usepgfplotslibrary{groupplots,fillbetween}
\usetikzlibrary{arrows.meta,positioning,calc,shapes.geometric,fit}
\usepackage[colorlinks=true,citecolor=blue!50!black,linkcolor=blue!50!black,urlcolor=blue!50!black]{hyperref}
\newcommand{\ket}[1]{\lvert #1\rangle}
\newcommand{\bra}[1]{\langle #1\rvert}
\newcommand{\Tr}{\operatorname{Tr}}

\newcommand{\CN}{\mathcal{CN}}
\hypersetup{pdftitle={CV-TF-QKD},pdfauthor={Stefano Pirandola}}
\begin{document}
\title{CV-TF-QKD}
\author{Stefano Pirandola}
\affiliation{Department of Computer Science, University of York, York YO10 5GH, United Kingdom}
\begin{abstract}
We introduce a continuous-variable version of twin-field quantum key distribution, using independent Gaussian-modulated coherent states and an untrusted single-photon relay. Full-record reconciliation and a binary phase-disclosure protocol achieve nearly equal asymptotic key rates against collective attacks. At fixed weak input-referred excess noise per arm, both rates per relay use scale as $\sqrt{\eta}$, where $\eta$ is the end-to-end transmissivity. For phase disclosure, we establish composable security against general coherent attacks using a bounded test of preparation intensities and sign-parity complementarity. For specified apparatus parameters, finite-key calculations incorporating detector inefficiency, dark counts, phase fluctuations, and reference overhead predict rates exceeding the pure-loss and matched thermal-loss repeaterless bounds. These results extend the twin-field advantage to Gaussian-modulated continuous-variable quantum cryptography.
\end{abstract}
\maketitle
\section{Introduction}

Gaussian-modulated coherent states provide simple optical transmitters for continuous-variable quantum key distribution (CV-QKD)~\cite{CVreview,Grosshans2002,Grosshans2003,Weedbrook2004}. Their communication range is constrained by photon loss: the pure-loss direct-channel secret-key capacity, also known as the Pirandola-Laurenza-Ottaviani-Banchi (PLOB) bound, is given by $K_0(\eta)=-\log_2(1-\eta)$ where $\eta$ is the end-to-end transmissivity~\cite{PLOB}. At high loss, this capacity scales as approximately $1.44\eta$ bits per channel use. Additional noise constraints can impose a maximum secure distance~\cite{QKDdistance}. These limitations motivate relay-assisted architectures for extending quantum communication beyond direct links.

Measurement-device-independent (MDI) protocols move detection to an untrusted relay~\cite{Braunstein2012,Lo2012,Pirandola2015}, while twin-field (TF)~\cite{Lucamarini2018} uses single-photon interference to achieve square-root loss scaling~\cite{Lucamarini2018,Ma2018,Wang2018,Yu2019SNS,Jiang2019SNS,Hu2019SNS,Xu2020SNS,Xu2020Hybrid}. A central question is whether this scaling can be combined with independent Gaussian coherent-state preparation at both users, retaining its simple source architecture and a security proof that supports finite experimental data.

We answer this question with CV-TF-QKD. Alice and Bob independently draw complex Gaussian amplitudes and send coherent states to a balanced relay, which accepts one photon in one output and vacuum in the other. The continuous variables are the preparation records; the heralding measurement is non-Gaussian. The non-Gaussian optical operation is confined to the untrusted relay, so neither user needs a local single-photon source, squeezing or a quantum memory. Earlier CV schemes surpass the repeaterless bound using a local single-photon resource~\cite{Winnel2021} or photon-number encoding~\cite{Erkilic2023}. Gaussian-modulated CV relay protocols with postselection instead use a homodyne Bell measurement~\cite{PostCV}. The present combination connects Gaussian modulation at both transmitters with single-photon TF heralding. As a relay-assisted scheme, it remains subject to the corresponding network capacity bounds~\cite{Network2019}.

The classical processing is essential to this connection. Full-record reconciliation, called CV-TF-Basic, uses the complete complex preparation records. Phase disclosure, called CV-TF-PD, extracts Alice's binary sign after she announces her intensity and phase modulo $\pi$, while Bob keeps his own record private. Although these protocols reveal different information to Eve, their asymptotic rates against collective attacks are nearly equal and retain square-root scaling at fixed weak input-referred excess noise. Binary phase disclosure also exposes a sign--parity complementarity that permits a direct composable proof against general coherent attacks.

Together, these results provide a route from Gaussian optical modulation to finite-key TF operation beyond direct-channel bounds. We first describe the two decoding procedures and their IID asymptotic rates, then compare these rates and finite CV-TF-PD predictions with pure-loss and matched thermal-loss benchmarks. The finite calculation includes detector inefficiency, dark events, residual phase fluctuations and reference overhead under specified apparatus assumptions. The main text states the resulting composable key-length bound; the appendices derive the optical states, information quantities, security estimates and finite decoding guarantees.

\section{General optical protocol}
\label{main:optical}
Alice and Bob independently draw complex amplitudes
\begin{equation}
 \alpha,\beta_{\rm em}\sim\CN(0,N),~~
 p_N(z)=\frac{e^{-|z|^2/N}}{\pi N},
 \label{main:gaussian}
\end{equation}
and send $\ket\alpha$ and $\ket{\beta_{\rm em}}$ to Charlie. The real and imaginary parts of each amplitude have variance $N/2$ and each signal has mean photon number $N$. The model of legitimate operation uses matched modes, a common phase reference and a balanced beam splitter. Charlie reports $s=+1$ for output counts $(1,0)$, $s=-1$ for $(0,1)$, and a failure symbol for every other event. The ideal asymptotic detector resolves photon number exactly; the finite honest model will also include loss and background counts.

After Charlie commits to the complete relay record, Bob stores the corrected amplitude
\begin{equation}
 \beta=s\beta_{\rm em},~~ A=|\alpha|^2,\quad B=|\beta|^2
 \label{main:correction}
\end{equation}
on each success. From this point, $\beta$ already includes the port correction. Alice and Bob select tests (for parameter estimation) using fresh randomness after the relay commitment. Their laboratories, preparations, private records and random generators are trusted, and classical communication is authenticated. The relay and both channels are untrusted. Its apparatus is an honest-performance model rather than an assumption restricting an adversary.

\begin{figure}[t]
\vspace{-0.5cm}
\centering
\begin{tikzpicture}[>=Latex,font=\small,box/.style={draw,rounded corners,align=center,minimum width=1.5cm,minimum height=.8cm}]
 \node[box,fill=blue!5] (A) at (0,1) {Alice\\$\ket\alpha$};
 \node[box,fill=blue!5] (B) at (0,-1) {Bob\\$\ket{\beta_{\rm em}}$};
 \node[box,fill=orange!8,minimum height=1.5cm] (C) at (3,0) {Charlie\\50:50 + PNR};
 \node[box,fill=green!5] (F) at (6,0) {Commit\\$s=\pm1,\varnothing$};
 \draw[->,thick] (A.east)--node[above,sloped] {arm} (C.north west);
 \draw[->,thick] (B.east)--node[below,sloped] {arm} (C.south west);
 \draw[->,thick] (C)--(F);
 \draw[->,dashed] (F.north) to[out=90,in=90] node[above] {public herald} (A.north);
\end{tikzpicture}
\caption{The common optical experiment. The prescribed success is one detected photon in one output and vacuum in the other. Both classical protocols use these same preparation and heralding steps. PNR denotes photon-number-resolving detection.}
\label{fig:optics}
\end{figure}
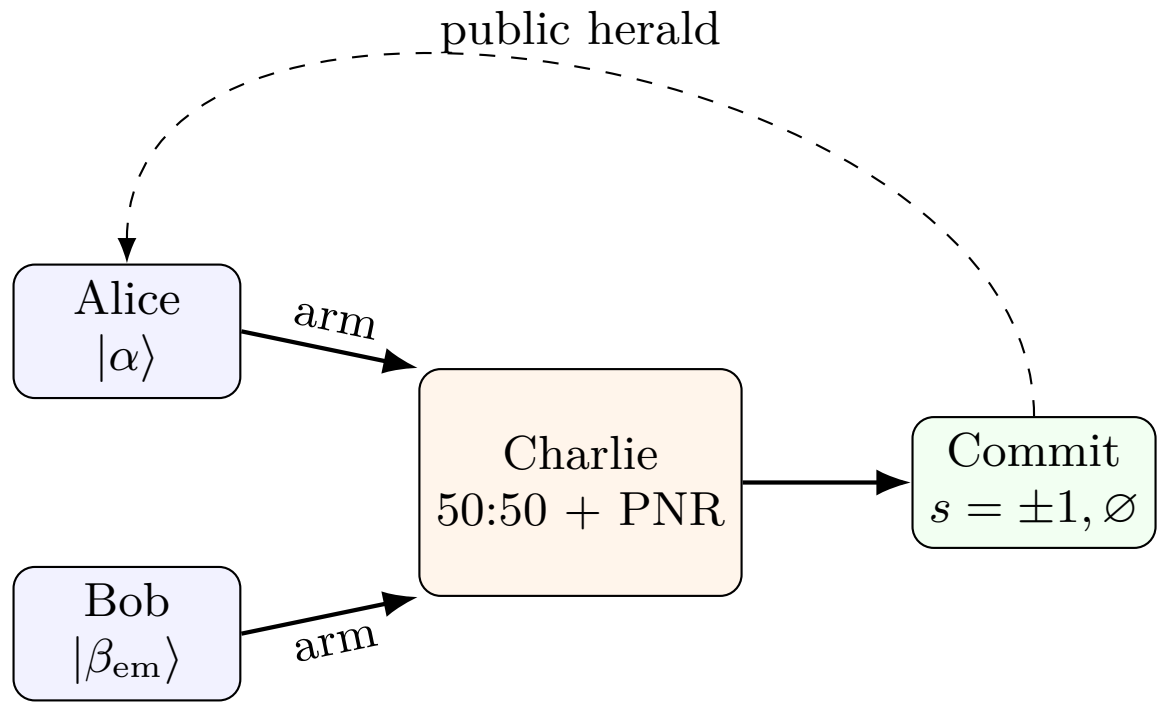

Let $M$ count all attempted simultaneous two-arm quantum transmissions, including failed heralds. If a passing execution outputs $\ell$ secret bits, its output rate is $\ell/M$. The expected honest rate includes the completion event $\Omega$:
\begin{equation}
 R=\Pr(\Omega)\frac{\ell}{M}.
 \label{main:rate_convention}
\end{equation}
Tests and honest aborts reduce the numerator. At nominal slot clock $f_0$ and quantum duty fraction $d_Q$, acquisition time is $T=M/(d_Qf_0)$ and expected throughput is $d_Qf_0R$. Reference allocation enters through the acquisition time and throughput. Direct-channel benchmarks use the end-to-end fiber transmissivity, excluding relay detector inefficiency.

\section{CV-TF-Basic}
\label{proc:basic}
CV-TF-Basic keeps the full complex preparation records private. Bob is the key owner and Alice is the decoder. The four decoding steps below use a common two-round example. All records include the relay-port correction. Both users' records are shown only to explain the arithmetic; each party uses only its own private data and the public transcript.

\subsection*{Step 1: quantize Bob's records}
The users agree on a quantizer $\mathcal Q$ that partitions the complex plane into labeled cells, with fixed rules for boundaries and amplitudes outside the chosen range. Bob forms the raw symbols $X_i=\mathcal Q(\beta_i)$. The cell label gives a finite classical description of each complex preparation record.

As an example, a simple quantizer just labels the four quadrants counterclockwise by $0,1,2,3$. For this example, suppose Bob's two corrected records give $X^2=(1,2)$.

\subsection*{Step 2: announce reconciliation data}

For $n$ candidate key rounds, Bob holds the quantized block
$X^n=(X_1,\ldots,X_n)$. The parties agree in advance on a publicly
known reconciliation map $\operatorname{Syn}$, which assigns each
possible block a shorter message, i.e., its \emph{syndrome}.
For a linear code, the syndrome consists of prescribed parity-check
sums of the symbols. Bob computes and sends
\begin{equation}
M_{\rm EC}=\operatorname{Syn}(X^n)
\end{equation}
over the authenticated public channel. 
The syndrome reveals information
about Bob's block, which is accounted for in privacy amplification.

For our example, choose the syndrome map
\begin{equation}
\operatorname{Syn}(x_1,x_2)=(x_1+x_2)\bmod 4.
\end{equation}
Bob's block $X^2=(1,2)$ gives $M_{\rm EC}=3$. This two-bit message
reduces the sixteen possible symbol pairs to four compatible
candidates:
\begin{equation}
(0,3),~~ (1,2),~~ (2,1),~~ (3,0).
\end{equation}
The syndrome alone does not distinguish these pairs; Alice's private
records supply the additional information used in the next step.

\subsection*{Step 3: reconstruct using Alice's private records}
Alice integrates the accepted conditional likelihood over each quantizer cell. Under the i.i.d. model her block reconstruction target is
\begin{equation}
 \widehat X^n=\underset{x^n:\,\operatorname{Syn}(x^n)=M_{\rm EC}}
 {\arg\max}\ \prod_{i=1}^n p(x_i\mid\alpha_i,s_i).
 \label{proc:basicdecoder}
\end{equation}

If several syndrome-compatible blocks attain the same maximum
conditional probability, Alice selects the first according to a
publicly agreed ordering. For example, lexicographic ordering compares
the first symbols, then the second symbols if the first ones coincide,
and continues until a difference is found. This rule makes the
decoder's output unique when equally likely candidates exist.

Continuing our example, suppose Alice obtains cell probabilities $(0.15,0.35,0.35,0.15)$ on the first round and $(0.15,0.15,0.35,0.35)$ on the second. The four compatible pairs, in the order above, have likelihoods $0.0525,0.1225,0.0525,0.0225$. She selects $(1,2)$.

\subsection*{Step 4: verify and extract}
After Alice has fixed her reconstructed block $\widehat X^n$, the
parties verify that it agrees with Bob's block $X^n$. They use a
publicly specified two-universal family of hash functions
\begin{equation}
V_r:\mathcal X^n\longrightarrow\{0,1\}^{v},
\end{equation}
where $\mathcal X$ is the quantizer's symbol alphabet, $r$ labels
the chosen function, and $v$ is the verification-tag length.
Two-universality means that, for any two distinct blocks $x^n$
and $y^n$,
\begin{equation}
\Pr_r\!\left[V_r(x^n)=V_r(y^n)\right]\le 2^{-v},
\end{equation}
when the seed $r$ is chosen uniformly.

Bob draws a fresh seed $r_{\rm ver}$, independently of the raw
records, and sends it together with the tag
\begin{equation}
t_B=V_{r_{\rm ver}}(X^n)
\end{equation}
over the authenticated public channel. Alice computes
$t_A=V_{r_{\rm ver}}(\widehat X^n)$ and compares the two tags.
A mismatch causes an abort. A match allows the protocol to
continue, with probability at most $2^{-v}$ of accepting unequal
raw blocks, assuming authenticated communication. The seed is
public; its purpose is to select an independent random check.

\begin{figure}[tbp]
\centering
\begin{tikzpicture}[
 bx/.style={draw=black!65,rounded corners=2pt,align=center,font=\footnotesize,inner xsep=4pt,inner ysep=5pt,line width=.45pt},
 wide/.style={bx,text width=7.35cm,fill=black!3},
 local/.style={bx,text width=3.25cm,fill=blue!4},
 flow/.style={-{Latex[length=1.6mm]},line width=.5pt},
 msg/.style={flow,dashed},lab/.style={font=\scriptsize,fill=white,inner sep=1.5pt}]
\node[wide] (start) at (0,0) {\textbf{Both: select accepted, non-test rounds}\\Bob's record already contains the relay-port correction.};
\node[local,anchor=north,minimum height=1.25cm] (alice) at ([xshift=-1.95cm,yshift=-.45cm]start.south) {\textbf{Alice: decoder}\\Keep $\alpha^n$; evaluate conditional cell probabilities.};
\node[local,anchor=north,minimum height=1.25cm] (bob) at ([xshift=1.95cm,yshift=-.45cm]start.south) {\textbf{Bob: owner}\\Keep $X_i=\mathcal Q(\beta_i)$; form $M_{\rm EC}$.};
\node[local,below=.55cm of alice] (decode) {Use private records and $M_{\rm EC}$ to reconstruct $\widehat X^n$.};
\node[wide,anchor=north] (verify) at ([xshift=1.95cm,yshift=-.5cm]decode.south) {\textbf{Both: compare fresh verification hashes}\\Bob hashes $X^n$; Alice hashes $\widehat X^n$.};
\node[local,fill=green!6,anchor=north] (key) at ([xshift=-1.95cm,yshift=-.6cm]verify.south) {\textbf{Privacy amplification}\\Output only a positive certified key length.};
\node[local,fill=black!6,anchor=north] (abort) at ([xshift=1.95cm,yshift=-.6cm]verify.south) {\textbf{Abort}\\No key output.};
\draw[flow] (start.south) -- ++(0,-.2) -| (alice.north);
\draw[flow] (start.south) -- ++(0,-.2) -| (bob.north);
\draw[flow] (alice) -- (decode);
\draw[msg] (bob.south) |- node[lab,pos=.75,above] {$M_{\rm EC}$} (decode.east);
\draw[flow] (decode.south) -- (decode.south |- verify.north);
\draw[flow] (bob.south) -- (bob.south |- verify.north);
\draw[flow] (verify.south -| key.north) -- node[lab,left] {match} (key.north);
\draw[flow] (verify.south -| abort.north) -- node[lab,right] {mismatch} (abort.north);
\end{tikzpicture}
\caption{CV-TF-Basic full-record reconstruction. The dashed arrow is the public reconciliation message; solid arrows are processing dependencies. Test acceptance and a positive security bound are additional requirements for output.}
\label{proc:basicflow}
\end{figure}
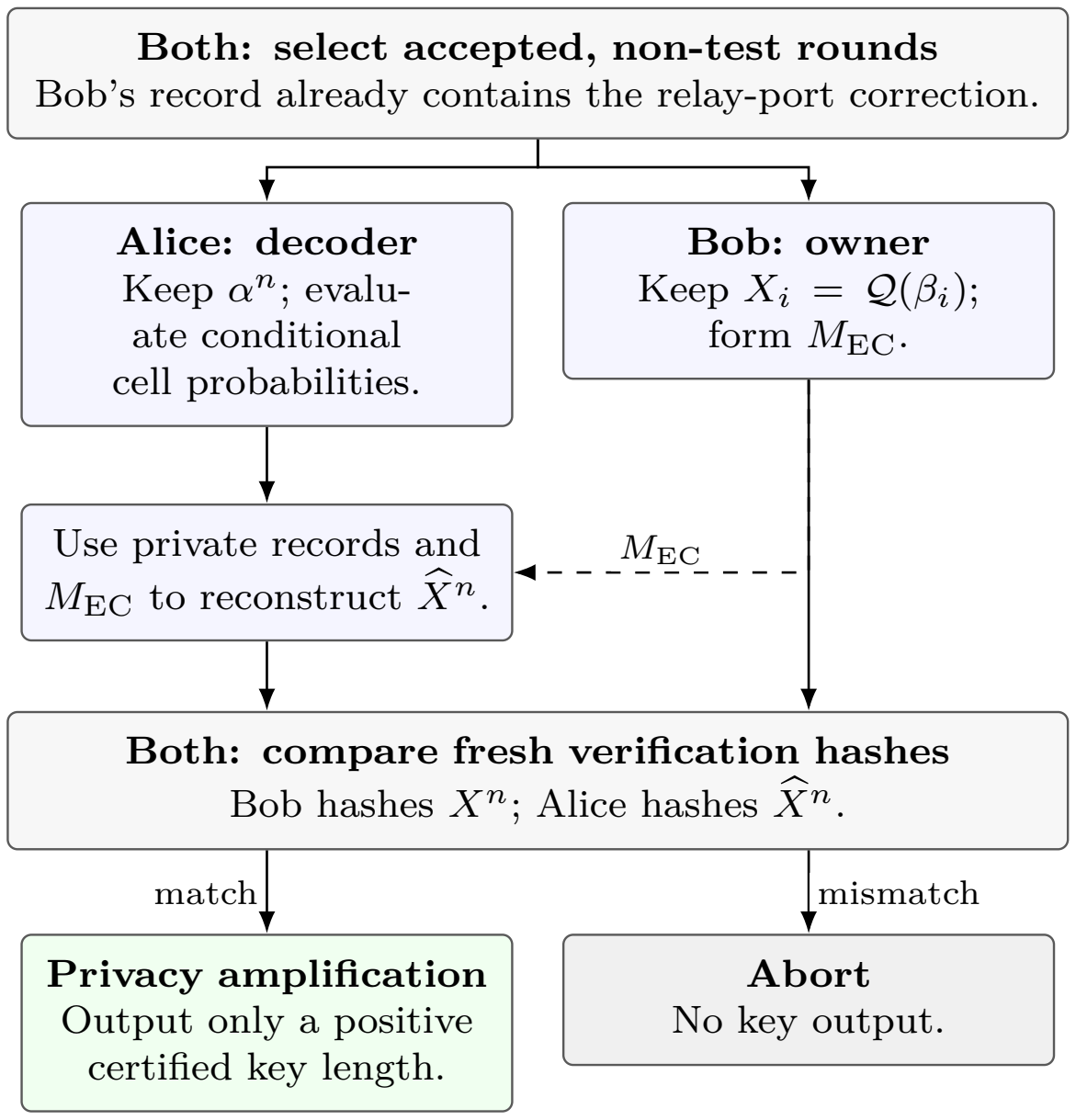

The parties next apply privacy amplification to reduce Eve's
information about the verified block. They choose an output length
$\ell$ allowed by the security analysis and abort if no positive
length is certified. Using a second, independent public seed
$r_{\rm PA}$, they select a two-universal hash function
\begin{equation}
F_{r_{\rm PA}}:\mathcal X^n\longrightarrow\{0,1\}^{\ell}
\end{equation}
and compute
\begin{equation}
K_B=F_{r_{\rm PA}}(X^n),~~
K_A=F_{r_{\rm PA}}(\widehat X^n).
\end{equation}
The admissible length $\ell$ accounts for Eve's quantum information
and the complete public transcript, including the reconciliation
message and verification tag. Verification checks agreement;
privacy amplification compresses the block to obtain a key whose
secrecy is justified by that bound. Neither operation requires
its random seed to remain secret.
Figure~\ref{proc:basicflow} summarizes the operations.

In our example, Bob's block and Alice's reconstructed block both
equal $(1,2)$, so their verification tags agree for every seed.
Equal blocks also produce equal privacy-amplification outputs.
In an actual run, however, the most likely reconstruction can be
incorrect, which is why verification remains necessary.

\section{CV-TF-PD: phase disclosure}
\label{proc:pd}
In CV-TF-PD, Alice owns a binary sign key and Bob reconstructs it from his full private preparation record. On key rounds Alice discloses her intensity and her phase modulo $\pi$. Bob discloses neither his intensity nor his phase. Every selected key round is retained.

\subsection*{Step 1: extract Alice's sign}
Alice writes her preparation record uniquely as
\begin{equation}
 \alpha=(-1)^{Z_A}\sqrt A\,e^{i\theta_A},~~
 Z_A\in\{0,1\},\quad \theta_A\in[0,\pi),
 \label{proc:signmap}
\end{equation}
where $A=|\alpha|^2$. For a nonzero amplitude, let $\psi_A=\arg\alpha$ be chosen in $[0,2\pi)$. The \emph{folded phase} $\theta_A=\psi_A\bmod\pi$ is the representative in $[0,\pi)$ obtained by identifying phases that differ by $\pi$. The bit $Z_A$ records which half of the full phase circle contains $\psi_A$: it is zero for $\psi_A<\pi$ and one otherwise. Disclosing $\theta_A$ therefore leaves the two antipodal amplitudes $\pm\sqrt A e^{i\theta_A}$ unresolved. Assign $Z_A=0$, $\theta_A=0$ at zero amplitude. This relabels the original Gaussian preparation records; the optical modulation remains Gaussian. Bob keeps his corrected complex record $\beta=s\beta_{\rm em}$ private.

As an example, consider three accepted plus-port rounds with $A_i=B_i=0.01$ and records
\begin{center}
\begin{tabular}{c|ccc}
$i$ & $Z_{A,i}$ & $\theta_{A,i}$ & Bob's private $\beta_i$\\\hline
1 & 1 & $5\pi/6$ & $0.1$\\
2 & 0 & $0$ & $0.1e^{i\arccos(3/5)}$\\
3 & 1 & $0$ & $-0.1e^{i\arccos(-1/5)}$
\end{tabular}
\end{center}
Alice's private sign block is $101$. For its first entry, the full phase $11\pi/6$ is folded to $5\pi/6$, while the removed half-turn is recorded by $Z_{A,1}=1$.

\subsection*{Step 2: disclose Alice's sign-independent labels}
Alice announces $A$ and $\theta_A$, leaving the antipodal alternatives $\pm\sqrt A e^{i\theta_A}$ unresolved. The public class is $C=(s,A,\theta_A)$. Bob keeps his entire corrected record private. Neither phase agreement nor intensity agreement is used to reject a selected key round.

For this example, Alice announces the intensity string $(0.01,0.01,0.01)$ and phase string $(5\pi/6,0,0)$. Together with the three plus-port heralds these specify the public classes. Her sign string $101$ is not among these labels, and Bob sends no key-round intensity or phase label.

\subsection*{Step 3: compute Bob's soft decoding inputs}
Bob evaluates the conditional bit probability and log-likelihood ratio
\begin{equation}
 p_i=\Pr(Z_{A,i}=1\mid\beta_i,C_i),~~
 L_i=\ln\frac{1-p_i}{p_i}.
 \label{proc:llr}
\end{equation}
A positive $L_i$ favors zero, a negative value favors one, and its magnitude measures confidence. The likelihood includes the channel and detector model used for reconciliation. A mismatch can lower decoding success; verification and the separate security test remain necessary.

For a concrete noisy example, take independent, identical thermal-loss arms of transmissivity $t$, ideal photon-number detection, and input-referred excess noise $\xi$ per arm in units of vacuum quadrature variance one. Write $w=t\xi/2$ for the output thermal occupation. With $B=|\beta|^2$, the exact fine-record posterior is
\begin{align}
 \kappa(A,\theta_A,\beta)
 &=\frac{2\sqrt A\,\operatorname{Re}(e^{-i\theta_A}\beta)}
 {A+B+\xi(1+w)},\label{proc:noisyposterior00} \\
 \Pr(Z_A=x\mid\beta,C)&=\frac{1+(-1)^x\kappa}{2}.
 \label{proc:noisyposterior}
\end{align}
The port correction is already included in $\beta$. The denominator contains the thermal background; at $\xi=0$ the formula reduces to the interference contrast of the two coherent amplitudes.

Continuing the example, let $t=10^{-4.5}$ and $\xi=10^{-5}$. The three soft inputs are
\begin{center}
\begin{tabular}{c|rrc}
$i$ & $p_i$ & $L_i$ & favored bit\\\hline
1 & $0.932796$ & $-2.630459$ & 1\\
2 & $0.200150$ & $ 1.385358$ & 0\\
3 & $0.400050$ & $ 0.405257$ & 0
\end{tabular}
\end{center}
Bob's separate guesses give $100$, whereas Alice has $101$. A posterior preference is not a deterministic inference. The third round remains in the block despite its incorrect individual guess.

\subsection*{Step 4: reconcile, verify and extract}
For an agreed binary matrix $H$, Alice sends the syndrome $M_{\rm EC}=HZ_A^n\pmod2$. Under the i.i.d. decoding model, Bob's reconstruction target is
\begin{equation}
 \widehat Z^n=\underset{x^n:\,Hx^n=M_{\rm EC}}{\arg\max}
 \prod_{i=1}^n p_i^{x_i}(1-p_i)^{1-x_i},
 \label{proc:signblock}
\end{equation}
with a fixed ordering for ties. A reconciliation code supplies an efficient implementation. Alice sends a fresh universal-hash verification tag and Bob aborts on a mismatch. If verification and the parameter test both pass, they apply privacy amplification to the certified length. Alice's public labels, reconciliation data and verification tag are all included in the security analysis.

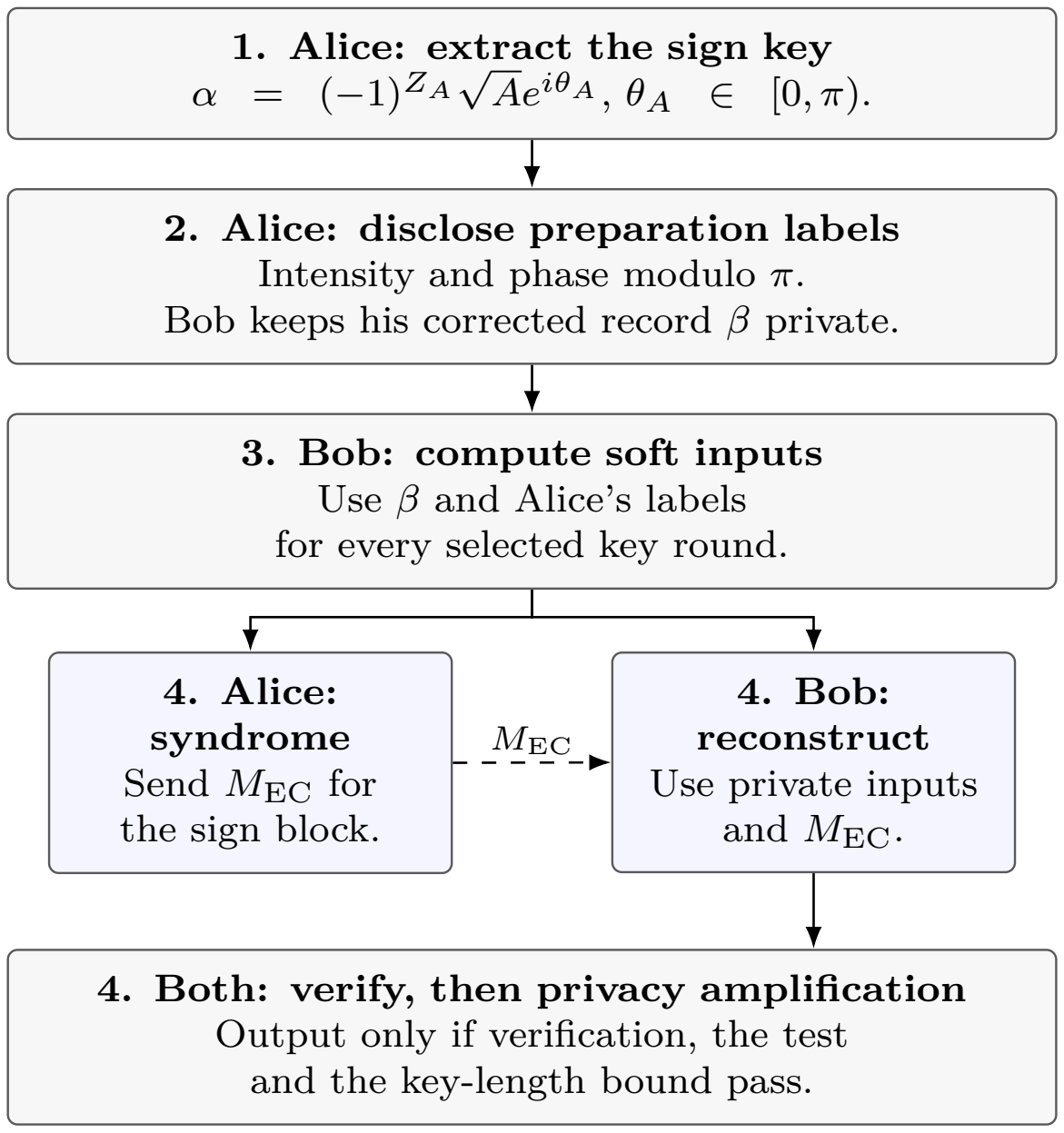
\begin{figure}[tbp]
\centering
\begin{tikzpicture}[
 bx/.style={draw=black!65,rounded corners=2pt,align=center,font=\footnotesize,inner xsep=4pt,inner ysep=5pt,line width=.45pt},
 wide/.style={bx,text width=7.35cm,fill=black!3},
 local/.style={bx,text width=2.65cm,fill=blue!4},
 flow/.style={-{Latex[length=1.6mm]},line width=.5pt},
 msg/.style={flow,dashed},lab/.style={font=\scriptsize,fill=white,inner sep=1.5pt}]
\node[wide] (one) at (0,0) {\textbf{1. Alice: extract the sign key}\\$\alpha=(-1)^{Z_A}\sqrt A e^{i\theta_A}$, $\theta_A\in[0,\pi)$.};
\node[wide,below=.35cm of one] (two) {\textbf{2. Alice: disclose preparation labels}\\Intensity and phase modulo $\pi$.\\Bob keeps his corrected record $\beta$ private.};
\node[wide,below=.35cm of two] (three) {\textbf{3. Bob: compute soft inputs}\\Use $\beta$ and Alice's labels\\for every selected key round.};
\node[local,anchor=north,minimum height=1.15cm] (owner) at ([xshift=-2.05cm,yshift=-.45cm]three.south) {\textbf{4. Alice:}\\\textbf{syndrome}\\Send $M_{\rm EC}$ for\\the sign block.};
\node[local,anchor=north,minimum height=1.15cm] (decoder) at ([xshift=2.05cm,yshift=-.45cm]three.south) {\textbf{4. Bob:}\\\textbf{reconstruct}\\Use private inputs\\and $M_{\rm EC}$.};
\node[wide,anchor=north] (end) at ([xshift=-2.05cm,yshift=-.55cm]decoder.south) {\textbf{4. Both: verify, then privacy amplification}\\Output only if verification, the test and the key-length bound pass.};
\draw[flow] (one) -- (two);
\draw[flow] (two) -- (three);
\draw[flow] (three.south) -- ++(0,-.2) -| (owner.north);
\draw[flow] (three.south) -- ++(0,-.2) -| (decoder.north);
\draw[msg] (owner.east) -- node[lab,above] {$M_{\rm EC}$} (decoder.west);
\draw[flow] (decoder.south) -- (decoder.south |- end.north);
\end{tikzpicture}
\caption{CV-TF-PD decoding with Alice as the fixed key owner. Only Alice's sign-independent preparation labels are disclosed on key rounds. Bob's full corrected record supplies his private decoding information.}
\label{proc:pdflow}
\end{figure}

For the three-bit example, choose
\begin{equation}
 H=\begin{pmatrix}1&1&0\\0&1&1\end{pmatrix},~~
 H(101)^{\mathsf T}=(11)^{\mathsf T}\pmod2.
 \label{proc:exampleH}
\end{equation}
The compatible strings are $101$ and $010$, with likelihoods $0.298476$ and $0.00806982$. Bob reconstructs $101$. If the fresh verification seed realizes
\begin{equation}
 T=\begin{pmatrix}1&0&1\\0&0&1\end{pmatrix},
 \label{proc:exampleT}
\end{equation}
both tags equal $T(101)^{\mathsf T}=01$. This is one realized hash, not a fixed verifier used on every run. The short block demonstrates reconstruction only: its two syndrome bits and two verification bits already exceed the three raw bits, so it certifies no secret output. Figure~\ref{proc:pdflow} summarizes the decoding.

In a finite-size implementation, Alice announces finite intensity and phase-cell labels, including underflow and overflow conventions. Bob integrates the likelihood over Alice's undisclosed within-cell record, or uses a cell-center approximation with a bounded error. He may quantize his own intensity, phase or resulting soft input locally; none of these local labels is announced. The finite performance calculation accounts for those approximation errors and retains all tail classes. On test rounds, both users disclose finite intervals containing their preparation intensities, as specified by the parameter test.

\section{IID asymptotic security and rates}
\label{asym:section}
We first assume collective attacks: accepted source pairs are independent copies of a joint state, while Eve can store her systems and measure them jointly after all announcements. Asymptotic tests identify the complete accepted preparation-record distribution. In the virtual source description this is an informationally complete two-mode heterodyne distribution. The thermal-loss model below is a prediction for those complete statistics; estimating only a covariance matrix or a value of excess noise does not certify the non-Gaussian accepted state. Appendix~\ref{coll:appendix} derives both rates from that state.

For the independent, identical thermal-loss arms, let $t$ be each arm's transmissivity, so the end-to-end transmissivity is $\eta=t^2$. In vacuum-variance-one units, let $\xi$ be the input-referred excess noise per arm. Define
\begin{equation}
 \begin{aligned}
 w&=t\xi/2,&m_{\rm ch}&=tN+w,\\
 \lambda&=\frac{tN}{m_{\rm ch}(1+w)},&
 \nu&=\frac{N(1-t+w)}{1+m_{\rm ch}},\\
 p_{\rm acc}&=\frac{2m_{\rm ch}}{(1+m_{\rm ch})^3}.&&
 \end{aligned}
 \label{asym:parameters}
\end{equation}
Here $p_{\rm acc}$ includes both accepted ports. The parameters $\lambda$ and $\nu$ specify the accepted virtual state: in balanced sum/difference modes it is $[(1-\lambda)\tau_\nu+\lambda\tau_\nu^+]\otimes\tau_\nu$, where $\tau_\nu$ is a thermal state of mean photon number $\nu$, and $\tau_\nu^+=\hat a^\dagger\tau_\nu\hat a/(1+\nu)$ is its normalized photon-added state; $\hat a$ annihilates that retained mode. Eve is assigned a purification of the accepted state, including the thermal noise.

Write $[x]_+=\max\{x,0\}$. For CV-TF-Basic, $I_{\rm full}$ is the mutual information between the two full corrected records and $\chi_{\rm full}$ is Eve's Holevo information about Bob's record. For CV-TF-PD, $I_{\rm sign}$ is the mutual information between Alice's sign and Bob's private record conditioned on the public data, and $\overline\chi_{\rm PD}$ is Eve's corresponding averaged Holevo information. Ideal one-way reconciliation gives
\begin{align}
 R_{\rm Basic}^{\rm iid}&=p_{\rm acc}[I_{\rm full}-\chi_{\rm full}]_+,\\
 R_{\rm PD}^{\rm iid}&=p_{\rm acc}[I_{\rm sign}-\overline\chi_{\rm PD}]_+.
 \label{asym:fixed_rates}
\end{align}
For CV-TF-PD, the public data comprise Alice's intensity and folded phase. Bob uses his full private complex record to decode her sign. Eve's conditional state is averaged over Bob's undisclosed record before its entropy is evaluated. The positive part is taken after averaging all retained rounds; no intensity or phase class is rejected.

For fixed $N$ and fixed input-referred $\xi$, the parameters approach $\lambda\to N/(N+\xi/2)$ and $\nu\to N$ at high loss, while $p_{\rm acc}\sim2(N+\xi/2)\sqrt\eta$. Every positive limiting secret fraction therefore gives $R\propto\sqrt\eta$. This statement keeps input-referred noise fixed; a fixed nonzero optical background is a different loss sweep.

\section{Asymptotic comparison and finite performance}
\label{main:performance}
\begin{figure*}[t]
\vspace{-0.0cm}
\centering
\includegraphics[width=\textwidth]{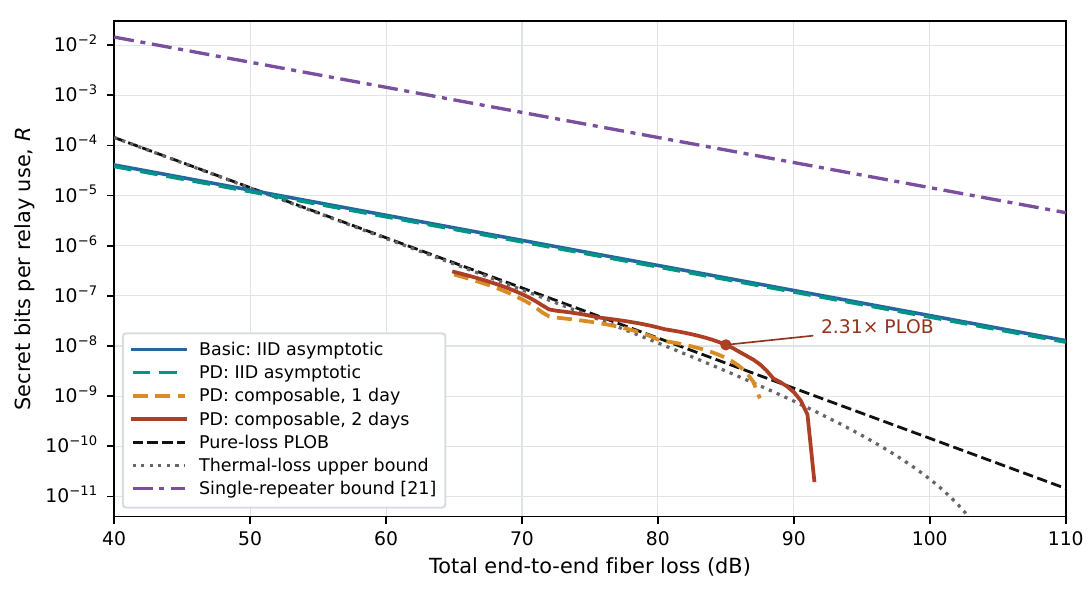}
\vspace{-0.8cm}
\caption{Secret bits per attempted relay use versus total fiber loss.
CV-TF-Basic and CV-TF-PD IID curves use $\xi=10^{-5}$ SNU per arm,
ideal detection and phase alignment, and the fixed modulations stated
in the text. CV-TF-Basic uses the continuous-record limit.
Composable CV-TF-PD curves are lower bounds on the expected rate for
one and two days with the apparatus in Table~\ref{tab:hardware};
they include testing, finite reconciliation and completion allowances,
with reference overhead charged through $M=d_Qf_0T$.
The IID and finite curves use different modulations and apparatus
models, so their separation is not solely a finite-size penalty.
The pure-loss and matched thermal-loss repeaterless bounds are shown
together with the pure-loss single-repeater bound
$K_1(\eta)=-\log_2(1-\sqrt{\eta})$ for an ideal repeater halfway along
the link~\cite{Network2019}. All three benchmarks use the physical
end-to-end fiber transmissivity $\eta=10^{-L/10}$, where $L$ is the
total fiber loss in dB, excluding relay detector inefficiency.
At $85$ dB, two-day CV-TF-PD reaches $2.31$ times pure-loss PLOB.}
\label{fig:performance}
\end{figure*}

\subsection{Pure-loss and thermal-loss benchmarks}
For total end-to-end fiber loss $L$ in decibels, put $\tau=10^{-L/20}$ and $\eta=\tau^2$. In the ideal asymptotic comparison each arm has $t=\tau$ and fixed input-referred excess noise $\xi=10^{-5}$ SNU. The matched direct thermal channel is the cascade of the two arms, with their common environment occupation $\bar n=\tau\xi/[2(1-\tau)]$. Define $g(x)=(x+1)\log_2(x+1)-x\log_2x$, with $g(0)=0$. In addition to pure-loss PLOB, $K_0(\eta)=-\log_2(1-\eta)$, we show the thermal-loss upper bound~\cite{PLOB}
\begin{equation}
 K_{\rm th}=\begin{cases}
 -\log_2[(1-\eta)\eta^{\bar n}]-g(\bar n),&\bar n<\eta/(1-\eta),\\
 0,&\bar n\ge\eta/(1-\eta).
 \end{cases}
 \label{main:thermal_plob}
\end{equation}
The direct input-referred noise is $(1+\tau)\xi/\tau$, not the per-arm value $\xi$. The thermal expression is an upper bound, while the pure-loss expression is the exact direct-channel capacity.

Figure~\ref{fig:performance} compares the two IID rates at fixed modulation across the loss sweep: $N=0.015$ for CV-TF-Basic and $N=0.016196$ for CV-TF-PD. These are separate design choices near useful high-loss operating points, and both curves use the full noisy accepted state. Their rates are very close: at $90$ dB, $R_{\rm Basic}^{\rm iid}=1.280\times10^{-7}$ and $R_{\rm PD}^{\rm iid}=1.195\times10^{-7}$ bits per relay use, a difference of about $7\%$ relative to CV-TF-Basic. Both exceed the direct-channel benchmarks at this operating point. CV-TF-PD retains nearly the same IID performance while providing a practical binary reconciliation procedure and the composable security extension developed below.

The CV-TF-Basic curve evaluates the ideal continuous-record expression. In terms of the decoding quantizer $\mathcal Q$, this corresponds to resolving the complex record arbitrarily finely while extending its covered range; the plotted calculation itself imposes no finite grid or range. A particular finite quantizer requires evaluating both the correlation and Eve-information terms for its actual cells, including overflow. 
CV-TF-PD instead reconciles Alice's sign, with her intensity and phase modulo $\pi$ public and Bob's complete record private.

\subsection{CV-TF-PD with the specified detector and phase model}
The finite CV-TF-PD curves in Fig.~\ref{fig:performance} are lower bounds on the expected rate under the apparatus conditions in Table~\ref{tab:hardware}. The prediction assumes independent noise during legitimate operation; security against arbitrary attacks is established separately in Sec.~\ref{main:composable}. Detector efficiency $\eta_d$ gives effective arm transmissivity $t=\eta_d\tau$. Each output independently adds Poisson dark events with mean $d=D_{\rm dark}\Delta t$ per gate. Residual Gaussian phase fluctuations have visibility $V_\phi=e^{-\sigma_\phi^2/2}$. They reduce Bob's decoding correlation but do not alter the preparation-intensity test. Dark events affect both the accepted-state mixture and the success probability and are treated separately from thermal excess noise. Appendix~\ref{pdimpl:appendix} derives this model and its completion probabilities.

\begin{table}[t]
\vspace{-0.2cm}
\caption{Apparatus parameters used to predict the finite CV-TF-PD curves in Fig.~\ref{fig:performance}. The calculation assumes ideal trusted sources and exact photon-number classification after loss and dark events; these parameters are joint design targets.}
\label{tab:hardware}
\begin{ruledtabular}
\begin{tabular}{lc}
Nominal slot clock $f_0$ & $1$ GHz\\
Quantum duty $d_Q$ & $0.90$\\
Effective detection efficiency $\eta_d$ & $0.80$\\
Dark rate per output $D_{\rm dark}$ & $1$ s$^{-1}$\\
Gate width $\Delta t$ & $1$ ns\\
Dark-event mean $d$ & $10^{-9}$\\
Thermal excess noise per arm $\xi_0$ & $10^{-5}$ SNU\\
Phase visibility $V_\phi$ & $0.995$\\
Phase RMS $\sigma_\phi$ & $\simeq0.10$ rad\\
\end{tabular}
\end{ruledtabular}
\end{table}

Phase stabilization and low-background detection are established experimental development directions~\cite{Liu2023,Clivati2022}. High-efficiency PNR devices nevertheless have nonzero backgrounds and imperfect photon-number response~\cite{Ding2025}. The present model assumes exact PNR classification after the specified efficiency and Poisson events, and ideal trusted preparations. It does not substitute these targets for a measured detector response matrix or a source-calibration error model.

The finite designs use explicit binary polar codes, with frame lengths $2^{22}$ or $2^{24}$ and a bounded reconstruction failure probability. For each loss and duration, the modulation, code, number of key frames and test threshold are fixed before transmission. The expected rates include insufficient-count, test, finite-input and decoding abort allowances. Acquisition lasts one or two days at quantum duty $d_Q=0.9$ and nominal clock $f_0=1$ GHz; this duration excludes decoder processing time. Appendix~\ref{pdimpl:appendix} specifies the finite algorithms and their bounds.

At $85$ dB, both durations use $N\simeq0.005418$ and frame length $2^{24}$. The one-day design uses one key frame and outputs $473\,508$ bits on completion; the two-day design uses three frames and outputs $1\,733\,181$ bits. The calculated lower bounds on their expected per-relay rates are approximately $5.78\times10^{-9}$ and $1.06\times10^{-8}$, respectively, corresponding to $1.27$ and $2.31$ times pure-loss PLOB and also exceeding the matched thermal-loss bound. The corresponding two-day throughput bound is approximately $9.50$ bits/s. These are predictions for the stated apparatus and noise model, with exact finite budgets given in Appendix~\ref{pdimpl:appendix}.

\section{Composable CV-TF-PD security against general attacks}
\label{main:composable}
CV-TF-PD admits a finite-key guarantee against arbitrary coherent attacks under the trusted-source and laboratory assumptions of Sec.~\ref{main:optical}. Alice remains the owner of every selected sign bit, and Bob keeps his key-round preparation record private. After the relay has committed its flags, the parties randomly select test positions and disclose finite intervals containing their preparation intensities. These tests are excluded from the key.

Let $n$ be the number of remaining sign bits and $k$ the statistically certified cap on their virtual parity errors. Here an error means equal photon-number parity in the two retained modes of the source-replacement description; these photon numbers are proof variables, not measurements added to the protocol. Appendix~\ref{app:composable} derives $k$ from the intensity test and random sampling, with explicit confidence allowances. Define the binary entropy $h_2(x)=-x\log_2x-(1-x)\log_2(1-x)$, with $0\log_2 0=0$. Let $L_{\rm EC}$ be the cap on key-dependent reconciliation communication, $v$ the verification-tag length in bits, and $\epsilon_{\rm PA}$ the privacy-amplification allowance. For $k<n/2$, a sufficient secret-key length is
\begin{equation}
 \begin{split}
 \ell=\max\biggl\{0,\biggl\lfloor n[1-h_2(k/n)]-L_{\rm EC}-v-1\\
 {}-2\log_2(1/\epsilon_{\rm PA})\biggr\rfloor\biggr\}.
 \end{split}
 \label{main:finite_length}
\end{equation}
The first term bounds the secret randomness after the parameter test and Alice's public intensity and phase disclosures. The remaining terms account for reconciliation, verification, one final acceptance bit and privacy amplification. The protocol aborts when $k\ge n/2$, the test or verification fails, or $\ell=0$.

The plotted designs use $v=36$ and $\epsilon_{\rm PA}=2\times10^{-11}$. With the remaining confidence and authentication allowances specified in Appendix~\ref{app:composable}, their total composable security parameter is below $6.46\times10^{-11}$. The expected rate follows from Eq.~\eqref{main:rate_convention}, using a lower bound on honest completion for the simulated apparatus in Table~\ref{tab:hardware}. The complete secrecy proof and finite implementation are given in Appendices~\ref{app:composable} and \ref{pdimpl:appendix}; generated key is reported before authentication-key replenishment.

\section{Conclusion}
CV-TF-QKD combines independent Gaussian-modulated coherent transmitters with single-photon interference at an untrusted relay. Full-record reconciliation and binary phase disclosure yield nearly equal IID asymptotic rates, with $\sqrt\eta$ scaling at fixed weak input-referred excess noise. This identifies an optical route to the TF advantage that preserves Gaussian preparation at both users and places the non-Gaussian measurement at the relay.

For phase disclosure, a bounded preparation-intensity test certifies the virtual parity correlations needed to protect the binary sign. This gives a direct composable key-length bound against general coherent attacks and links the asymptotic advantage to a finite processing procedure. Under the specified apparatus conditions, one- and two-day calculations predict rates beyond both pure-loss and matched thermal-loss repeaterless bounds, including detector loss, dark events, phase fluctuations and reference overhead. These predictions define an experimental target. Their quantitative realization requires calibrated source and detector responses and a benchmarked reconciliation implementation; the acquisition times do not include decoding.

The same optical architecture also offers scope for further optimization of the classical processing. Choices of public disclosure, data selection and reconciliation may be refined jointly to make fuller use of the available correlations and tailored to channel conditions and practical decoding constraints. Appendix~\ref{app:opd} illustrates this possibility with an optimized phase-disclosure variant and evaluates its rate advantage under IID assumptions. Exploring these choices more broadly, and assessing the resulting protocols within a composable finite-key treatment, are natural directions for improving the practical performance of CV-TF-QKD.

The broader implication is that Gaussian-modulated continuous-variable cryptography can access the TF loss advantage while retaining finite-key security against unrestricted attacks. This opens a path to long-distance CV quantum communication with simple transmitters and an untrusted non-Gaussian relay.

\smallskip
\textbf{Acknowledgments.~}
Under the author's direction, OpenAI's ChatGPT (GPT-6 Astra) and Anthropic's Claude (Opus 5.5) were used to generate and develop candidate mathematical derivations and proofs and to improve the manuscript's presentation. The author independently verified the AI-assisted mathematical content, revised it where necessary, and takes full responsibility for the accuracy and scientific content of the work.

\appendix
\section{IID collective-attack rates and conditional optical states}
\label{coll:appendix}

This appendix derives the two IID rates from the optical state and the actual public announcements. The calculation has three parts: identify the accepted state of two virtual retained modes, calculate the decoder's classical information, and subtract Eve's quantum information. Multiplication by the relay acceptance probability converts the secret fraction per accepted event into a rate per attempted relay use. CV-TF-Basic and CV-TF-PD share the optical state but use different key variables and disclosures, so their information terms differ.

A collective attack acts identically and independently on successive source pairs; Eve may retain all her outputs and measure them jointly after public communication. We evaluate the rates when complete asymptotic accepted preparation statistics identify the state associated with the stated thermal-loss model. Estimating only transmissivity, excess noise, or a covariance matrix does not identify this non-Gaussian postselected state. The source is trusted, but neither the relay nor a physical implementation of the thermal channel is trusted. This appendix establishes IID rates under that statistical identification; it does not reduce coherent attacks to collective attacks.

\subsection{States, source replacement, and certification}
For a bosonic mode, $\ket j$ denotes the state of $j$ photons and $\hat a\ket j=\sqrt j\ket{j-1}$. Coherent and thermal states are
\begin{equation}
 \begin{aligned}
 \ket z&=e^{-|z|^2/2}\sum_{j=0}^{\infty}\frac{z^j}{\sqrt{j!}}\ket j,\\
 \tau_\nu&=\sum_{j=0}^{\infty}\frac{\nu^j}{(1+\nu)^{j+1}}\ket j\bra j.
 \end{aligned}
 \label{coll:states}
\end{equation}
The number operator is $\hat n=\hat a^\dagger\hat a$, with $\hat n\ket j=j\ket j$ and $[\hat a,\hat a^\dagger]=1$. The coherent state obeys $\hat a\ket z=z\ket z$; its photon-number probabilities are Poisson with mean $|z|^2$. The thermal state is diagonal with geometric photon-number probabilities and mean $\nu$. Thus a coherent amplitude is a complex field coordinate, whereas $\nu$ is an average occupation, not an amplitude. All information logarithms are base two. We use $S(\rho)=-\Tr\rho\log_2\rho$, $H(\mathbf p)=-\sum_jp_j\log_2p_j$, $[x]_+=\max\{x,0\}$, and
\begin{equation}
 \begin{aligned}
 h_2(p)&=-p\log_2p-(1-p)\log_2(1-p),\\
 g(v)&=(v+1)\log_2(v+1)-v\log_2v,
 \end{aligned}
 \label{coll:entropies}
\end{equation}
with $0\log_2 0=0$. In particular, $S(\tau_\nu)=g(\nu)$: inserting the geometric probabilities into $H$ gives this expression using their normalization and mean $\nu$. For a general state, $S$ is the Shannon entropy of its eigenvalues, rather than of arbitrary diagonal matrix entries.

Each transmitter prepares an independent circular-Gaussian coherent amplitude with density $p_N(\alpha)=e^{-|\alpha|^2/N}/(\pi N)$ and mean photon number $N>0$. For source replacement, set $q=N/(1+N)$ and use
\begin{equation}
 \ket{\Psi_N}_{Aa}=\sqrt{1-q}\sum_{j=0}^{\infty}q^{j/2}\ket j_A\ket j_a.
 \label{coll:source}
\end{equation}
Here uppercase $A$ labels a virtual retained mode and lowercase $a$ the physical traveling mode; Bob has an independent copy $Bb$. The heterodyne measurement has effect density $\ket{a_0}\bra{a_0}/\pi$ with respect to $d^2a_0=d\operatorname{Re}a_0\,d\operatorname{Im}a_0$. The resolution of the identity and the associated Husimi probability density of a state $\omega$ are
\begin{equation}
 \begin{aligned}
 \int_{\mathbb C}\frac{d^2a_0}{\pi}\ket{a_0}\bra{a_0}&=I,\\
 Q_\omega(a_0)&=\frac{\bra{a_0}\omega\ket{a_0}}{\pi}.
 \end{aligned}
\end{equation}
Unlike a photon-number measurement, heterodyne detection has a complex-valued outcome. Projecting $A$ onto a coherent state with outcome $a_0$ gives
\begin{equation}
 {}_A\!\bra{a_0}\Psi_N\rangle_{Aa}
 \propto\sum_{j=0}^{\infty}
 \frac{(\sqrt q\,a_0^*)^j}{\sqrt{j!}}\ket j_a
 \propto\ket{\sqrt q\,a_0^*}_a.
 \label{coll:source_heterodyne}
\end{equation}
Thus the physical preparation record is $\alpha=\sqrt q\,a_0^*$. Tracing either mode out of Eq.~\eqref{coll:source} gives the thermal probabilities $(1-q)q^j$ and mean $q/(1-q)=N$. Consequently the heterodyne outcome density and the coordinate Jacobian are
\begin{equation}
 \begin{aligned}
 p_{\rm het}(a_0)&=\frac{1-q}{\pi}e^{-(1-q)|a_0|^2},\\
 d^2\alpha&=q\,d^2a_0.
 \end{aligned}
\end{equation}
Dividing this density by $q$ and inserting $|a_0|^2=|\alpha|^2/q$ gives exactly $p_N(\alpha)$. The virtual construction therefore reproduces both the emitted coherent states and their preparation probabilities. The virtual measurement can be deferred because it commutes with the relay operation; no squeezed source or quantum memory is needed in the implementation.

The relay accepts one photon in exactly one output port. Its public sign is $s=\pm1$, and Bob replaces his emitted record by $\beta=s\beta_{\rm em}$. Hence $\beta$ and $b_0=\beta^*/\sqrt q$ already include the port correction. Let $p(\alpha,\beta\mid s)$ be the accepted corrected preparation distribution. The retained two-mode Husimi density is
\begin{equation}
 Q_s(a_0,b_0)=q^2p(\sqrt q\,a_0^*,\sqrt q\,b_0^*\mid s).
 \label{coll:tomography}
\end{equation}
The factor $q^2$ is the Jacobian for two complex coordinates. To see how this probability density identifies a quantum state, write $\omega_{jk}=\bra j\omega\ket k$ for one mode. The coherent-state expansion gives
\begin{equation}
 \pi e^{|a_0|^2}Q_\omega(a_0)
 =\sum_{j,k\geq0}\frac{\omega_{jk}(a_0^*)^j a_0^k}{\sqrt{j!k!}}.
\end{equation}
Its power-series coefficients recover every matrix element. The two-mode version follows by using both coordinates and multiplying $Q_s$ by $\pi^2e^{|a_0|^2+|b_0|^2}$. Thus the full Husimi function is informationally complete; knowledge of its full shape, including its tails, fixes the retained state. Complete asymptotic accepted statistics, obtainable by revealing randomly selected preparation records, identify $\rho_{AB\mid s}$. A covariance matrix gives only a few moments of this function and cannot replace it after photon-counting postselection.

We give Eve a purification of the identified state; this includes any purification of the thermal environment and is conservative if Eve holds less information. All purifications are related by an isometry on the purifying system and yield the same conditional entropies. Eve need not implement a thermal-loss channel: any collective attack producing the identified accepted state is covered. The identification here is an ideal infinite-data condition: exact agreement of the full accepted distribution is stronger than fitting a few channel parameters. It cannot by itself certify a state from finitely many observed records. A practical finite test need not perform this full tomography; the finite-key proof uses a separate, explicitly bounded test.

For clarity, let $X$ be the owner's raw symbol, $Y$ the decoder's private record, and $C$ the public data before reconciliation. For a discrete public variable, Eve's conditional Holevo information is
\begin{equation}
 \begin{split}
 \chi(X:E\mid C)=\sum_c p(c)\biggl[
 S(\rho_{E\mid c})\\
 {}-\sum_xp(x\mid c)S(\rho_{E\mid x,c})\biggr],
 \end{split}
 \label{coll:Holevo_definition}
\end{equation}
with sums replaced by integrals for continuous records. Here $\rho_{E\mid c}=\sum_xp(x\mid c)\rho_{E\mid x,c}$ is formed before its entropy is evaluated. For each $c$, the state containing the classical symbol and Eve's system is $\sum_xp(x\mid c)\ket x\bra x\otimes\rho_{E\mid x,c}$. Its block-diagonal form gives the conditional entropy $H(X\mid E,C)=H(X\mid C)-\chi(X:E\mid C)$. This is the private randomness available before reconciliation. For a discrete raw key, ideal one-way reconciliation discloses $H(X\mid Y,C)$ bits per accepted event, which must be subtracted before privacy amplification. Their difference is $I(X:Y\mid C)-\chi(X:E\mid C)$~\cite{DW}. This gives the common rate rule: acceptance probability times the nonnegative information difference. The continuous-record CV-TF-Basic expression is the corresponding ideal fine-quantization benchmark. Both information terms must use the same raw variable and transcript: CV-TF-Basic uses Bob's record and the public port; CV-TF-PD uses Alice's sign and additionally her disclosed intensity and folded phase.

\subsection{Thermal loss and the accepted retained state}
The optical prediction uses independent, identical thermal-loss arms, each with transmissivity $0<t<1$, so the end-to-end transmissivity is $\eta=t^2$. Their environment modes are independent thermal states of the same mean photon number $\bar n$. Equality of the two single-arm covariance matrices alone would not imply this independence. Let $V_{\rm in}$ and $V_{\rm out}$ denote the input and output quadrature covariance matrices, and let $I_2$ be the two-dimensional identity. In units where vacuum quadrature variance is one, an arm with bath photon number $\bar n$ acts as
\begin{equation}
 V_{\rm out}=tV_{\rm in}+(1-t)(2\bar n+1)I_2.
 \label{coll:channel}
\end{equation}
Define the input-referred excess noise $\xi=2(1-t)\bar n/t$, the output thermal contribution $w$, and the received mean photon number $m_{\rm ch}$ by
\begin{equation}
 \begin{aligned}
 w&=t\xi/2,\\
 m_{\rm ch}&=tN+w,\\
 \lambda&=\frac{tN}{m_{\rm ch}(1+w)},\\
 \nu&=\frac{N(1-t+w)}{1+m_{\rm ch}}.
 \end{aligned}
 \label{coll:parameters}
\end{equation}
Averaging the coherent preparation gives $V_{\rm in}=(2N+1)I_2$; Eq.~\eqref{coll:channel} then gives $V_{\rm out}=(2m_{\rm ch}+1)I_2$, with $w=(1-t)\bar n$. The parameter $\nu$ is the residual thermal occupation in a retained mode after a vacuum projection at the receiver. The weight $\lambda$ quantifies the photon-added component after a one-photon projection. Both will enter the non-Gaussian accepted state. Ideal detection, a stable common phase reference, and no dark counts are assumed in this appendix. The probability of either accepted output is
\begin{equation}
 p_{\rm acc}=\frac{2m_{\rm ch}}{(1+m_{\rm ch})^3}.
 \label{coll:acceptance}
\end{equation}
Indeed, each received mode is thermal with vacuum probability $(1+m_{\rm ch})^{-1}$ and one-photon probability $m_{\rm ch}(1+m_{\rm ch})^{-2}$. It is the product of the two thermal density operators that is invariant under the balanced beam splitter. Explicitly, putting $\zeta=m_{\rm ch}/(1+m_{\rm ch})$, that product is $(1-\zeta)^2\zeta^{\hat n_1+\hat n_2}$. A passive beam splitter preserves the total number $\hat n_1+\hat n_2$, so its outputs remain independent thermal states with the same occupation. Each outcome $(1,0)$ or $(0,1)$ therefore has probability $m_{\rm ch}/(1+m_{\rm ch})^3$, and their sum gives Eq.~\eqref{coll:acceptance}.

Write $\tau=\tau_\nu\otimes\tau_\nu$, and let $\hat a,\hat b$ act on the retained Alice and Bob modes. After the port correction, the accepted retained state is
\begin{equation}
 \rho_{\lambda,\nu}=(1-\lambda)\tau+
 \frac{\lambda}{2(1+\nu)}(\hat a^\dagger+\hat b^\dagger)
 \tau(\hat a+\hat b).
 \label{coll:rho}
\end{equation}
Let $a_0$ and $c_0$ be heterodyne coordinates of the retained and received modes, respectively. The zero-mean complex Gaussian variable $(a_0,c_0^*)$ has covariance
\begin{equation}
 G=\begin{pmatrix}N+1&\sqrt{tN(N+1)}\\
 \sqrt{tN(N+1)}&1+m_{\rm ch}\end{pmatrix}.
\end{equation}
The diagonal entries are heterodyne variances, and the off-diagonal entry is the source correlation reduced by $\sqrt t$. Inverting this matrix gives the joint Husimi density $e^{-W/D}/(\pi^2D)$, where
\begin{equation}
 \begin{aligned}
 D&=(N+1)(1+w),\\
 c&=\sqrt{tN(N+1)},\\
 W&=(1+m_{\rm ch})|a_0|^2+(N+1)|c_0|^2\\
  &{}-c(a_0c_0+a_0^*c_0^*).
 \end{aligned}
 \label{coll:joint_husimi}
\end{equation}
Indeed, $\det G=(N+1)(1+m_{\rm ch})-tN(N+1)=D$. The two projections can be evaluated directly from this Gaussian density. Setting $c_0=0$ selects the received vacuum and gives a retained Husimi function proportional to $\exp[-|a_0|^2/(1+\nu)]$, because $1+\nu=D/(1+m_{\rm ch})$. Its normalized state is $\tau_\nu$. To select one photon, recall that the coherent-state expansion contains $|c_0|^2\bra1\rho\ket1$ after multiplication by $e^{|c_0|^2}$. Extracting the $c_0c_0^*$ coefficient therefore implements precisely the one-photon projection. To perform the extraction explicitly, denote the joint density by $Q(a_0,c_0)$ and write
\begin{equation}
 \begin{split}
 e^{|c_0|^2}Q(a_0,c_0)=Q(a_0,0)\exp\biggl[
 \left(1-\frac{N+1}{D}\right)|c_0|^2\\
 {}+\frac{c}{D}(a_0c_0+a_0^*c_0^*)\biggr].
 \end{split}
\end{equation}
The desired coefficient receives one contribution from the quadratic term of the exponent and another from the product of its two linear terms. The factor multiplying the same retained Gaussian is
\begin{equation}
 1-\frac{N+1}{D}+\frac{c^2|a_0|^2}{D^2}.
 \label{coll:one_photon_coefficient}
\end{equation}
The constant term is thermal. The term proportional to $|a_0|^2$ is photon-added thermal because $\bra{a_0}\hat a^\dagger\tau_\nu\hat a\ket{a_0}=|a_0|^2\bra{a_0}\tau_\nu\ket{a_0}$. Define $\tau_\nu^+=\hat a^\dagger\tau_\nu\hat a/(1+\nu)$; the denominator is its trace before normalization. The trace follows from $\Tr(\hat a^\dagger\tau_\nu\hat a)=\Tr[\tau_\nu(\hat n+1)]=1+\nu$. In Eq.~\eqref{coll:one_photon_coefficient}, averaging over the normalized retained thermal Husimi density uses $\mathbb E|a_0|^2=1+\nu$. The integrated constant and quadratic contributions are $w/(1+w)$ and $tN/[(1+w)(1+m_{\rm ch})]$, respectively. Their sum is $m_{\rm ch}/(1+m_{\rm ch})$, so the normalized quadratic weight is exactly $\lambda$ in Eq.~\eqref{coll:parameters}. The one-photon-conditioned state is consequently $(1-\lambda)\tau_\nu+\lambda\tau_\nu^+$.

Because the two source--channel pairs are independent and identical, the same balanced rotation on the received coordinates and on the retained coordinates produces two independent collective pairs with the same covariance $G$. The successful relay result is then one photon in one collective mode and vacuum in the other. Applying the two conditional states just derived, and transforming the retained coordinates back to Alice and Bob, proves Eq.~\eqref{coll:rho}.

In the collective retained coordinates $A_\pm=(A\pm B)/\sqrt2$, the same state is
\begin{equation}
 \rho_{A_+A_-}=[(1-\lambda)\tau_\nu+\lambda\tau_\nu^+]\otimes\tau_\nu.
 \label{coll:factorization}
\end{equation}
This rotation is only a calculation device. The users do not combine their retained systems. As a normalization check, the photon-added term in Eq.~\eqref{coll:rho} has trace $2(1+\nu)$ before division; its cross terms have zero trace because a thermal mode has zero mean amplitude. The physical parameters give $0\leq\lambda\leq1$ and $\nu\geq0$, so the state is positive and normalized. In the lossless, noiseless limit, $\lambda=1$ and $\nu=0$, and the retained state becomes $(\ket{10}+\ket{01})/\sqrt2$, as expected from a single-photon relay projection.

The same conditioning has a direct prepare-and-measure interpretation. Conditional on the emitted amplitudes, the corrected relay output displacements are $\gamma_\pm=\sqrt{t/2}(\alpha\pm\beta)$. Each output is a displaced thermal state with thermal occupation $w$. Its vacuum and one-photon probabilities are
\begin{equation}
 \begin{aligned}
 P_0(\gamma)&=\frac{e^{-|\gamma|^2/(1+w)}}{1+w},\\
 P_1(\gamma)&=P_0(\gamma)\left[
 \frac{w}{1+w}+\frac{|\gamma|^2}{(1+w)^2}\right].
 \end{aligned}
\end{equation}
These expressions follow by extracting the vacuum and one-photon coefficients of its displaced Gaussian Husimi function, by the same method used above. At $w=0$ they reduce to the Poisson probabilities of a coherent state. Independence of the two thermal noise modes makes the probability of the designated accepted port $P_1(\gamma_+)P_0(\gamma_-)$. Bayes' rule therefore gives
\begin{equation}
 p(\alpha,\beta\mid s)=
 \frac{p_N(\alpha)p_N(\beta)P_1(\gamma_+)P_0(\gamma_-)}{p_{\rm acc}/2}.
\end{equation}
The denominator is the probability of this particular port; the correction makes the expression identical for both values of $s$. This also explains physically why acceptance correlates otherwise independent preparations: the one-photon likelihood contains the interference intensity $|\alpha+\beta|^2$. Averaging the likelihood reproduces the acceptance probability, while its dependence on the records agrees with the retained-state construction.

\subsection{CV-TF-Basic full-record mutual information}
The accepted classical density follows by heterodyning Eq.~\eqref{coll:rho}. The thermal factor gives
$e^{-(|a_0|^2+|b_0|^2)/(1+\nu)}/[\pi^2(1+\nu)^2]$, while the photon-added factor multiplies it by $|a_0+b_0|^2/[2(1+\nu)]$. Put $\kappa=(1+m_{\rm ch})/[N(1+w)]$ and define normalized corrected records $z_A=\sqrt\kappa\,\alpha$ and $z_B=\sqrt\kappa\,\beta$. Since $\kappa q=1/(1+\nu)$, these are $a_0^*/\sqrt{1+\nu}$ and $b_0^*/\sqrt{1+\nu}$. The Jacobian cancels the thermal prefactor and gives the normalized density
\begin{equation}
 f_\lambda(z_A,z_B)=\frac{e^{-|z_A|^2-|z_B|^2}}{\pi^2}
 \left(1-\lambda+\frac\lambda2|z_A+z_B|^2\right).
 \label{coll:record_density}
\end{equation}
For continuous records the mutual information is the relative entropy of the joint density and the product of its marginals:
\begin{equation}
 I=\int_{\mathbb C^2}d^2z_A\,d^2z_B\,
 f_\lambda\log_2\frac{f_\lambda}{f_Af_B}.
\end{equation}
An invertible local rescaling does not change this quantity: its Jacobians cancel in the density ratio and in the probability measure. Define
\begin{equation}
 \begin{aligned}
 F_\lambda(x)&=1-\lambda+\lambda x,\\
 Z_\lambda(x)&=1-\lambda/2+\lambda x/2.
 \end{aligned}
 \label{coll:FZ}
\end{equation}
To obtain either local marginal, integrate over the other complex record: its cross term vanishes by rotational symmetry, and its Gaussian mean squared magnitude is one. The result is $e^{-|z|^2}Z_\lambda(|z|^2)/\pi$. Both radial weights are normalized: $\int_0^\infty e^{-x}F_\lambda(x)dx=\int_0^\infty e^{-x}Z_\lambda(x)dx=1$.

Let $z_\pm=(z_A\pm z_B)/\sqrt2$. This unit-Jacobian change of variables factorizes the joint density into $e^{-|z_-|^2}/\pi$ and $e^{-|z_+|^2}F_\lambda(|z_+|^2)/\pi$. More directly, the density ratio entering the mutual information is
\begin{equation}
 \frac{f_\lambda(z_A,z_B)}{f_A(z_A)f_B(z_B)}
 =\frac{F_\lambda(|z_+|^2)}{Z_\lambda(|z_A|^2)Z_\lambda(|z_B|^2)}.
 \label{coll:information_ratio}
\end{equation}
The Gaussian exponentials and factors of $\pi$ cancel pointwise. Averaging the logarithm leaves one joint expectation of $\log_2F_\lambda$ and two identical marginal expectations of $\log_2Z_\lambda$. For a rotationally invariant complex coordinate, $d^2z=d(|z|^2)d\arg z/2$; the angular integral cancels its $\pi$ normalization. This gives
\begin{equation}
 \begin{split}
 I_{\rm full}(\lambda)
 &=\int_0^\infty e^{-x}F_\lambda(x)\log_2F_\lambda(x)\,dx\\
 &{}-2\int_0^\infty e^{-x}Z_\lambda(x)\log_2Z_\lambda(x)\,dx.
 \end{split}
 \label{coll:I_full}
\end{equation}
For $\lambda=0$, both weights are one, the two records are independent, and $I_{\rm full}=0$. The dependence on $\nu$ has disappeared from this classical information because it only sets the local amplitude scale; it will still enter the quantum entropy. At $\lambda=1$ this equals $[2\ln2-\gamma_E-\mathrm e E_1(1)]/\ln2\simeq0.30690644$, where $\gamma_E$ is Euler's constant and $E_1(x)=\int_x^\infty e^{-u}\,du/u$.

\subsection{CV-TF-Basic full-record entropy and rate}
Let $x=\nu/(1+\nu)$. The spectrum $\mathbf r=(r_0,r_1,\ldots)$ of the first factor in Eq.~\eqref{coll:factorization} is
\begin{equation}
 \begin{aligned}
 r_0&=(1-\lambda)(1-x),\\
 r_j&=(1-\lambda)(1-x)x^j
       +\lambda j(1-x)^2x^{j-1},~ j\ge1.
 \end{aligned}
 \label{coll:spectrum}
\end{equation}
For clarity, $\hat a^\dagger\tau_\nu\hat a$ has $j$th diagonal entry $j(1-x)x^{j-1}$ for $j\ge1$ and zero vacuum entry. Division by $1+\nu=1/(1-x)$ produces the second term in Eq.~\eqref{coll:spectrum}. The photon-added probabilities sum to one because $\sum_{j\geq1}jx^{j-1}=(1-x)^{-2}$; the thermal probabilities also sum to one. Hence $\mathbf r$ is a convex mixture of two normalized spectra. At $\nu=0$ its continuous limit gives $r_0=1-\lambda$, $r_1=\lambda$, and all other entries zero. The collective-mode unitary changes no entropy, and the two factors in Eq.~\eqref{coll:factorization} are independent. Consequently, $S(\rho_{\lambda,\nu})=H(\mathbf r)+g(\nu)$.

Bob owns the full corrected record and Alice reconstructs it. Phase covariance lets Bob's virtual heterodyne amplitude be taken real when computing entropy. Write $r=|b_0|^2/(1+\nu)$ and $b_r=\sqrt{(1+\nu)r}$. For a complex outcome $b_0$, the thermal projection factor is
\begin{equation}
 \gamma_{b_0}=\bra{b_0}\tau_\nu\ket{b_0}
 =\frac{e^{-|b_0|^2/(1+\nu)}}{1+\nu}.
\end{equation}
Using $\hat b\ket{b_0}=b_0\ket{b_0}$ on the right and its adjoint on the left gives the unnormalized contraction
\begin{equation}
 \begin{split}
 {}_B\!\bra{b_0}\rho_{\lambda,\nu}\ket{b_0}_B
 =\gamma_{b_0}\biggl[(1-\lambda)\tau_\nu\\
 {}+\frac{\lambda}{2(1+\nu)}
 (\hat a^\dagger+b_0^*)\tau_\nu(\hat a+b_0)\biggr].
 \end{split}
\end{equation}
Changing the phase of $b_0$ produces a unitary phase rotation of Alice's conditional state, which leaves its eigenvalues unchanged. We may therefore set $b_0=b_r$ for the entropy calculation without discarding any classical phase information. Its common thermal projection factor cancels upon normalization; the remaining trace is $Z_\lambda(r)$. Alice's conditional state and its average entropy are
\begin{equation}
 \sigma_A(r)=
 \frac{(1-\lambda)\tau_\nu+
 \frac{\lambda}{2(1+\nu)}(\hat a^\dagger+b_r)\tau_\nu(\hat a+b_r)}
 {Z_\lambda(r)},
 \label{coll:conditional}
\end{equation}
\begin{equation}
 \mathcal C(\nu,\lambda)=\int_0^\infty e^{-r}Z_\lambda(r)S[\sigma_A(r)]\,dr.
 \label{coll:conditional_average}
\end{equation}
To evaluate this entropy, put $p_n=(1-x)x^n$, $p_{-1}=0$, and $a_\lambda=\lambda/[2(1+\nu)]$. The only nonzero entries are
\begin{equation}
 \begin{aligned}
 \bigl[\sigma_A(r)\bigr]_{nn}
 &=\frac{(1-\lambda+a_\lambda b_r^2)p_n+a_\lambda n p_{n-1}}
 {Z_\lambda(r)},\\
 \bigl[\sigma_A(r)\bigr]_{n+1,n}
 &=\frac{a_\lambda b_r\sqrt{n+1}\,p_n}{Z_\lambda(r)},
 \end{aligned}
 \label{coll:tridiagonal}
\end{equation}
with the conjugate off-diagonal entry fixed by Hermiticity. These entries follow by expanding $(\hat a^\dagger+b_r)\tau_\nu(\hat a+b_r)$. Its $\hat a^\dagger\tau_\nu\hat a$ term contributes $np_{n-1}$ on the diagonal; the two terms linear in $b_r$ connect only adjacent photon numbers. The numerator's trace is $1-\lambda+a_\lambda(1+\nu+b_r^2)=Z_\lambda(r)$, which checks normalization. The radial probability weight in Eq.~\eqref{coll:conditional_average} is Bob's marginal derived above, not the original unconditioned Gaussian preparation density.

For a purification $ABE$, a rank-one heterodyne outcome on $B$ leaves $AE$ pure. The two reduced states of a pure bipartite state have the same nonzero spectrum, so $S(E)=S(AB)$ before Bob's measurement and $S(E\mid b_0)=S(A\mid b_0)$ after it. This avoids constructing Eve's purification explicitly. Averaging the second equality over Bob's outcomes gives
\begin{equation}
 \chi_{\rm full}=H(\mathbf r)+g(\nu)-\mathcal C(\nu,\lambda).
 \label{coll:Basic_Holevo}
\end{equation}
Subtracting this from the decoder's mutual information gives the achievable rate per attempted relay use
\begin{equation}
 \begin{split}
 R_{\rm Basic}^{\rm iid}=p_{\rm acc}
 \bigl[I_{\rm full}(\lambda)-H(\mathbf r)-g(\nu)
 +\mathcal C(\nu,\lambda)\bigr]_+.
 \end{split}
 \label{coll:basic_rate}
\end{equation}
Both accepted ports are included in $p_{\rm acc}$.

The plotted CV-TF-Basic rate evaluates the continuous-record expression directly. For a specified finite quantizer $\mathcal Q$, including an overflow symbol, let $X=\mathcal Q(\beta)$. Ideal one-way reconciliation instead gives
\begin{equation}
 R_{\mathcal Q}^{\rm iid}=p_{\rm acc}
 [I(X:\alpha\mid s)-\chi(X:E\mid s)]_+,
 \label{coll:quantized_rate}
\end{equation}
where $E$ is Eve's quantum system and the information terms are conditioned on the public relay flag $s$. Both terms must be evaluated for the chosen quantization cells. If $\Omega_x$ is the region assigned symbol $x$, its probability and Eve's conditional state obey
\begin{equation}
 \begin{aligned}
 p(x\mid s)&=\int_{\Omega_x}d^2\beta\,p(\beta\mid s),\\
 \rho_{E\mid x,s}&=\frac{1}{p(x\mid s)}
 \int_{\Omega_x}d^2\beta\,p(\beta\mid s)\rho_{E\mid\beta,s}.
 \end{aligned}
\end{equation}
The analogous integral of $p(\beta\mid\alpha,s)$ determines the decoder's symbol likelihood. A finite cell combines many heterodyne outcomes, so Eve's states must be mixed before evaluating their entropy. In particular, the rank-one conditional-entropy identity used for a single outcome cannot simply be applied to a cell. Overflow regions are part of these integrals, not discarded events. The continuous curve is an ideal fine-quantization benchmark and does not specify a finite quantizer or a reconciliation code. The numerical photon-number cutoff used to evaluate the retained state's spectrum is unrelated to this classical quantization.
\subsection{Alice's phase disclosure and Bob's sign likelihood}
CV-TF-PD uses Alice's sign $Z\in\{0,1\}$ as the raw key. Write her preparation record as $\alpha=(-1)^Z\sqrt A e^{i\theta_A}$, where $A=|\alpha|^2$ and $\theta_A\in[0,\pi)$. The folded phase $\theta_A$ is the ordinary phase reduced modulo $\pi$: the two antipodal amplitudes $+\sqrt A e^{i\theta_A}$ and $-\sqrt A e^{i\theta_A}$ share it, and $Z$ distinguishes them. Only $A$ and $\theta_A$ are disclosed on a key round. Bob keeps his complete corrected record $\beta$ private. Define
\begin{equation}
 \begin{aligned}
 s_0&=\frac{N(1+w)}{1+m_{\rm ch}}=q(1+\nu),\\
 u&=\frac{A}{s_0}, v=\frac{|\beta|^2}{s_0}.
 \end{aligned}
 \label{coll:scaled_intensities}
\end{equation}
Thus $u$ and $v$ are dimensionless squared magnitudes of the normalized records in Eq.~\eqref{coll:record_density}. For Bob's local calculation, let $\psi=\arg\beta-\theta_A$ modulo $2\pi$. This relative angle is available to Bob because he knows both $\beta$ and the announced $\theta_A$; it is not an additional announcement. Summing over Alice's two signs and integrating the phases gives the accepted intensity density
\begin{equation}
 \begin{aligned}
 f_{uv}(u,v)&=e^{-u-v}W_\lambda(u+v),\\
 W_\lambda(T)&=1-\lambda+\lambda T/2,\\
 V_\lambda(u,v)&=\frac{2\lambda\sqrt{uv}}{2(1-\lambda)+\lambda(u+v)}.
 \end{aligned}
 \label{coll:intensity_density}
\end{equation}
To see both the weights and the posterior, substitute $z_A=(-1)^Z\sqrt u e^{i\theta_A}$ and $z_B=\sqrt v e^{i(\theta_A+\psi)}$ into Eq.~\eqref{coll:record_density}. The factor in parentheses becomes $W_\lambda(u+v)+(-1)^Z\lambda\sqrt{uv}\cos\psi$. The polar-coordinate measure supplies a factor $1/2$ for each complex record: $d^2z=d(|z|^2)d\arg z/2$. On either fixed-sign branch, the change from the two ordinary phases to $(\theta_A,\psi)$ has unit Jacobian. Therefore
\begin{equation}
 d^2z_A\,d^2z_B=\frac14\,du\,dv\,d\theta_A\,d\psi.
\end{equation}
The density is independent of the common folded phase, whose range has length $\pi$. Its integration changes the prefactor $1/(4\pi^2)$ into $1/(4\pi)$. Thus the joint density of $u,v,\psi,Z=z$, after integrating the common folded phase, is
\begin{equation}
 \frac{e^{-u-v}}{4\pi}
 \left[W_\lambda(u+v)+(-1)^z\lambda\sqrt{uv}\cos\psi\right].
 \label{coll:sign_joint_density}
\end{equation}
Summing over the signs cancels the interference term. The remaining joint density is $f_{uv}(u,v)/(2\pi)$, which proves Eq.~\eqref{coll:intensity_density} and shows that the sign-averaged $\psi$ is uniform on $[0,2\pi)$ independently of $(u,v)$. Normalization follows from
\begin{equation}
 \int_0^\infty\!\int_0^\infty
 e^{-u-v}\left[1-\lambda+\frac\lambda2(u+v)\right]du\,dv=1.
\end{equation}
Dividing an individual sign's density by the sum over both signs gives Bob's posterior
\begin{equation}
 \Pr(Z=z\mid u,\theta_A,\beta)
 =\frac{1+(-1)^zV_\lambda(u,v)\cos\psi}{2}.
 \label{coll:class_density}
\end{equation}
The contrast $V_\lambda$ lies between zero and one: $2\sqrt{uv}\le u+v$ bounds its numerator by its denominator for $0\le\lambda\le1$. The posterior is therefore a valid binary probability. Bob can compute it without announcing $v$ or $\psi$. The isolated zero-denominator point at $\lambda=1$, $u=v=0$ has probability zero and may be assigned any posterior. Integrating over Bob's private record gives Alice's public intensity density
\begin{equation}
 \begin{aligned}
 f_u(u)&=e^{-u}Z_\lambda(u),\\
 Z_\lambda(u)&=1-\lambda/2+\lambda u/2.
 \end{aligned}
 \label{coll:alice_marginal}
\end{equation}
Indeed, integrating $W_\lambda(u+v)$ against $e^{-v}dv$ replaces $v$ by its exponential mean one. The interference term integrates to zero over $\psi$, separately for either sign. Each sign therefore has half of the public-label density: Alice's sign is uniform for every disclosed intensity and folded phase, although Bob's private record generally makes its posterior nonuniform.

\subsection{Sign mutual information for CV-TF-PD}
The mutual information is the reduction of Alice's binary uncertainty when Bob uses his private record. It is useful to summarize that record by its reliability, the absolute difference between the two probabilities in Eq.~\eqref{coll:class_density}:
\begin{equation}
 \mathcal R=V_\lambda(u,v)|\cos\psi|.
\end{equation}
Reliability zero means equal posterior probabilities, whereas reliability one means a certain sign. It has the representation
\begin{equation}
 \begin{aligned}
 \mathcal R&=\frac{\lambda T}{2(1-\lambda)+\lambda T}U,\\
 U&\sim\operatorname{Unif}[0,1],
 \end{aligned}
 \label{coll:reliability}
\end{equation}
where $T=u+v$ has density $e^{-T}T W_\lambda(T)$ and $U$ is independent of $T$. To see this, change variables from $(u,v)$ to $(T,x)$ with $x=u/T$. The inverse transformation is $u=Tx$, $v=T(1-x)$, with Jacobian $T$. Thus
\begin{equation}
 f_{T,x}(T,x)=e^{-T}T W_\lambda(T)
\end{equation}
for $T\ge0$ and $0\le x\le1$. The absence of $x$ makes it uniform and independent of $T$. The sign-averaged relative phase $\psi$ is also uniform and independent. These two uniform variables have a useful geometric interpretation. Put $\cos\vartheta=2x-1$. Uniformity of $\cos\vartheta$ and of the azimuth $\psi$ gives the uniform surface measure on a unit sphere, so
\begin{equation}
 (\sin\vartheta\cos\psi,\sin\vartheta\sin\psi,\cos\vartheta)
\end{equation}
is a uniformly distributed unit vector. By rotational symmetry, each Cartesian component has the same uniform distribution on $[-1,1]$ as $\cos\vartheta$. Taking the absolute value of its first component gives
$U=2\sqrt{x(1-x)}|\cos\psi|$, uniform on $[0,1]$ and independent of $T$. This gives Eq.~\eqref{coll:reliability} without a phase-window approximation.
For $0\le r<1$ and $0<\lambda\le1$, define $a(T)=\lambda T/[2(1-\lambda)+\lambda T]$. Conditional on $T$, the variable $\mathcal R=a(T)U$ is uniform between zero and $a(T)$. Its probability of exceeding $r$ is therefore $[1-r/a(T)]_+$. A contribution is possible only when $a(T)>r$, equivalently $T>T_r$ with $T_r=2(1-\lambda)r/[\lambda(1-r)]$. Averaging this conditional probability against $f_T(T)=e^{-T}T W_\lambda(T)$, and using $a(T)=\lambda T/[2W_\lambda(T)]$, gives the survival probability
\begin{equation}
 \Pr(\mathcal R>r)=\int_{T_r}^{\infty}e^{-T}
 \left[T W_\lambda(T)-\frac{2r}{\lambda}W_\lambda(T)^2\right]dT.
 \label{coll:survival_integral}
\end{equation}
For transparency, with $b=1-\lambda$ the bracket is
\begin{equation}
 \frac{\lambda(1-r)}2T^2+b(1-2r)T-\frac{2rb^2}{\lambda}.
\end{equation}
The required exponential tail integrals are
\begin{equation}
 \begin{aligned}
 \int_{T_r}^\infty e^{-T}dT&=e^{-T_r},\\
 \int_{T_r}^\infty T e^{-T}dT&=(T_r+1)e^{-T_r},\\
 \int_{T_r}^\infty T^2 e^{-T}dT&=(T_r^2+2T_r+2)e^{-T_r}.
 \end{aligned}
\end{equation}
Substitution cancels the terms depending polynomially on $T_r$ and reduces the result to
\begin{equation}
 \Pr(\mathcal R>r)=(1-\lambda r)
 \exp\!\left[-\frac{2(1-\lambda)r}{\lambda(1-r)}\right].
 \label{coll:survival}
\end{equation}
At reliability $r$, Bob's posterior binary entropy is $h_2((1-r)/2)$. Alice's bit is uniform given her public labels, so the information obtained from that observation is $1-h_2((1-r)/2)$. Write this information gain as $G(r)=1-h_2((1-r)/2)$. It satisfies $G(0)=0$ and
\begin{equation}
 G'(r)=\frac1{2\ln2}\ln\frac{1+r}{1-r}
       =\frac{\operatorname{atanh}(r)}{\ln2}.
\end{equation}
The average can be written directly in terms of a survival probability:
\begin{equation}
 \begin{aligned}
 \mathbb E[G(\mathcal R)]
 &=\mathbb E\!\left[\int_0^{\mathcal R}G'(r)dr\right]\\
 &=\int_0^1G'(r)\Pr(\mathcal R>r)dr.
 \end{aligned}
\end{equation}
This is integration by parts for a nonnegative random variable; the second equality also follows by interchanging two nonnegative integrals. It avoids differentiating the survival distribution. Substituting Eq.~\eqref{coll:survival} yields the mutual information available to Bob,
\begin{equation}
 \begin{split}
 I_{\rm sign}(\lambda)=\frac1{\ln2}\int_0^1
 (1-\lambda r)\exp\!\left[-\frac{2(1-\lambda)r}{\lambda(1-r)}\right]
 \operatorname{atanh}(r)\,dr.
 \end{split}
 \label{coll:I_sign}
\end{equation}
The limiting values are $I_{\rm sign}(0)=0$ and $I_{\rm sign}(1)=1-1/(2\ln2)\simeq0.27865248$. All accepted intensity and phase values contribute; there is no phase-window rejection.

\subsection{Eve's conditional states and the CV-TF-PD rate}
Let $\rho=\rho_{\lambda,\nu}$ be the accepted retained state in Eq.~\eqref{coll:rho}. Eve holds a purification of this state, so the information available to her is determined by the measurement events on the retained modes and the labels made public. An effect is the positive operator that determines an event's probability by the Born rule. For a continuous outcome, it is an effect density, and its trace against $\rho$ is a probability density rather than the probability of one exact outcome. For an effect or effect density $F$ satisfying $\Tr(\rho F)>0$, define
\begin{equation}
 \Gamma_\rho(F)=\frac{\sqrt\rho F\sqrt\rho}{\Tr(\rho F)}.
 \label{coll:Gamma}
\end{equation}
To see why this determines Eve's entropy, choose an orthonormal basis $\{\ket i\}$ for the two retained modes, an identical basis for a purifying system, and write $\ket{\Psi_\rho}=\sum_{ij}(\sqrt\rho)_{ij}\ket i_{AB}\ket j_E$. This vector is normalized because $\Tr\rho=1$, and tracing Eve gives $\rho$. Tracing the retained modes after effect $F$ leaves Eve with the unnormalized matrix $(\sqrt\rho F\sqrt\rho)^T$, where the transpose is in the chosen purification basis. Its trace is $\Tr(\rho F)$. Transposition preserves eigenvalues, so normalization gives entropy $S[\Gamma_\rho(F)]$. This construction avoids assigning Eve a particular measurement and includes purification of the thermal noise.

Coarse graining a measurement outcome means adding positive effects, with their probability weights still present. If a disclosed class contains outcomes with effects $F_j$, the unnormalized operator entering Eq.~\eqref{coll:Gamma} is $\sqrt\rho(\sum_jF_j)\sqrt\rho$. One must form this sum before normalizing the conditional state and evaluating its entropy. Averaging the entropies of the separately normalized states would give Eve the finer outcome labels as well. The same rule applies to continuous classes, with sums replaced by integrals.

By simultaneous phase covariance of the accepted state, Alice's announced phase can be set to zero when computing spectra. This is a change of optical coordinates, not a phase selection: a unitary change leaves every entropy unchanged. Her two retained-mode coherent amplitudes are then $\pm a_u$, where $a_u=\sqrt{(1+\nu)u}$. Define
\begin{equation}
 \begin{aligned}
 A_z(u)&=\ket{(-1)^z a_u}\bra{(-1)^z a_u},\\
 F_z(u)&=A_z(u)\otimes I_B,\\
 F_\Sigma(u)&=[A_0(u)+A_1(u)]\otimes I_B.
 \end{aligned}
 \label{coll:effects}
\end{equation}
Here $I_B$ is the identity on Bob's retained optical mode. It appears because Bob reveals neither his key-round intensity nor his phase. His unannounced heterodyne outcome is summed using $\int d^2b_0\,\ket{b_0}\bra{b_0}/\pi=I_B$, \emph{before} taking Eve's entropy. Bob still uses this outcome for decoding; keeping it private does not mean discarding it locally. Averaging entropies conditioned on his individual outcomes would instead give those outcomes to Eve and describe a different protocol. Likewise, $F_\Sigma$ adds Alice's two signs because the public pair $(A,\theta_A)$ does not reveal the sign. These are two distinct heterodyne alternatives whose effects are added; they are not projected onto a coherent superposition of $\ket{a_u}$ and $\ket{-a_u}$. Their identical Jacobians and heterodyne prefactors cancel in Eq.~\eqref{coll:Gamma}.

Simultaneous photon parity leaves $\rho$ and $F_\Sigma$ invariant and interchanges $F_0$ and $F_1$. Consequently $\Tr(\rho F_0)=\Tr(\rho F_1)$, so both key values are equiprobable, and their normalized states have the same spectrum. In particular,
\begin{equation}
 \Gamma_\rho(F_\Sigma)
 =\frac12\Gamma_\rho(F_0)+\frac12\Gamma_\rho(F_1).
\end{equation}
The Holevo information is the entropy of this mixture minus the average entropy of its two sign-conditioned states. Their equal entropies let us write it as a single difference. Define
\begin{equation}
 \begin{aligned}
 \chi_{\rm PD}(u)&=S[\Gamma_\rho(F_\Sigma(u))]
                 -S[\Gamma_\rho(F_0(u))],\\
 \overline\chi_{\rm PD}(\lambda,\nu)
   &=\int_0^\infty e^{-u}Z_\lambda(u)\chi_{\rm PD}(u)\,du.
 \end{aligned}
 \label{coll:average_Holevo}
\end{equation}
Because Alice's sign is uniform in each public class, its conditional entropy before Eve's information is included is one bit. Equation~\eqref{coll:average_Holevo} therefore gives $H(Z\mid E,C)=1-\overline\chi_{\rm PD}$, where $C$ denotes Alice's disclosed intensity, folded phase and the relay flag. Ideal reconciliation reveals $H(Z\mid Y,C)=1-I_{\rm sign}$ bits per accepted sign, with $Y$ Bob's private record. Subtracting this leakage gives
\begin{equation}
 R_{\rm PD}^{\rm iid}=p_{\rm acc}
 [I_{\rm sign}(\lambda)-\overline\chi_{\rm PD}(\lambda,\nu)]_+.
 \label{coll:simplified_rate}
\end{equation}
The positive part is taken after averaging all intensities. Placing it inside the integral would add intensity-dependent rejection and change the protocol.

\subsection{A small-matrix evaluation of the CV-TF-PD entropies}
The operator formula above is exact but appears to require a square root of a two-mode density operator. A Gram matrix replaces this task by matrices on Bob's single mode. Only eigenvalues are needed: the von Neumann entropy of a state is the Shannon entropy of its eigenvalues. We first state the matrices and then derive their connection to Eve's conditional states.

Let $x=\nu/(1+\nu)$, $p_n=(1-x)x^n$, $p_{-1}=0$, and $c=\lambda/[2(1+\nu)]$. At a fixed $u$, abbreviate $a=a_u$ and $Z=Z_\lambda(u)$. Here the scalar $Z$ is a normalization, distinct from Alice's binary sign, and the matrix $C$ below is distinct from the public transcript in conditional-information expressions. Define two real symmetric tridiagonal matrices $D$ and $C$, indexed by photon number $n\ge0$, by
\begin{equation}
 \begin{aligned}
 D_{nn}&=(1-\lambda+ca^2)p_n+cnp_{n-1},\\
 D_{n,n+1}&=ca\sqrt{n+1}\,p_n,\\
 C_{nn}&=(-1)^n[(1-\lambda-ca^2)p_n+cnp_{n-1}],\\
 C_{n,n+1}&=(-1)^{n+1}D_{n,n+1}.
 \end{aligned}
 \label{coll:PD_tridiagonal}
\end{equation}
The spectrum conditioned on Alice's full record is that of $D/Z$. The spectrum conditioned only on her public labels is the union of the spectra of
\begin{equation}
 \frac{D+e^{-2\nu u}C}{2Z},
 \frac{D-e^{-2\nu u}C}{2Z}.
 \label{coll:PD_public_blocks}
\end{equation}
Their combined trace is one; each block is generally subnormalized. Their eigenvalues are concatenated before computing the public conditional entropy, rather than normalizing the blocks separately.

Here is a derivation of these small matrices. Define a linear map $V$ from two copies of Bob's Hilbert space into the two-mode retained Hilbert space by
\begin{equation}
 V(\ket{\phi_0}\oplus\ket{\phi_1})
 =\sqrt\rho\left(\ket a\otimes\ket{\phi_0}
                    +\ket{-a}\otimes\ket{\phi_1}\right).
 \label{coll:Gram_map}
\end{equation}
The direct sum gives each sign alternative its own copy of Bob's space. Applying $V$ maps both copies into the common retained space, with the state weight $\sqrt\rho$ included. Then $VV^\dagger=\sqrt\rho F_\Sigma\sqrt\rho$, whereas $V^\dagger V$ is a two-by-two operator matrix with Bob-mode blocks $\langle(-1)^z a|\rho|(-1)^{z'}a\rangle_A$. The nonzero eigenvalues of these two products agree: if $V^\dagger V\ket\phi=\zeta\ket\phi$ with $\zeta>0$, then $V\ket\phi$ is an eigenvector of $VV^\dagger$ with eigenvalue $\zeta$. Thus one can compute Eve's spectrum from the smaller operator Gram matrix without constructing $\sqrt\rho$.

Let $P\ket n=(-1)^n\ket n$ be Bob's parity operator and let $\gamma=\langle a|\tau_\nu|a\rangle=e^{-u}/(1+\nu)$. For the positive amplitude, the coherent-state eigenvalue rule gives
\begin{equation}
 \begin{aligned}
 \gamma^{-1}\langle a|\rho|a\rangle_A
 &=(1-\lambda)\tau_\nu\\
 &{}+c(a+\hat b^\dagger)\tau_\nu(a+\hat b)\\
 &=D.
 \end{aligned}
\end{equation}
The term $\hat b^\dagger\tau_\nu\hat b$ contributes $np_{n-1}$ to the $n$th diagonal entry. The terms linear in $a$ connect adjacent photon numbers and contribute $ca\sqrt{n+1}p_n$ to the adjacent entries of $D$. Its trace is
\begin{equation}
 \Tr D=1-\lambda+ca^2+c(1+\nu)=Z,
\end{equation}
which explains the normalization $D/Z$. Replacing $a$ by $-a$ changes the signs of the adjacent-number entries, exactly as conjugation by parity does. The two diagonal Gram blocks are therefore $\gamma D$ and $\gamma PDP$. For the cross block, expansion of Eq.~\eqref{coll:rho} gives
\begin{equation}
 \begin{aligned}
 &\langle a|\rho|-a\rangle_A
 =\langle a|\tau_\nu|-a\rangle K,\\
 K&=(1-\lambda-ca^2)\tau_\nu+c\hat b^\dagger\tau_\nu\hat b
       +ca(\tau_\nu\hat b-\hat b^\dagger\tau_\nu).
 \end{aligned}
 \label{coll:Gram_cross}
\end{equation}
In the Fock basis, $KP=C$ with the entries in Eq.~\eqref{coll:PD_tridiagonal}; the parity makes the cross block real symmetric. The coherent thermal overlaps obey
\begin{equation}
 \frac{\langle a|\tau_\nu|-a\rangle}
 {\langle a|\tau_\nu|a\rangle}
 =e^{-2xa^2}=e^{-2\nu u}.
 \label{coll:thermal_overlap_ratio}
\end{equation}
For example, the two sums are
\begin{equation}
 \begin{aligned}
 \langle a|\tau_\nu|a\rangle
   &=(1-x)e^{-a^2}\sum_{n=0}^\infty\frac{(xa^2)^n}{n!}\\
   &=(1-x)e^{-(1-x)a^2},\\
 \langle a|\tau_\nu|-a\rangle
   &=(1-x)e^{-a^2}\sum_{n=0}^\infty\frac{(-xa^2)^n}{n!}\\
   &=(1-x)e^{-(1+x)a^2}.
 \end{aligned}
\end{equation}
Their ratio is Eq.~\eqref{coll:thermal_overlap_ratio}, since $xa^2=\nu u$. Conjugating the second row and column of the Gram matrix by $P$ is the unitary transformation $I_B\oplus P$. It makes both diagonal blocks equal to $\gamma D$ and both off-diagonal blocks equal to $\gamma e^{-2\nu u}C$. Dividing by the trace $2\gamma Z$ therefore gives
\begin{equation}
 \mathcal G=\frac1{2Z}
 \begin{pmatrix}D&e^{-2\nu u}C\\e^{-2\nu u}C&D\end{pmatrix}.
 \label{coll:PD_Gram}
\end{equation}
The remaining reduction uses vectors $(\ket\phi,\ket\phi)/\sqrt2$ and $(\ket\phi,-\ket\phi)/\sqrt2$. Acting with $\mathcal G$ on these symmetric and antisymmetric combinations leaves the respective subspaces invariant and gives the operators $(D+e^{-2\nu u}C)/(2Z)$ and $(D-e^{-2\nu u}C)/(2Z)$. This proves the spectral decomposition in Eq.~\eqref{coll:PD_public_blocks}; it does not assume that $D$ and $C$ commute. If Alice's sign is also fixed, only one diagonal Gram block remains, yielding $D/Z$. By exchange symmetry, its entropy agrees with the full-record conditional entropy used for CV-TF-Basic at the same radius. This provides a check on the calculation. The symbol $C$ in these matrix expressions denotes the cross matrix, not the public transcript used in the conditional information notation.

\subsection{High-loss scaling}
The high-loss limit is $t\to0$ with $N>0$ and input-referred $\xi\ge0$ fixed. Substituting $w=t\xi/2$ and $m_{\rm ch}=t(N+\xi/2)$ into the optical parameters gives
\begin{equation}
 \begin{aligned}
 \lambda&=\frac{N}{(N+\xi/2)(1+t\xi/2)},\\
 \nu&=\frac{N(1-t+t\xi/2)}{1+t(N+\xi/2)},\\
 p_{\rm acc}&=\frac{2t(N+\xi/2)}{[1+t(N+\xi/2)]^3}.
 \end{aligned}
\end{equation}
The first two quantities have finite limits, whereas acceptance is first order in $t$:
\begin{equation}
 \begin{aligned}
 \lambda&\longrightarrow\lambda_0=\frac{N}{N+\xi/2},\\
 \nu&\longrightarrow N,\\
 p_{\rm acc}&=2(N+\xi/2)t+O(t^2).
 \end{aligned}
 \label{coll:highloss_parameters}
\end{equation}
Let $\mathcal K_j(\lambda,\nu)$ denote the nonnegative secret fraction in the CV-TF-Basic or CV-TF-PD rate above. These fractions are continuous along this limit. In the entropy expressions, the thermal photon weights have geometrically decreasing tails uniformly for $\nu$ near $N$, and multiplication by the photon-number factors preserves summable entropy tails. The radial averages have exponentially decreasing weights. The PD Holevo term is bounded by one bit. For Basic, the conditional mean photon number obeys
\begin{equation}
 \begin{aligned}
 \Tr[\hat a^\dagger\hat a\,\sigma_A(r)]
 &=\nu+\frac{\lambda(1+\nu)}{2Z_\lambda(r)}\\
 &\le\nu+\frac{\lambda(1+\nu)}{2-\lambda}.
 \end{aligned}
\end{equation}
A thermal state maximizes entropy at fixed mean photon number, so applying $g$ to this last bound gives a finite, radius-independent bound on the conditional entropy. These bounds allow the limiting parameters to be taken through the averages. Multiplying the limiting secret fraction by the acceptance expansion gives
\begin{equation}
 \begin{aligned}
 R_j^{\rm iid}&=2(N+\xi/2)\mathcal K_j(\lambda_0,N)\sqrt\eta\\
 &{}+o(\sqrt\eta)
 \end{aligned}
 \label{coll:highloss_rate}
\end{equation}
as $\eta\to0$. Thus $R_j^{\rm iid}/\sqrt\eta$ tends to $2(N+\xi/2)\mathcal K_j(\lambda_0,N)$. When this limit is positive, the rate has the claimed square-root scaling. This leading statement needs continuity of the secret fraction, without assuming a differentiable entropy expansion. Physically, the useful acceptance probability is first order in the transmission of one arm, while the secret fraction approaches a positive constant. Since one arm has transmission $t=\sqrt\eta$, this yields the square-root scaling. It is a statement about ideal detectors and a family of channels with fixed input-referred noise, not an unlimited-distance prediction for detectors with fixed dark-count probability. Fixed input-referred noise is also not fixed bath energy: $\bar n=t\xi/[2(1-t)]$ decreases as $t\to0$. At fixed nonzero $\bar n$, instead, $\lambda\to0$ and background heralds eventually remove the positive key rate. The channel-noise convention is therefore essential when interpreting the asymptotic scaling.

\subsection{Numerical evaluation}
The fixed modulations for the IID curves in Fig.~\ref{fig:performance} are $N_{\rm Basic}=0.015$ and $N_{\rm PD}=0.016196$. They are held fixed across the loss sweep. The CV-TF-PD value is chosen by a one-dimensional optimization of the high-loss noisy rate, without intensity-dependent rejection. Both IID curves use $\xi=10^{-5}$ in vacuum-noise units per arm; the finite experimental design uses a different modulation and apparatus model.

At end-to-end loss $L$ dB, set $\eta=10^{-L/10}$ and $t=10^{-L/20}$, then evaluate $m_{\rm ch},\lambda,\nu$ and $p_{\rm acc}$. For CV-TF-Basic, compute $I_{\rm full}$, the geometric and photon-added spectrum $\mathbf r$, and the radial average of the tridiagonal conditional entropy. For CV-TF-PD, compute $I_{\rm sign}$, diagonalize $D/Z$ and the two public blocks at each integration radius, take the difference of their entropies, and integrate with weight $e^{-u}Z_\lambda(u)$. In each case, subtract the information terms before applying the positive part and multiplying by $p_{\rm acc}$. This procedure computes bits per attempted relay use and includes both accepted ports.

The accepted state and conditional entropies are evaluated anew for losses $40,45,\ldots,110$ dB. The radial entropy averages use adaptive Gauss--Kronrod integration with absolute tolerance $10^{-11}$ and relative tolerance $10^{-10}$. The photon basis contains levels $n=0,\ldots,47$ for CV-TF-PD and $n=0,\ldots,63$ for CV-TF-Basic. The sign-information integral uses absolute tolerance $10^{-13}$ and relative tolerance $10^{-12}$. At 40, 90 and 110 dB, increasing CV-TF-PD to 80 levels and tolerance $10^{-13}$ changes its rate by less than $6\times10^{-16}$ relatively. A separate 256-point Gauss--Laguerre quadrature, using 80 photon levels, agrees with the refined adaptive calculation within $3.9\times10^{-7}$ relatively. The CV-TF-Basic 48-level and 80-level calculations agree to the reported precision at the same losses.

The photon cutoffs enter numerical entropy evaluation only; they do not restrict Eve's state space or remove preparation-intensity classes. The full infinite-dimensional expressions define the IID rates. Increasing the cutoff checks the truncation in the state spectra, while changing the quadrature tests the radial integration. These convergence checks support the reported numerical rates.
\section{Composable security of CV-TF-PD}
\label{app:composable}
The aim is to certify a finite secret key without assuming that successive channel uses are independent. The proof has four steps. First, a score calculated from disclosed preparation intensities bounds virtual photon-parity errors in the test sample. Second, random sampling transfers that bound to the key positions. Third, an uncertainty relation converts the parity bound into secret randomness in Alice's sign string. Finally, reconciliation leakage is subtracted and privacy amplification produces the key.

Eve may apply an arbitrary joint attack to all transmitted signals and control the relay. The assumptions are independent trusted Gaussian preparations, accurate private preparation records, isolated laboratories and authenticated classical communication. The detector and noise parameters used to predict honest completion do not restrict Eve in this proof.

\subsection{Protocol, source replacement and the reference parity relation}
Let $M$ be the number of attempted relay uses. Each sender independently draws a coherent amplitude of variance $N>0$, so each signal has mean photon number $N$. Alice's amplitude is $\alpha$; Bob's emitted amplitude is $\beta_{\rm em}$. On a successful relay flag $s=\pm1$, Bob stores the corrected record $\beta=s\beta_{\rm em}$.

To analyze the protocol, put $q=N/(1+N)$ and replace each classical preparation by the equivalent entangled source
\begin{equation}
 \lvert\Psi_q\rangle=\sqrt{1-q}\sum_{j=0}^{\infty}
 q^{j/2}\lvert j\rangle_{\rm ret}\lvert j\rangle_{\rm sig}.
 \label{pdsec:source}
\end{equation}
Here $\lvert j\rangle$ is the $j$-photon Fock state, and ``ret'' and ``sig'' denote the retained and transmitted modes. The geometric weights are normalized because $(1-q)\sum_{j\ge0}q^j=1$. Heterodyne detection has effects $\lvert a_0\rangle\langle a_0\rvert d^2a_0/\pi$. Contracting the retained mode with its coherent-state outcome gives
\begin{equation}
 {}_{\rm ret}\langle a_0|\Psi_q\rangle
 =\sqrt{1-q}e^{-(1-q)|a_0|^2/2}
 |\sqrt q a_0^*\rangle_{\rm sig}.
\end{equation}
Consequently, outcome $a_0$ prepares the transmitted mode in $\lvert\sqrt q\,a_0^*\rangle$. The heterodyne outcome has a circular Gaussian distribution of variance $1/(1-q)$, so the prepared amplitude has variance $q/(1-q)=N$, exactly as in the actual protocol. Thus
\begin{equation}
 a_0=\frac{\alpha^*}{\sqrt q},~~
 b_0=\frac{\beta^*}{\sqrt q}
 \label{pdsec:virtual_records}
\end{equation}
are the virtual retained-mode records, including Bob's port correction. Retained-mode measurements commute with every operation on the signals. They can therefore be postponed until after the relay announcement in the proof, without changing the implemented preparation statistics or Eve's conditional state. This change of measurement order is a proof construction: the users still prepare coherent states directly and need no entangled source or quantum memory in their laboratories.

Before transmission, fix positive test and key counts $m,n$, all finite quantizers, score threshold $g_*\le1$, reconciliation cap $L_{\rm EC}$ and verification length $v$. After the relay commits all flags, choose $m+n$ successful positions uniformly and choose a uniform $m$-position test subset $T$ from them. The remaining $n$ positions form the key set $K$. Abort on a shortfall. Neither the selection nor the block parameters depend on the undisclosed key records. The relay must fix its successes before learning the test subset. Thus it can influence which systems are available, but it cannot assign their subsequent roles as tests or keys. The fresh random partition is the basis of the sampling argument.

On a test position, both users disclose finite intervals containing their normalized preparation intensities
\begin{equation}
 u_A=\frac{|\alpha|^2}{q}=|a_0|^2,~~
 u_B=\frac{|\beta|^2}{q}=|b_0|^2.
 \label{pdsec:test_intensities}
\end{equation}
For example, a uniform interval label $j$ of width $\Delta$ means that the local value lies in $[j\Delta,(j+1)\Delta)$; a separate label covers overflow. These intervals are sometimes called radial cells because $|a_0|$ is the radius of a complex amplitude and $|a_0|^2$ its squared radius. The normalization by $q$ converts the physical preparation intensity into the squared magnitude of a retained-mode heterodyne outcome; it does not change the transmitted state. The interval labels discretize an already known classical record and require no additional optical measurement. Test positions are excluded from the key, so disclosure of their intensities does not reveal a key symbol.

On a key position Alice writes
\begin{equation}
 \begin{gathered}
 \alpha=(-1)^{Z_A}\sqrt A\,e^{i\theta_A},~~ A=|\alpha|^2,\\
 Z_A\in\{0,1\},~~\theta_A\in[0,\pi).
 \end{gathered}
 \label{pdsec:sign_definition}
\end{equation}
The folded phase $\theta_A$ is the ordinary phase reduced modulo $\pi$: phases separated by $\pi$ have the same folded value and differ only in the private sign $Z_A$. Alice discloses finite cells of $A$ and $\theta_A$. Each label therefore leaves two antipodal regions, associated with $Z_A=0$ and $Z_A=1$, unresolved. At zero amplitude set $Z_A=0$ and $\theta_A=0$. Bob keeps his full corrected complex record private, and every selected key intensity and phase class is retained.

Photon parity records only whether a Fock number is even or odd: the operator $(-1)^{\hat n}$ has eigenvalue $+1$ on $|0\rangle,|2\rangle,\ldots$ and $-1$ on $|1\rangle,|3\rangle,\ldots$. The relevant reference correlation is photon-parity anticorrelation. In the ideal lossless, noiseless case, a single-photon relay herald projects the retained modes, before Bob's virtual port correction, onto
\begin{equation}
 \lvert\psi_s\rangle_{AB}
 =\frac{\lvert1,0\rangle+s\lvert0,1\rangle}{\sqrt2}.
 \label{pdsec:reference_state}
\end{equation}
This follows because the source in Eq.~\eqref{pdsec:source} pairs equal retained and transmitted photon numbers, while a one-photon beam-splitter outcome coherently combines the two possible input origins. In either term of Eq.~\eqref{pdsec:reference_state}, one retained mode has even photon number and the other has odd photon number. Bob's virtual port correction maps either outcome to $\lvert\psi_+\rangle$. The correction changes a relative phase and preserves the parity relation. Equal local parities give an even total photon number, whereas opposite local parities give an odd total. This is why the error definition depends only on the total retained photon number, although the two local photon numbers need not be known individually.

Let $j_A,j_B$ be virtual retained photon numbers and $J=j_A+j_B$. We therefore define a parity error by
\begin{equation}
 D(J)=\mathbf1\{j_A\bmod2=j_B\bmod2\}
     =\mathbf1\{J\text{ is even}\}.
 \label{pdsec:parity_error}
\end{equation}
These are proof variables, not photon numbers measured by Alice or Bob. Under loss, noise or an attack, a relay success does not imply the reference state. The test must bound deviations from its parity correlation. These virtual errors are also distinct from Bob's sign-reconstruction errors, which are checked by verification.

\subsection{Step 1: preparation intensities bound test parity errors}
For each test position define the intensity sum and score
\begin{equation}
 \begin{split}
 u&=u_A+u_B,\\
 Q(u)&=\left(\frac32\right)^5e^{-u/2}
 \left(-u^2+\frac{14u}{3}-\frac{32}{9}\right).
 \end{split}
 \label{pdsec:score}
\end{equation}
The score is a fixed classical function of the disclosed intensities. It is not an estimate of the photon number in a single round. Its coefficients are chosen so that its expectation provides a conservative witness of the parity relation.

To establish that property, consider a virtual Fock state $\lvert j\rangle$. A heterodyne outcome $a$ has density $e^{-|a|^2}|a|^{2j}/(\pi j!)$. Integrating over its phase gives the density $e^{-r}r^j/j!$ of the squared magnitude $r=|a|^2$. For a Fock pair $\lvert j_A,j_B\rangle$, the two heterodyne outcomes are independent. The density of their intensity sum is the convolution
\begin{equation}
 \begin{split}
 p(u\mid j_A,j_B)
 &=\frac{e^{-u}}{j_A!j_B!}
   \int_0^u r^{j_A}(u-r)^{j_B}dr\\
 &=\frac{e^{-u}u^{j_A+j_B+1}}{(j_A+j_B+1)!}.
 \end{split}
\end{equation}
In the second line, substituting $r=ut$ reduces the integral to the beta integral $\int_0^1t^{j_A}(1-t)^{j_B}dt=j_A!j_B!/(j_A+j_B+1)!$. Thus the sum has density
\begin{equation}
 p(u\mid J)=\frac{e^{-u}u^{J+1}}{(J+1)!},~~ u\ge0.
 \label{pdsec:radial_density}
\end{equation}
This is a Gamma density with shape $J+2$ and unit scale. Its dependence on the total $J$, rather than on how the photons are divided between the modes, is useful for the parity test.

Denote the conditional mean score by $\mu_J$. Multiplying the density by the exponential in $Q$ changes its decay factor from $e^{-u}$ to $e^{-3u/2}$. For $r=0,1,2$, the required weighted moments are
\begin{equation}
 \mathbb E[u^r e^{-u/2}\mid J]
 =\frac{(J+1+r)!}{(J+1)!}
 \left(\frac23\right)^{J+2+r}.
\end{equation}
Combining these three moments with the polynomial coefficients of $Q$, or using $\int_0^\infty u^k e^{-a u}\,du=k!/a^{k+1}$ directly, gives
\begin{equation}
 \begin{split}
 \mu_J&:=\mathbb E[Q(u)\mid J]
       =\int_0^\infty Q(u)p(u\mid J)\,du\\
 &=J(2-J)\left(\frac32\right)^{1-J}.
 \end{split}
 \label{pdsec:score_mean}
\end{equation}
In particular, $\mu_0=\mu_2=0$, $\mu_1=1$, and $\mu_J<0$ for $J\ge3$. Thus, for every photon number,
\begin{equation}
 \mu_J\le1-D(J).
 \label{pdsec:witness_eigenvalue}
\end{equation}
Neither $Q$ nor $\mu_J$ is a probability. Higher odd photon numbers are penalized too, although they are not parity errors; this makes the witness conservative.

The same statement holds for an arbitrary two-mode quantum state. Let $P_A=(-1)^{\hat n_A}$ and $P_B=(-1)^{\hat n_B}$ be the local photon-parity operators and $\Pi_{\rm err}=(I+P_AP_B)/2$ the equal-parity projector. Define the score operator
\begin{equation}
 \begin{split}
 W_Q&=\int_{\mathbb C^2}Q(|a|^2+|b|^2)
 \lvert a,b\rangle\langle a,b\rvert\,
 \frac{d^2a\,d^2b}{\pi^2}\\
 &=\sum_{j_A,j_B\ge0}\mu_{j_A+j_B}
 \lvert j_A,j_B\rangle\langle j_A,j_B\rvert.
 \end{split}
 \label{pdsec:score_operator}
\end{equation}
To see why this operator is diagonal, a matrix element between distinct local Fock numbers contains an angular factor $e^{i(j-j')\theta}$. Its integral over a full phase period vanishes unless $j=j'$. The two phase integrations impose this condition independently on Alice's and Bob's modes. The remaining diagonal element is precisely the mean $\mu_{j_A+j_B}$ already calculated. The eigenvalue of $I-\Pi_{\rm err}$ on the same Fock pair is $1-D(j_A+j_B)$, so Eq.~\eqref{pdsec:witness_eigenvalue} is an eigenvalue-by-eigenvalue comparison and implies $W_Q\le I-\Pi_{\rm err}$. For any state $\rho$, its error probability obeys $\operatorname{Tr}(\rho\Pi_{\rm err})\le1-\operatorname{Tr}(\rho W_Q)=1-\mathbb E_\rho Q$. This operator inequality explains why an intensity score can constrain a virtual photon-parity error without assuming a Gaussian state.

The users have finite interval labels rather than exact intensities. They calculate a lower score $Q^-$ from those labels such that
\begin{equation}
 Q(u)-\zeta\le Q^-\le Q(u),~~\zeta=10^{-10},
 \label{pdsec:score_rounding}
\end{equation}
including overflow. The lower score can be obtained by bounding the minimum of $Q$ over the interval of possible intensity sums, with an explicit bound for the overflow class. The test passes when $\widehat Q^-:=m^{-1}\sum_{i\in T}Q_i^-\ge g_*$. Since $Q_i^-\le Q(u_i)$, passing also implies that the exact sample average $\widehat Q:=m^{-1}\sum_{i\in T}Q(u_i)$ is at least $g_*$. Downward rounding cannot create an artificially favorable score.

The remaining task is to allow for a sample mean that fluctuates above its expectation. The needed constants are
\begin{equation}
 \begin{gathered}
 V_0=\frac{270459}{8192},~~ V_1=\frac{223457}{16384},\\
 B=V_0-V_1=\frac{317461}{16384}.
 \end{gathered}
 \label{pdsec:variance_constants}
\end{equation}
Here $V_0$ and $V_1$ are the conditional variances of $Q$ at $J=0$ and $J=1$. The following bounds hold uniformly for all $J$:
\begin{equation}
 \operatorname{Var}(Q\mid J)\le V_0-B\mu_J,
 ~~ Q-\mu_J<10.
 \label{pdsec:variance_envelope}
\end{equation}
The variance envelope depends on the conditional mean, rather than replacing it by one worst-case variance for all photon numbers. The upper bound on the centered score is one-sided; it does not assert $|Q-\mu_J|<10$. The scalar bounds are established below. Their uniformity is essential: an honest-channel variance cannot replace this bound in a proof against general attacks.

To apply concentration without an IID assumption, condition on the committed relay record and the selected systems. Let $\rho$ be their arbitrary joint retained-mode state. The selected modes can be entangled across positions and with Eve. The test-intensity effects, including finite radial intervals, and the virtual key-parity effects are diagonal in local photon number. In the trace giving their joint probabilities, off-diagonal matrix elements of $\rho$ therefore make no contribution. Write $\mathbf j$ for a full string of local photon numbers, with total $J_i$ at position $i$. These probabilities admit the classical representation
\begin{equation}
 p(\mathbf j)=\langle\mathbf j|\rho|\mathbf j\rangle.
\end{equation}
One may first draw this arbitrarily correlated string, choose the independent uniform test subset, and then draw the test outcomes conditional on the entire string. In that conditional calculation the Fock state is a product over positions, so the test intensity sums are independent with their respective densities in Eq.~\eqref{pdsec:radial_density}. Unconditionally the outcomes can remain strongly correlated. This is only a representation of the probabilities needed for the parity test, not an instruction to dephase the actual key state or a claim that Eve's attack is a mixture of Fock preparations. The sign coherence needed for key generation is left intact.

For a fixed string and test subset, let
\begin{equation}
 D_T=\sum_{i\in T}D(J_i),~~
 \mu_T=\frac1m\sum_{i\in T}\mu_{J_i}.
 \label{pdsec:test_population}
\end{equation}
Averaging Eq.~\eqref{pdsec:witness_eigenvalue} gives $D_T/m\le1-\mu_T$. Introduce the centered random variables $X_i=Q(u_i)-\mu_{J_i}$ and the average-variance bound $v_T=V_0-B\mu_T$. Conditional on the string and subset, the $X_i$ are independent, have mean zero, satisfy $X_i<10$, and obey $\sum_{i\in T}\mathbb E X_i^2\le mv_T$. They need not have identical distributions.

For completeness, the one-sided Bernstein argument bounds the moment-generating function at $0<\lambda<3/10$. For a mean-zero random variable bounded above by $10$, the scalar exponential bound gives
\begin{equation}
 \log\mathbb E e^{\lambda X_i}
 \le\frac{\lambda^2\mathbb E X_i^2}{2(1-10\lambda/3)}.
\end{equation}
This bound needs no corresponding lower bound of $-10$: for a negative argument the quadratic upper bound on the exponential remainder is already sufficient. Independence allows the logarithms to be summed. Markov's inequality then gives
\begin{equation}
 \Pr\left\{\sum_{i\in T}X_i\ge mz\right\}
 \le\exp\left[-\lambda mz+
 \frac{m\lambda^2v_T}{2(1-10\lambda/3)}\right].
\end{equation}
Choosing $\lambda=z/(v_T+10z/3)$ gives, for $z>0$,
\begin{equation}
 \Pr\{\widehat Q-\mu_T\ge z\}
 \le\exp\left[-\frac{mz^2}
 {2(V_0-B\mu_T)+(20/3)z}\right].
 \label{pdsec:bernstein}
\end{equation}
All logarithms in these concentration formulas are natural. Choose a radial-inference error allowance $\epsilon_r\in(0,1)$ and define
\begin{equation}
 L_r=\ln(1/\epsilon_r),~~
 C_r=B+10/3,~~ V_g=V_0-Bg_*.
 \label{pdsec:margin_parameters}
\end{equation}
Since $g_*\le1$, one has $V_g\ge V_1>0$. If $\mu_T<g_*$, passing requires an upward fluctuation $z=g_*-\mu_T$. Substituting $\mu_T=g_*-z$ also changes the variance bound: $V_0-B\mu_T=V_g+Bz$. Thus the exponent in Eq.~\eqref{pdsec:bernstein} becomes $mz^2/(2V_g+2C_rz)$, an increasing function of $z>0$. The term $Bz$ must be kept when inverting the tail bound; the variance is not a fixed value evaluated only at the threshold. Setting the exponent to $L_r$ gives
\begin{equation}
 mz^2-2C_rL_rz-2L_rV_g=0.
\end{equation}
Its positive root is the required margin,
\begin{equation}
 \begin{split}
 \delta_r&=\frac{C_rL_r+
 \sqrt{C_r^2L_r^2+2mL_rV_g}}{m},\\
 e_T^U&=1-g_*+\delta_r.
 \end{split}
 \label{pdsec:radial_margin}
\end{equation}
Suppose the virtual test-error fraction exceeds $e_T^U$. Since $D_T/m\le1-\mu_T$, this implies $\mu_T<g_*-\delta_r$. The probability of nevertheless passing is at most $\epsilon_r$ by Eq.~\eqref{pdsec:bernstein}. Averaging the conditional result over the arbitrary Fock string and random test subset yields
\begin{equation}
 \Pr\{\text{test passes},\ D_T>me_T^U\}\le\epsilon_r.
 \label{pdsec:test_error}
\end{equation}
This bounds the joint event of passing and having too many virtual errors. It is not a statement that the conditional error probability given passing is at most $\epsilon_r$. Keeping the joint-event form avoids a division by a potentially small passing probability.

As an example, the one-day design at $85$ dB uses $m=20\,942\,187$, $g_*=0.976839718177$ and $\epsilon_r=2.5\times10^{-23}$. Equation~\eqref{pdsec:radial_margin} gives $\delta_r\simeq0.00842426$ and $e_T^U\simeq0.03158454$. Thus passing with more than approximately $3.1585\%$ virtual errors in the test positions has probability at most $\epsilon_r$. A separate sampling argument is still needed for the key positions.

\subsection{Step 2: random sampling bounds key parity errors}
Abort if $e_T^U\notin[0,1)$. Otherwise set $a=\lfloor me_T^U\rfloor$ and $x=a/m$. Outside the exceptional event in Eq.~\eqref{pdsec:test_error}, a passing test has at most $a$ virtual test errors. Let $D_K$ be the error count in the $n$ complementary key positions. We seek an integer $k$ for which the event $D_T\le a$ but $D_K>k$ is sufficiently unlikely.

Fix any error pattern on the selected $m+n$ positions and let $w$ be its total number of errors. Sampling a subset of fixed size means sampling without replacement: a position cannot appear in both the test and key sets. Every size-$m$ subset is equally likely, so counting subsets containing $d$ errors gives the hypergeometric law
\begin{equation}
 \Pr_w(D_T=d)=
 \frac{\binom wd\binom{m+n-w}{m-d}}{\binom{m+n}{m}}.
\end{equation}
A binomial coefficient is zero when its lower argument is outside its allowed integer range. Since $D_K=w-D_T$, the bad event has the equivalent form
\begin{equation}
 D_T\le\min\{a,w-k-1\}.
\end{equation}
If $w\le a+k+1$, this is exactly the key upper-tail event $D_K\ge k+1$: that many key errors automatically leave at most $a$ test errors. Its probability increases with $w$. This monotonicity follows by coupling two populations that differ only by changing one nonerror position into an error and selecting the same random subset in each. The key error count cannot decrease. If $w\ge a+k+1$, the event is exactly the test lower-tail event $D_T\le a$, whose probability decreases with $w$ by the same coupling. The maximum is therefore attained at $w=a+k+1$; no distributional assumption on the original error pattern is required.

For a trial cap $0\le k<n$, put $p_k=(a+k+1)/(m+n)$. Define the binary relative entropy in natural units by
\begin{equation}
 D_{\rm bin}(x\|p)=x\ln\frac xp+(1-x)\ln\frac{1-x}{1-p},
 \label{pdsec:binary_divergence}
\end{equation}
using the continuous convention when $x=0$. At the maximizing population, the error fraction is $p_k$. The exponential moment of sampling without replacement is no larger than that for $m$ independent Bernoulli draws with the same error fraction~\cite{Hoeffding1963}. In particular, for $\lambda<0$,
\begin{equation}
 \mathbb E e^{\lambda D_T}
 \le(1-p_k+p_ke^\lambda)^m.
\end{equation}
Because $\lambda$ is negative, the event $D_T\le mx$ implies $e^{\lambda D_T}\ge e^{\lambda mx}$. Markov's inequality therefore bounds its probability by $e^{-\lambda mx}(1-p_k+p_ke^\lambda)^m$. For $0<x<p_k$, minimizing this expression gives
\begin{equation}
 e^\lambda=\frac{x(1-p_k)}{p_k(1-x)}<1.
\end{equation}
Substitution produces the exponent $-mD_{\rm bin}(x\|p_k)$; the case $x=0$ follows by taking a limit. Combining this lower-tail bound with the preceding maximization over $w$, when $p_k>x$ we obtain
\begin{equation}
 \Pr\{D_T\le a,D_K>k\}\le e^{-mD_{\rm bin}(x\|p_k)}.
 \label{pdsec:sampling}
\end{equation}
For a sampling allowance $\epsilon_s\in(0,1)$, the function $D_{\rm bin}(x\|p)$ increases from zero to infinity on $x<p<1$. Indeed,
\begin{equation}
 \frac{\partial}{\partial p}D_{\rm bin}(x\|p)
 =\frac{p-x}{p(1-p)}>0.
\end{equation}
Since $a<m$, one has $x<1$ and the divergence tends to infinity as $p$ approaches one. There is therefore a unique solution to the monotone equation
\begin{equation}
 mD_{\rm bin}(x\|p_*)=\ln(1/\epsilon_s),~~ p_*>x,
 \label{pdsec:sampling_root}
\end{equation}
using a certified upper bracket for its root, and choose
\begin{equation}
 k=\left\lceil(m+n)p_*-a\right\rceil.
 \label{pdsec:key_error_cap}
\end{equation}
If $k\ge n/2$, abort; otherwise $p_k>p_*$, so Eq.~\eqref{pdsec:sampling} is at most $\epsilon_s$. The rounding is conservative because the bad event contains at least $k+1$ key errors. Explicitly, the ceiling ensures $a+k\ge(m+n)p_*$, and the additional $+1$ in $p_k$ gives strict inequality. A numerical upper bracket for $p_*$ can only increase the chosen cap and hence cannot underestimate the allowed number of errors.

A passing execution with $D_K>k$ must either have $D_T>a$ or belong to the event bounded by Eq.~\eqref{pdsec:sampling}. The union bound therefore gives
\begin{equation}
 \Pr\{\text{test passes},D_K>k\}\le\epsilon_r+\epsilon_s.
 \label{pdsec:joint_error}
\end{equation}
The conditional arguments are uniform in the virtual photon-number string, so averaging over its arbitrary correlated distribution proves the same bound for coherent attacks. It bounds the total probability weight of passing executions with excessive key errors. It does not assert the same bound after normalizing the passing branch, or separately for every later public disclosure. This aggregate form is the one needed for composable secrecy. The random sample transfers a statement about disclosed tests to a statement about the undisclosed keys; the users never measure the key photon numbers.
\subsection{Step 3: parity control gives secret sign randomness}
We now convert the bound on virtual parity errors into a bound on Eve's information about Alice's actual sign string. The two measurements are alternatives on the same retained systems: parity is used only in the proof, whereas the sign is obtained from the preparation record. Their incompatibility is the resource quantified by an entropic uncertainty relation.

We first specify the entropy and distance conventions. A classical--quantum state has the form $\rho_{ZE}=\sum_z\lvert z\rangle\langle z\rvert\otimes\rho_E^z$, where $Z$ is a classical register and the positive operators $\rho_E^z$ contain both the probabilities and Eve's conditional states. Its conditional min-entropy is
\begin{equation}
 H_{\min}(Z\mid E)_\rho=-\log_2p_{\rm guess}(Z\mid E)_\rho,
\end{equation}
where $p_{\rm guess}=\max_{\{D_z\}}\sum_z\operatorname{Tr}(D_z\rho_E^z)$ and $\{D_z\}$ ranges over Eve's POVMs. Thus min-entropy measures the difficulty of guessing the complete string, rather than a typical individual bit. The complementary quantity is the conditional max-entropy. For quantum systems $U,V$, its definition is
\begin{equation}
 2^{H_{\max}(U\mid V)_\rho}
 =\max_{\sigma_V:\operatorname{Tr}\sigma_V=1}
 \bigl\|\sqrt{\rho_{UV}}\sqrt{I_U\otimes\sigma_V}\bigr\|_1^2,
\end{equation}
with the maximization over density operators. Below, its classical specialization will reduce to a bound on the number of compatible parity strings.

We retain passing probabilities in the states themselves. If an event has probability $p$ and normalized conditional state $\widehat\rho$, its subnormalized state is $\rho=p\widehat\rho$, with $\operatorname{Tr}\rho=p\le1$. Guessing probabilities computed from this state are also weighted by $p$. For two such states define the generalized fidelity and purified distance by
\begin{equation}
 \begin{split}
 \overline F(\rho,\sigma)
 &=\|\sqrt\rho\sqrt\sigma\|_1
 +\sqrt{(1-\operatorname{Tr}\rho)(1-\operatorname{Tr}\sigma)},\\
 d_{\rm pur}(\rho,\sigma)&=\sqrt{1-\overline F(\rho,\sigma)^2}.
 \end{split}
\end{equation}
Here $\|A\|_1=\operatorname{Tr}\sqrt{A^\dagger A}$. Smooth min-entropy $H_{\min}^{\epsilon}$ is the largest min-entropy of a state within purified distance $\epsilon$ of the original state. Smoothing will allow us to remove the small weight on which the statistical inference fails. The finite-dimensional definitions suffice initially; the source approximation is removed at the end of this appendix.

Let $c$ denote Alice's public intensity and folded-phase cell on one key position. In the virtual heterodyne plane, the two regions associated with $Z_A=0$ and $Z_A=1$ are antipodal. Let $M_{0c},M_{1c}$ be their heterodyne POVM effects. Photon parity $P=(-1)^{\hat n}$ acts as $P\lvert a\rangle=\lvert-a\rangle$, so
\begin{equation}
 \begin{gathered}
 M_{1c}=PM_{0c}P,\\
 F_c=M_{0c}+M_{1c},\ [F_c,P]=0.
 \end{gathered}
 \label{pdsec:antipodal}
\end{equation}
The public outcome has effect $F_c$. Equivalently, Alice first applies the filter $F_c^{1/2}$, announces $c$, and then measures her sign using
\begin{equation}
 B_{zc}=F_c^{-1/2}M_{zc}F_c^{-1/2},
 \label{pdsec:normalized_sign}
\end{equation}
for $z\in\{0,1\}$, where the inverse is taken on the support of $F_c$. Indeed, $F_c^{1/2}B_{zc}F_c^{1/2}=M_{zc}$, so both descriptions give the same outcome probabilities and the same conditional state of Eve after Alice's system is traced out. On this support, $B_{0c}+B_{1c}=I$ and $B_{1c}=PB_{0c}P$. The public label can be correlated with Eve; no independence of $c$ is assumed.

Let $\Pi_r=[I+(-1)^rP]/2$, $r=0,1$, be the even/odd parity projectors restricted to the same support. Since $P\Pi_r=(-1)^r\Pi_r$, conjugation by $P$ leaves either diagonal parity block unchanged. Thus $\Pi_rB_{0c}\Pi_r=\Pi_rB_{1c}\Pi_r$, and their sum is $\Pi_r$. Consequently,
\begin{equation}
 \begin{gathered}
 \Pi_rB_{zc}\Pi_r=\tfrac12\Pi_r,\\
 \|\sqrt{B_{zc}}\Pi_r\|_\infty^2\le\tfrac12.
 \end{gathered}
 \label{pdsec:overlap}
\end{equation}
The operator norm $\|A\|_\infty$ is the largest singular value. Its squared value here is the norm of $\Pi_rB_{zc}\Pi_r$, so equality holds for every nonempty parity sector. The upper bound also covers an empty sector. The maximum overlap is $1/2$: sign and parity remain complementary after the public label is known. This is a property of the two measurements, not an assertion that every attacked state has an unbiased sign distribution. The uncertainty relation will combine this property with the measured constraint on parity errors. Bob announces no key-round preparation label, so there is no corresponding public filter on his key modes.

For the $n$ key pairs, let $\mathsf P_{\le k}$ project onto the subspace with at most $k$ equal-parity positions. Let $\rho$ be the subnormalized test-passing state, including the public test transcript. Equation~\eqref{pdsec:joint_error} states that the removed weight is
\begin{equation}
 \delta=\operatorname{Tr}[(I-\mathsf P_{\le k})\rho]
 \le\epsilon_r+\epsilon_s.
\end{equation}
Set $\widetilde\rho=\mathsf P_{\le k}\rho\mathsf P_{\le k}$ and $p=\operatorname{Tr}\rho$. A purification of $\rho$ and its projected vector have overlap $\operatorname{Tr}(\mathsf P_{\le k}\rho)=p-\delta$. Tracing out the purifying system cannot decrease their fidelity. Since $\operatorname{Tr}\widetilde\rho=p-\delta$, the generalized fidelity therefore obeys
\begin{equation}
 \begin{split}
 \overline F(\rho,\widetilde\rho)
 &\ge p-\delta+\sqrt{(1-p)(1-p+\delta)}\\
 &\ge p-\delta+(1-p)=1-\delta.
 \end{split}
\end{equation}
Thus $d_{\rm pur}(\rho,\widetilde\rho)\le\sqrt{2\delta-\delta^2}$, and it suffices to use
\begin{equation}
 \epsilon_{\rm sm}=\sqrt{2(\epsilon_r+\epsilon_s)}.
 \label{pdsec:smoothing}
\end{equation}
This projection form of quantum sampling~\cite{Bouman2010} replaces the actual passing state by a nearby state with controlled-error support. It does not assert that the actual state has a hard parity-error cutoff, and it involves no division by the passing probability.

Alice's filters commute with local parity and hence with $\mathsf P_{\le k}$. Moreover, $\sum_cF_c=I$ because all key cells, including tails, are retained. Summing over the public filter outcomes therefore preserves the aggregate bad weight, while purified distance contracts under this complete instrument. The projected state's support restriction holds in every public-label branch. In contrast, the original state's bad weight need not be small in each normalized branch: a rare label could concentrate that weight. We use only the aggregate distance bound. No physical dephasing or parity measurement is applied to the key systems.

Write $X_A^n,X_B^n$ for the virtual parity strings, $B^n$ for Bob's retained systems, and $C_0$ for relay flags, selected positions, test messages and Alice's key-cell labels. Fixing $X_B^n$ on the projected state leaves at most
\begin{equation}
 \sum_{j=0}^k\binom nj\le2^{nh_2(k/n)}
 \label{pdsec:hamming_ball}
\end{equation}
possible $X_A^n$: Alice's parity differs from the bitwise complement of Bob's at no more than $k$ positions. The binomial coefficient counts the choices of those positions. Here $h_2(p)=-p\log_2p-(1-p)\log_2(1-p)$, with $0\log_2 0=0$, and $k<n/2$.

For completeness, put $x=k/n$ when $k>0$. Since $x<1/2$, each factor $x^j(1-x)^{n-j}$ with $j\le k$ is at least $x^k(1-x)^{n-k}$. The binomial expansion gives
\begin{equation}
 \begin{split}
 1&\ge\sum_{j=0}^k\binom nj x^j(1-x)^{n-j}\\
 &\ge x^k(1-x)^{n-k}\sum_{j=0}^k\binom nj.
 \end{split}
\end{equation}
Dividing by the prefactor yields Eq.~\eqref{pdsec:hamming_ball}. For $k=0$, the sum equals one and the bound holds directly.

To see how this counting controls entropy, let $D$ denote the sum in Eq.~\eqref{pdsec:hamming_ball}. For normalized classical variables $X,Y$,
\begin{equation}
 2^{H_{\max}(X\mid Y)}
 =\sum_y p(y)\left(\sum_x\sqrt{p(x\mid y)}\right)^2.
\end{equation}
If each conditional support has size at most $D$, Cauchy--Schwarz bounds every squared sum by $D$. Measuring Bob's quantum system to obtain $X_B^n$ can only discard conditioning information. Hence, on the projected state,
\begin{equation}
 H_{\max}(X_A^n\mid B^n,C_0)
 \le H_{\max}(X_A^n\mid X_B^n,C_0)
 \le nh_2(k/n).
 \label{pdsec:max_entropy}
\end{equation}
The support argument holds in each normalized projected public branch, and therefore also after combining their weights.

In any such branch, the unsmoothed entropic uncertainty relation~\cite{Tomamichel2011} gives
\begin{equation}
 H_{\min}(Z_A^n\mid E)+H_{\max}(X_A^n\mid B^n)\ge n.
\end{equation}
The right side is minus the logarithm of the product overlap, at most $2^{-n}$ by Eq.~\eqref{pdsec:overlap}. Combining this relation with Eq.~\eqref{pdsec:max_entropy} bounds Eve's guessing probability in each normalized projected branch by $2^{-n[1-h_2(k/n)]}$. To combine branches, write $w_c$ for their subnormalized weights and $p_{\rm guess}^{(c)}$ for their normalized guessing probabilities. Because the transcript is classical, Eve can choose a different guessing measurement in every branch, and
\begin{equation}
 \begin{split}
 p_{\rm guess}(Z_A^n\mid E,C_0)_{\widetilde\rho}
 &=\sum_c w_c p_{\rm guess}^{(c)}\\
 &\le 2^{-n[1-h_2(k/n)]}\sum_cw_c\\
 &\le 2^{-n[1-h_2(k/n)]}.
 \end{split}
\end{equation}
Here $c$ labels the complete public branch, and $\sum_cw_c=\operatorname{Tr}\widetilde\rho\le1$. Finally, its aggregate distance from the actual state supplies the smoothing:
\begin{equation}
 H_{\min}^{\epsilon_{\rm sm}}(Z_A^n\mid E,C_0)
 \ge n[1-h_2(k/n)].
 \label{pdsec:entropy}
\end{equation}
No common smoothing allowance is asserted for separately normalized actual branches. Bob may now measure his retained systems to obtain his private decoder records: a local trace-preserving operation on Bob leaves the Alice--Eve marginal unchanged. The virtual parity measurement was only a means of bounding a complementary entropy. In particular, the uncertainty relation compares two alternative measurements on the same underlying state; it never requires the users to measure both sign and parity in one run.

\subsection{Step 4: subtract disclosures and extract the key}
The entropy bound concerns secrecy before reconciliation, not agreement between the users. For example, $k=0$ certifies $n$ bits of sign randomness against Eve before further messages. It does not ensure that Bob can reconstruct all $n$ signs without communication. Reconciliation leakage and verification must still be included, and they can make the final length nonpositive. This separates the role of the intensity test from that of the decoder.

Reconciliation and verification reveal further information. Let $D_{\rm pub}$ contain all key-dependent messages sent after $C_0$, apart from independent random seeds. Cap reconciliation at $L_{\rm EC}$ bits, using fixed-length or padded messages. Once Bob has fixed his reconstruction, Alice sends a fresh $v$-bit two-universal verification hash; charge one further accept/abort bit. Then $D_{\rm pub}$ has at most $2^{L_{\rm EC}+v+1}$ possible values. The classical leakage chain rule subtracts the logarithm of this alphabet size from smooth min-entropy, leaving at least
\begin{equation}
 h_{\rm rem}=n[1-h_2(k/n)]-L_{\rm EC}-v-1.
 \label{pdsec:remaining_entropy}
\end{equation}
Additional key-dependent diagnostics, retransmission messages or variable-length termination information must fit the same cap. Alice's preparation-cell labels already enter $C_0$ in Eq.~\eqref{pdsec:entropy}; subtracting their lengths again would count that disclosure twice. Fresh independent public seeds incur no additional leakage when generated. The subsequent messages that depend on them are charged as above.

Two-universality means that distinct strings collide with probability at most $2^{-v}$ over a uniformly chosen verification seed. Choosing that seed after reconstruction is fixed makes this bound apply to any incorrect pair produced by Bob's decoder, including under attack. Thus the probability of accepting unequal strings is at most $2^{-v}$. This is an unconditional false-accept probability, not the conditional error probability given a matching tag. After charging the acceptance bit, restricting to the successful subnormalized branch cannot increase Eve's weighted guessing probability; purified distance also contracts under that restriction.

Let $C_{\rm pre}$ be the complete public transcript through verification, and let $F$ be an independent public seed for a two-universal privacy-amplification hash. It maps Alice's sign string to an $\ell$-bit key $K_A$. With $\tau_\ell$ uniform, the leftover-hash lemma~\cite{TomamichelHash2011} gives, on the accepted subnormalized state,
\begin{equation}
 \begin{split}
 \tfrac12\|\rho_{K_AEC_{\rm pre}F}
 -\tau_\ell\otimes\rho_{EC_{\rm pre}F}\|_1\\
 \le2\epsilon_{\rm sm}+\tfrac12\sqrt{2^{\ell-h_{\rm rem}}}.
 \end{split}
\end{equation}
The ideal comparison uses the actual marginal of Eve and the public transcript. The two smoothing terms account for replacing the real state by its nearby entropy-bounded state and replacing the corresponding ideal marginal. For $\epsilon_{\rm PA}\in(0,1)$, a sufficient length is
\begin{equation}
 \begin{split}
 \ell=\max\biggl\{0,\biggl\lfloor n[1-h_2(k/n)]-L_{\rm EC}\\
 {}-v-1-2\log_2\frac1{\epsilon_{\rm PA}}\biggr\rfloor\biggr\}.
 \end{split}
 \label{pdsec:length}
\end{equation}
Abort when $\ell=0$. An upper-rounded bound $e_U$ satisfying $k/n\le e_U<1/2$ may replace the exact ratio: $h_2$ increases on this interval, so the replacement only decreases the certified length. For positive output the floor in Eq.~\eqref{pdsec:length} ensures
\begin{equation}
 \frac12\sqrt{2^{\ell-h_{\rm rem}}}\le\frac{\epsilon_{\rm PA}}2.
\end{equation}
The secrecy allowance is therefore $\epsilon_{\rm sec}=2\epsilon_{\rm sm}+\epsilon_{\rm PA}/2$. Adding the verification and authentication-failure allowances gives
\begin{equation}
 \epsilon_{\rm tot}\le
 2\sqrt{2(\epsilon_r+\epsilon_s)}
 +\frac{\epsilon_{\rm PA}}2+2^{-v}+\epsilon_{\rm auth}.
 \label{pdsec:error_budget}
\end{equation}
For $\epsilon_r=\epsilon_s=2.5\times10^{-23}$, $\epsilon_{\rm PA}=\epsilon_{\rm auth}=2\times10^{-11}$ and $v=36$, this is less than $6.4552\times10^{-11}$.

Explicitly, let $C$ include the complete public transcript and both hash seeds, and let $\Omega$ denote complete protocol acceptance. The secrecy criterion before adding authentication failure is
\begin{equation}
 \tfrac12\|\rho_{K_AEC,\Omega}
 -\tau_\ell\otimes\rho_{EC,\Omega}\|_1
 \le\epsilon_{\rm sec}.
 \label{pdsec:secrecy_criterion}
\end{equation}
Both states have trace $\Pr(\Omega)$. Equivalently, this bounds the conditional secrecy distance multiplied by the passing probability; it does not promise the same conditional error for arbitrarily rare passes. The proof certifies security for any reconstruction algorithm obeying the disclosure cap. Positive honest throughput separately requires a completion bound. For fixed positive $\ell$, $R=\Pr(\Omega)\ell/M$, and acquisition at clock $f_0$ and quantum duty $d_Q$ takes approximately $M/(d_Qf_0)$ seconds. Authentication consumes preshared key and is debited separately when quoting net expansion.

\subsection{Verification of the scalar bounds and finite intensity messages}
This subsection supplies the arithmetic used in Step 1. Differentiating the score gives
\begin{equation}
 Q'(u)=\left(\frac32\right)^5e^{-u/2}
 \left(\frac{u^2}{2}-\frac{13u}{3}+\frac{58}{9}\right).
\end{equation}
Its stationary points are $(13\pm\sqrt{53})/3$. Evaluating $Q$ there, at zero and at infinity proves $-27\le Q<6$. To compute $s_J=\mathbb E[Q^2\mid J]$, expand the squared polynomial and use the Gamma density in Eq.~\eqref{pdsec:radial_density}. Each monomial is integrated through
\begin{equation}
 \mathbb E[u^r e^{-u}\mid J]
 =\frac{\Gamma(J+r+2)}{\Gamma(J+2)2^{J+r+2}}.
\end{equation}
Only $r=0,\ldots,4$ occur. Collecting terms yields
\begin{equation}
 \begin{split}
 s_J&=\frac{(3/2)^{10}}{81\,2^{J+6}}P(J),\\
 P(J)&=81J^4-378J^3+1503J^2\\
 &{}-1542J+2968.
 \end{split}
 \label{pdsec:second_moment}
\end{equation}
The variances $s_J-\mu_J^2$ at $J=0,1$ equal $V_0,V_1$ in Eq.~\eqref{pdsec:variance_constants}. Direct substitution gives $s_J<13$ for $J=2,3,4,5$. For every larger integer, write $J=y+5$, $y\ge0$, and observe
\begin{equation}
 \begin{split}
 2P(J)-P(J+1)={}&81y^4+918y^3+3771y^2\\
 &+5622y+1264>0.
 \end{split}
 \label{pdsec:moment_monotonicity}
\end{equation}
Because $s_{J+1}/s_J=P(J+1)/[2P(J)]$, this proves decreasing second moments thereafter. Since $\mu_J\le0$ for $J\ge2$, the variance obeys $s_J-\mu_J^2\le s_J<13<V_0\le V_0-B\mu_J$. This proves the variance envelope in Eq.~\eqref{pdsec:variance_envelope} for all photon numbers, with equality at $J=0,1$.

The mean formula in Eq.~\eqref{pdsec:score_mean} gives $\mu_0=\mu_2=0$ and $\mu_1=1$. For $J\ge3$, the ratio of consecutive magnitudes is $2(J+1)(J-1)/[3J(J-2)]$. It exceeds one through $J=5$ and is below one from $J=6$, placing the most negative mean at $\mu_6=-256/81$. Thus $Q-\mu_J<6+256/81<10$. Only this upper bound is needed for the upper-tail concentration inequality; it is not an absolute-deviation bound.

For finite test messages, recall $u_A=|\alpha|^2/q$, $u_B=|\beta|^2/q$ and $u=u_A+u_B$. Partition each local intensity into intervals of width $\Delta=2^{-42}$ on $[0,128)$, with a further overflow label. There are $2^{49}$ regular intervals plus overflow, so fifty bits per label suffice. Both users announce these interval labels on tests, and no intensity class is discarded. For regular labels let $u_0$ be the sum of their lower endpoints; the actual sum lies in $[u_0,u_0+2\Delta)$.

An elementary global derivative bound follows from maximizing $u^r e^{-u/2}$ for $r=1,2$: its maximum occurs at $u=2r$. Applying the triangle inequality to the displayed expression for $Q'$ gives
\begin{equation}
 |Q'(u)|\le\left(\frac32\right)^5
 \left(\frac8{e^2}+\frac{26}{3e}+\frac{58}{9}\right)<82<100.
\end{equation}
Hence $|Q(u)-Q(u_0)|<200\Delta$ throughout the disclosed interval. Compute a certified lower value $L\le Q(u_0)$ with $Q(u_0)-L\le\Delta$, then round downward once more:
\begin{equation}
 Q^-=\Delta\left\lfloor\frac{L-200\Delta}{\Delta}\right\rfloor.
 \label{pdsec:finite_score}
\end{equation}
The numerical and final rounding errors give $Q(u_0)-202\Delta<Q^-\le Q(u_0)-200\Delta$. Consequently $Q^-\le Q(u)$ and $0\le Q(u)-Q^-<402\Delta<10^{-10}$. If either local label overflows, $u\ge128$; set $Q^-=-10^{-10}$. On this tail $Q$ is negative and increases to zero, with $|Q(128)|<10^{-20}$. The same pointwise rounding requirement in Eq.~\eqref{pdsec:score_rounding} therefore holds on overflow as well. Finite messages preserve the witness inequality without excluding rare large intensities.

\subsection{Removing a finite-dimensional source cutoff}
The entropy argument may first be proved in finite dimension. For $q=N/(1+N)<1$, truncating the source in Eq.~\eqref{pdsec:source} to photon numbers $0,\ldots,d-1$ discards probability $q^d$. Its normalized truncated vector is
\begin{equation}
 \lvert\Psi_q^{(d)}\rangle
 =\sqrt{\frac{1-q}{1-q^d}}\sum_{j=0}^{d-1}
 q^{j/2}\lvert j\rangle_{\rm ret}\lvert j\rangle_{\rm sig}.
\end{equation}
The squared overlap with the original source is $1-q^d$. There are two independent sources for each of $M$ relay uses, so the total squared overlap is $(1-q^d)^{2M}$. The purified distance is therefore
\begin{equation}
 \delta_{\rm src}=\sqrt{1-(1-q^d)^{2M}}
 \le\sqrt{2Mq^d},
 \label{pdsec:source_distance}
\end{equation}
where the inequality follows from $(1-x)^{2M}\ge1-2Mx$.

This comparison is made before Eve acts. Apply the same arbitrary attack and protocol to both sources, including every public outcome and abort flag. Contractivity keeps the output distance at most $\delta_{\rm src}$. The two sources can have different passing probabilities; retaining those probabilities as traces is what makes the comparison valid without dividing by either of them. Compressing the retained-mode effects to the finite Fock support preserves parity and the antipodal identities. It also preserves the score bounds, which were proved separately for every photon number. Hence the finite-dimensional key-length and error bounds do not depend on $d$.

To transfer secrecy, compare the actual output to the truncated-source output, apply the latter's secrecy bound, and compare their ideal uniform-key counterparts. Each comparison costs at most $\delta_{\rm src}$ in trace distance: the second also follows by tracing out the key and adjoining a uniform key. The triangle inequality thus adds at most $2\delta_{\rm src}$. Taking $d\to\infty$ at fixed finite $M$ removes this cost. This is a limit used in the proof, not photon-number rejection imposed on the laboratory source. It retains ideal trusted Gaussian preparation and does not cover unspecified source deviations.
\section{Finite implementation and the experimental model for CV-TF-PD}
\label{pdimpl:appendix}
This appendix derives the finite CV-TF-PD curves in Fig.~\ref{fig:performance}. The composable theorem bounds the secrecy error of the accepted output, including its acceptance probability. A positive expected rate additionally requires a lower bound on how often the protocol passes under the intended experimental conditions. We therefore specify an \emph{honest model}: the statistical description of the sources, channels and detectors during normal operation, without an adversarial modification. Its independence and noise assumptions are used to predict completion, not to restrict the attacks covered by the security theorem.

Alice owns every raw sign bit. Writing $\alpha=(-1)^{Z_A}|\alpha|e^{i\theta_A}$ with $Z_A\in\{0,1\}$, her folded phase is $\theta_A=\arg\alpha\pmod\pi$ in $[0,\pi)$. At zero amplitude set $Z_A=0$ and $\theta_A=0$. Folding identifies phases separated by $\pi$, and therefore does not distinguish the two antipodal amplitudes $\alpha$ and $-\alpha$. Alice announces finite cells of her intensity and folded phase. Bob keeps his full preparation record private and incorporates the relay-port correction in $\beta$. All selected key intensity and phase classes are retained.

The calculation has three parts. The apparatus model determines the herald probability and Bob's binary decoding channel. A fixed finite polar code then supplies an explicit reconciliation message length and a decoding-failure bound. Finally, count, test and numerical-error allowances are combined with the composable key-length formula. The parameters specify a prospective configuration under the trusted-source and detector assumptions stated below.

\subsection{Loss, thermal noise, dark events and phase fluctuations}
Let $L$ be the end-to-end fiber loss in dB, $\tau=10^{-L/20}$ the transmissivity of each fiber arm, $\eta_d=0.80$ the combined relay collection and detection efficiency, and $t=\eta_d\tau$. Each sender prepares a circular Gaussian coherent-state amplitude of mean photon number $N$. Put $\xi_0=10^{-5}$ SNU per arm, $w=t\xi_0/2$, and $m_{\rm ch}=tN+w$. Independent Poisson dark events have mean $d=D_{\rm dark}\Delta t=10^{-9}$ per detector gate, using $D_{\rm dark}=1~\mathrm{s}^{-1}$ and $\Delta t=1$ ns. The detector model resolves the exact photon number after loss. An observed success has exactly one count across the two outputs, with probability
\begin{align}
 p_0&=(1+m_{\rm ch})^{-2},~~p_1=\frac{2m_{\rm ch}}{(1+m_{\rm ch})^3},\\
 p_{\rm obs}&=e^{-2d}(p_1+2dp_0).
 \label{pdimpl:herald}
\end{align}
Here $p_0$ is the probability that both optical output modes are in vacuum, and $p_1$ is the probability of exactly one optical photon across them. To obtain these expressions, recall that a thermal mode of mean $m_{\rm ch}$ has photon probabilities $P(k)=m_{\rm ch}^k/(1+m_{\rm ch})^{k+1}$. The two incoming averaged states are identical thermal states, so the balanced beam splitter leaves their product invariant. Thus $p_0=P(0)^2$ and $p_1=2P(0)P(1)$. The total number of independent dark events is Poisson with mean $2d$. Exactly one observed count therefore requires either one optical photon and zero dark events, or zero optical photons and one dark event. These alternatives are disjoint and give Eq.~\eqref{pdimpl:herald}. The factor $e^{-2d}$ excludes all other dark events. The two origins are not distinguished by the public herald. Conditioning on their mixture dilutes the photon-added part of the retained state by the optical fraction $p_1/(p_1+2dp_0)$. The accepted state is described by
\begin{align}
 \nu&=\frac{N(1-t+w)}{1+m_{\rm ch}},&
 \Lambda&=\frac{tN}{(1+w)[m_{\rm ch}+d(1+m_{\rm ch})]}.
 \label{pdimpl:parameters}
\end{align}
Here $\nu$ is the residual thermal occupation in each virtual retained mode and $\Lambda$ is the useful photon-added weight. The optical one-photon branch has photon-added weight
\begin{equation}
\lambda_{\rm opt}=\frac{tN}{(1+w)m_{\rm ch}}.
\end{equation}
The optical-vacuum branch has only the thermal-product component. Since the relay does not distinguish an optical photon from a dark event, the observed photon-added weight is their conditional mixture,
\begin{equation}
\Lambda=\frac{p_1}{p_1+2dp_0}\lambda_{\rm opt}.
\end{equation}
Substituting $p_0$ and $p_1$ gives Eq.~\eqref{pdimpl:parameters}. Both branches have the same residual thermal scale $\nu$, which is why their mixture changes this weight without introducing a second scale. Independent zero-mean Gaussian relative phase fluctuations have visibility $V_\phi=e^{-\sigma_\phi^2/2}=0.995$, or $\sigma_\phi\simeq0.100$ rad. They attenuate the interference term, while leaving the herald and radial-test laws unchanged. With $\tau_\nu$ the one-mode thermal state of mean $\nu$, $\rho_0=\tau_\nu\otimes\tau_\nu$, and retained-mode annihilation operators $a,b$, the accepted state is
\begin{align}
 \rho={}&(1-\Lambda)\rho_0+\frac{\Lambda}{2(1+\nu)}
 \bigl[a^\dagger\rho_0a+b^\dagger\rho_0b\nonumber\\
 &\hspace{34mm}+V_\phi(a^\dagger\rho_0b+b^\dagger\rho_0a)\bigr].
 \label{pdimpl:state}
\end{align}
The Gaussian visibility follows by averaging a relative phase $\varphi$ with mean zero and variance $\sigma_\phi^2$: its characteristic function is $\mathbb E[e^{i\varphi}]=e^{-\sigma_\phi^2/2}$. Only the cross terms carry this phase. The normalization can also be checked directly: $\operatorname{Tr}(a^\dagger\rho_0a)=\operatorname{Tr}(b^\dagger\rho_0b)=1+\nu$, whereas the cross terms have zero trace. The photon-added component is positive because its two-origin coefficient matrix has eigenvalues $1+V_\phi$ and $1-V_\phi$.

The diagonal photon-added terms describe the two possible photon origins. Their cross terms contain the coherence responsible for phase-sensitive decoding; averaging the relative phase multiplies these terms by $V_\phi$. Phase fluctuations do not change photon-number diagonal statistics, which explains why the intensity test has the same law. Dark events, in contrast, change both the herald probability and the weight $\Lambda$. These distinct effects are therefore kept separate from thermal excess noise.

A polar information set, or mask, identifies which transformed coordinates Bob must reconstruct rather than receive publicly. We precompute masks for three reference binary channels, indexed by a target photon-added weight $\Lambda_0\in\{0.990,0.995,0.999\}$. For each distance, the modulation solving $\Lambda=\Lambda_0$ is
\begin{equation}
 N_* =\frac{\Lambda_0(1+w)[w+d(1+w)]}
 {t[1-\Lambda_0(1+w)(1+d)]}.
 \label{pdimpl:modulation}
\end{equation}
This solution is positive provided $\Lambda_0(1+w)(1+d)<1$, as holds for the stated parameters. Indeed, at fixed $t,w,d$, the expression for $\Lambda$ increases with $N$ toward $1/[(1+w)(1+d)]$; the target must lie below this limiting value.

For numerical certification, $N$ is a rational value immediately above an outward enclosure of $N_*$ on a $10^{-65}$ grid, with $\Lambda\ge\Lambda_0$. The difference from the reference channel contributes the coupling allowance $n|\Lambda-\Lambda_0|$ for $n$ key positions. The coupling allowance follows from the affine state decomposition. At the same $\nu$ and $V_\phi$, changing only $\Lambda$ to $\Lambda_0$ changes a single accepted state by trace distance at most $|\Lambda-\Lambda_0|$. Measurements cannot increase trace distance, and replacing one factor at a time gives at most $n|\Lambda-\Lambda_0|$ for $n$ independent records. This compares the honest implementation with the channel used to construct the mask; it is not a restriction on an attack. The grid specifies the arithmetic precision of the model. Physical source-calibration errors require a separate source model.

\subsection{The honest binary channel and explicit polar information sets}
Let $q=N/(1+N)$ and $s=q(1+\nu)$. In this subsection, $s$ is an intensity scale, and $A=|\alpha|^2/s$, $B=|\beta|^2/s$ are dimensionless normalized preparation intensities. Their sum is $T=A+B$. The normalization removes the common accepted-state intensity scale from the decoder law; it does not change which data are public. Alice supplies her intensity label, while Bob evaluates $B$ from his private corrected record. Let $\theta_A\in[0,\pi)$ be Alice's phase modulo $\pi$, and define Bob's local relative angle $\vartheta=\arg\beta-\theta_A$ modulo $2\pi$. Write $Y$ for Bob's full private record together with the ideal fine versions of Alice's public intensity and phase labels. In the reference channel with mixture weight $\Lambda_0$, his fine-record posterior is
\begin{align}
 p&=\Pr(Z_A=0\mid Y),\nonumber\\
 &=\frac12\left[1+V_\phi
 \frac{2\Lambda_0\sqrt{AB}\cos\vartheta}
 {2(1-\Lambda_0)+\Lambda_0(A+B)}\right].
 \label{pdimpl:posterior}
\end{align}
For $z\in\{0,1\}$, the record likelihoods, up to a factor common to the two signs, are
\begin{equation}
\begin{aligned}
L_z={}&(1-\Lambda_0)+\frac{\Lambda_0}{2}(A+B)\\
 &+(-1)^z\Lambda_0V_\phi\sqrt{AB}\cos\vartheta.
\end{aligned}
\end{equation}
Alice's two signs have equal prior probability. Therefore $p=L_0/(L_0+L_1)$; the interference cancels in the denominator and gives Eq.~\eqref{pdimpl:posterior}. This is a likelihood calculation for the honest channel. The security proof does not assume that Eve implements this channel.

The scalar $\sqrt B\cos\vartheta=\operatorname{Re}(\beta e^{-i\theta_A})/\sqrt s$ is computed locally; no Bob intensity or phase message is needed. In the reference channel with mixture weight $\Lambda_0$, the reliability $r=|2p-1|$ has the representation
\begin{align}
 f_T(t)&=e^{-t}\left[(1-\Lambda_0)t+\frac{\Lambda_0t^2}{2}\right],\\
 r&=V_\phi\frac{\Lambda_0T}{2(1-\Lambda_0)+\Lambda_0T}\,U,\nonumber\\
 &\hspace{12mm}U\sim\operatorname{Unif}[0,1],\quad U\perp T.
 \label{pdimpl:reliability}
\end{align}
The reliability $r$ is zero for an uninformative observation, and a larger value means a more confident bit prediction. Here $0\le r<V_\phi$, because the visibility limits the available sign correlation. Its accompanying hard decision specifies which bit is favored. Under the independent honest model, different accepted records give a memoryless channel. Simultaneously reversing Alice's sign and Bob's corrected amplitude preserves the record distribution and exchanges the two bit hypotheses, so the channel is binary-input symmetric. These properties permit a polar-code construction for the reference channel.

The distribution in Eq.~\eqref{pdimpl:reliability} follows by integrating the record likelihoods. Summing over Alice's two signs gives a joint intensity density proportional to
\begin{equation}
e^{-(A+B)}\left[(1-\Lambda_0)+\frac{\Lambda_0}{2}(A+B)\right].
\end{equation}
Changing variables to $T=A+B$ and $X=A/T$ introduces Jacobian $T$. Thus $X$ is uniform on $[0,1]$, independent of $T$, and $T$ has the stated density. The relative angle is also uniform. The vector with components $2\sqrt{X(1-X)}\cos\vartheta$, $2\sqrt{X(1-X)}\sin\vartheta$ and $2X-1$ is consequently uniform on the unit sphere. Each component is uniform on $[-1,1]$, so the absolute value of its first component is the independent uniform variable $U$ in Eq.~\eqref{pdimpl:reliability}.

The following functions describe the reliability distribution. For $0\le x<1$, define
\begin{align}
 S_0(x)&=(1-\Lambda_0x)
 e^{-2(1-\Lambda_0)x/[\Lambda_0(1-x)]},\\
 J_0(x)&=\frac{\Lambda_0(1-x)^2}{2}
 e^{-2(1-\Lambda_0)x/[\Lambda_0(1-x)]}.
\end{align}
To see why these closed forms arise, write $a(T)=\Lambda_0T/[2(1-\Lambda_0)+\Lambda_0T]$. Conditional on $T$, the probability that $r>V_\phi x$ is $[1-x/a(T)]_+$, where $[y]_+=\max\{y,0\}$. It is nonzero only for
\begin{equation}
T>T_x=\frac{2(1-\Lambda_0)x}{\Lambda_0(1-x)}.
\end{equation}
Integrating $f_T(T)[1-x/a(T)]$ from $T_x$ to infinity gives $S_0(x)$. The second expression satisfies $-J_0'(x)=S_0(x)$ and $\lim_{x\to1^-}J_0(x)=0$, so it is the integral of this survival function. A survival function records the probability of exceeding a threshold, making it convenient for bin probabilities and moments.

For $r<V_\phi$, set $S(r)=S_0(r/V_\phi)$ and $J(r)=V_\phi J_0(r/V_\phi)$; set both to zero beyond that point. Then $S(r)=\Pr(\mathcal R>r)$ is the survival function of the random reliability $\mathcal R$, and $J(r)=\int_r^1S(x)\,dx$. Integrating this survival function over a cell determines its mean reliability. The decoder observes the hard posterior sign and one of $K=256$ uniform reliability intervals $[a_j,b_j)$ in $[0,1]$. Their exact weights, first moments and crossovers are
\begin{align}
 w_j&=S(a_j)-S(b_j),\\
 m_j&=a_jS(a_j)-b_jS(b_j)+J(a_j)-J(b_j),\\
 e_j&=\frac12(1-m_j/w_j).
 \label{pdimpl:bins}
\end{align}
The moment formula follows by integration by parts:
\begin{equation}
\begin{aligned}
\int_{a_j}^{b_j}r[-S'(r)]dr
 ={}&a_jS(a_j)-b_jS(b_j)\\
 &+\int_{a_j}^{b_j}S(r)dr.
\end{aligned}
\end{equation}
A record with reliability $r$ has conditional hard-decision error $(1-r)/2$. Averaging this identity within the bin gives $e_j$. In particular, $w_j$ is the probability of reliability cell $j$, $m_j=\mathbb E[\mathcal R\mathbf1\{\mathcal R\in[a_j,b_j)\}]$, and $e_j$ is the average probability that the favored bit is wrong in that cell. Thus the finite decoder retains the hard decision and a controlled amount of soft information. Zero-weight cells use a neutral placeholder. The decoding likelihoods use these true conditional crossovers, not bin centers or the degraded construction channel.

For a frame of $b=2^r$ raw sign bits, Alice applies $G_b=B_b\bigl(\begin{smallmatrix}1&0\\1&1\end{smallmatrix}\bigr)^{\otimes r}$, where $B_b$ reverses the $r$ binary digits of each coordinate index, and discloses all transformed coordinates outside an information set $\mathcal I$~\cite{Arikan2009,Arikan2010}. We use row vectors: if $Z_A^b$ is Alice's sign block, the disclosed and reconstructed coordinates belong to $U^b=Z_A^bG_b$, with arithmetic modulo two. Bob eventually inverts this transform to recover $Z_A^b$. The polar transform is invertible over the binary field. It separates coordinates that Bob can predict reliably from his observation and preceding coordinates from those that need to be supplied by Alice. The set $\mathcal I$ contains the reliably reconstructible coordinates; the complementary values form the public reconciliation message. Its length is exactly $b-|\mathcal I|$ bits per frame, and every one of these bits is charged as leakage.

This is source reconciliation: Alice already has a random block and sends selected linear functions of it. It is not necessary to replace that block by a prescribed channel-code word. The sign symmetry makes the honest raw block uniform, and an invertible binary transform preserves uniformity. Consequently the usual synthesized polar channels describe the prediction of each transformed bit from the observation and preceding transformed bits. This is the basis for using channel-polarization error bounds for the present source-reconstruction task.

A channel degradation discards information from the decoder output and can therefore only worsen the decoding problem. Constructing the information set on a deliberately degraded finite channel gives conservative error bounds for the actual reference channel. A symmetric output pair can be described by two nonnegative conditional masses: the probabilities of its two outputs for input zero are $A_j/2^{28}$ and $B_j/2^{28}$, and these probabilities are exchanged for input one. Its total weight is $(A_j+B_j)/2^{28}$ and its crossover probability is $B_j/(A_j+B_j)$. Reducing the weight is a sign-independent erasure; increasing the crossover toward $1/2$ is additional random bit flipping. Both operations can be performed using only the observed output. This explicitly realizes a degraded channel. 

Let $Z_i$ denote the Bhattacharyya parameter of synthesized coordinate $i$, a standard upper bound on its decision difficulty: for equal input priors its optimal bit-error probability is at most $Z_i/2$. For a binary channel with conditional output laws $W(y|0)$ and $W(y|1)$, this parameter is $Z=\sum_y\sqrt{W(y|0)W(y|1)}$. With equal priors, the optimal error is $\frac12\sum_y\min\{W(y|0),W(y|1)\}\le Z/2$. This proves the stated error comparison directly. 

The following construction controls each discretization and arithmetic step~\cite{Tal2013}.
\begin{enumerate}
 \item Evaluate Eq.~\eqref{pdimpl:bins} with directed 90-digit intervals. Represent each symmetric output pair by dyadic conditional masses $(A_j,B_j)/2^{28}$, $A_j\ge B_j$, normalized by $\sum_j(A_j+B_j)=2^{28}$. Round its total weight down to an even multiple of $2^{-28}$, round its crossover up, and assign removed weight to an uninformative output.
 \item Polar minus and plus transforms use exact integer products. Minus pairs are $(A_iA_j+B_iB_j,A_iB_j+B_iA_j)$; plus pairs are $(A_iA_j,B_iB_j)$ and $(A_iB_j,B_iA_j)$, reordered to put the larger mass first. The product denominator is $2^{56}$. Merge into 128 bins according to the reliability angle $\arccos[(A-B)/(A+B)]$; any such grouping is a channel degradation. For a merged pair $(a,b)$, put $T=a+b$, $W=2\lfloor T/2^{29}\rfloor$, $B'=\lceil W\min(a,b)/T\rceil$ and $A'=W-B'$. Replace the removed mass by a neutral pair. These are exact integer operations, so mass loss and crossover rounding only degrade the channel.
 \item Repeat to depth 12. Integer ceiling square roots give upper bounds on $Z=2\sum_j\sqrt{A_jB_j}/2^{28}$ with denominator $2^{60}$. Complete the tree to depth 22 or 24 using outward-rounded bounds $Z^+=Z^2$ and $Z^-\le2Z-Z^2$, clipped at one. Binary64 products are enclosed by adjacent representable values, including subnormals.
 \item Select coordinates with the smallest certified upper bounds. Sum the selected binary64 bounds exactly as integer multiples of $2^{-1074}$ and verify $\mathcal Z:=\sum_{i\in\mathcal I}Z_i<10^{-5}$.
\end{enumerate}
For the merge operation, $W/2^{28}\le T/2^{56}$ ensures that no output weight is increased. The ceiling defining $B'$ ensures $B'/W\ge\min(a,b)/T$ whenever $W>0$. An output with $W=0$ is erased entirely. Thus merging and rounding cannot create a better channel. The grouping angle affects the tightness of the construction but not this degradation argument.

The ceiling-square-root step also has an explicit integer form. For a channel with masses $A_j,B_j$, an upper bound is
\begin{equation}
\widehat Z=\min\left\{1,2^{-60}\sum_j
       2\left\lceil\sqrt{A_jB_j2^{64}}\right\rceil\right\}.
\end{equation}
At subsequent scalar steps, both $z^2$ and $2z-z^2$ increase on $[0,1]$, so an upper bound remains an upper bound after the stated outward rounding. Finally, every finite binary64 bound is an integer multiple of $2^{-1074}$; summing those integers avoids an unaccounted floating-point accumulation error. Denoting these certified bounds by $\widehat Z_i$, the construction verifies $\sum_{i\in\mathcal I}Z_i\le\sum_{i\in\mathcal I}\widehat Z_i<10^{-5}$.

The sum $\mathcal Z$ therefore bounds the accumulated decision difficulty without assuming a reconciliation efficiency. The resulting information-set sizes are listed in Table~\ref{pdimpl:masks}. Coordinates follow the standard minus-child/plus-child polar-tree order, starting at index zero.
\begin{table}[ht]
\centering
\caption{Explicit conservative polar sets, with $V_\phi=0.995$ and $\mathcal Z<10^{-5}$ for every row.}
\label{pdimpl:masks}
\begin{tabular}{crrr}
\hline
$\Lambda_0$ & $b$ & $|\mathcal I|$ & $b-|\mathcal I|$\\\hline
0.999 & $2^{22}$ & 945653 & 3248651\\
0.995 & $2^{22}$ & 935169 & 3259135\\
0.990 & $2^{22}$ & 922506 & 3271798\\
0.999 & $2^{24}$ & 3982955 & 12794261\\
0.995 & $2^{24}$ & 3939906 & 12837310\\
0.990 & $2^{24}$ & 3887490 & 12889726\\\hline
\end{tabular}
\end{table}

\subsection{Finite records and soft-input accuracy}
The reference likelihoods above use exact real-valued preparation records. An implementation uses finite public messages and finite arithmetic. We compare these two procedures on the same preparation records: if their finite decoder inputs agree, their subsequent decoding decisions can be compared directly. The probability that an input changes is charged as an honest completion allowance.

A cell is an interval containing a recorded value, whose index is communicated or stored instead of that value. Alice's public normalized intensity has logarithmic cells of width $2^{-48}$ over $[2^{-60},64)$, plus underflow and overflow labels; her folded phase has $2^{48}$ equal cells in $[0,\pi)$. The logarithmic cells are uniform in $\ln A$. Bob keeps his intensity record private and uses the same tail cutoffs. For bounded-argument arithmetic, he may represent his angle as $\vartheta=z_B\pi+\phi$ modulo $2\pi$, with $\phi\in[-\pi/3,2\pi/3)$ and $z_B\in\{0,1\}$; then $\cos\vartheta=(-1)^{z_B}\cos\phi$. This is only a local representation of his complex record. All intensity and phase labels are invariant under Alice's private sign. Regular Alice public-cell centers change the fine posterior by at most
\begin{equation}
 \delta_{\rm pub}=\left(\frac1{16}+\frac\pi4\right)2^{-48}.
\end{equation}
Indeed, set $c=2(1-\Lambda_0)/\Lambda_0$. The posterior is $p=1/2+V_\phi\sqrt{AB}\cos\vartheta/(A+B+c)$, and
\begin{equation}
\frac{\partial p}{\partial\ln A}
 =\frac{V_\phi\sqrt{AB}\cos\vartheta(B+c-A)}{2(A+B+c)^2}.
\end{equation}
With $x=A/(B+c)$, its absolute value is at most $\sqrt{x}|1-x|/[2(1+x)^2]\le1/8$. Similarly, $|\partial p/\partial\vartheta|\le\sqrt{AB}/(A+B+c)\le1/2$. Moving from a cell point to its center changes $\ln A$ by at most $2^{-49}$ and the phase by at most $\pi2^{-49}$. The sum of the two resulting errors gives the displayed $\delta_{\rm pub}$. A directed 90-digit frontend evaluates the center posterior; exponentials and square roots use outward neighbors, and a degree-32 cosine polynomial has remainder at most $\pi^{34}/34!$. Bob's binary64 intensity is enclosed with relative tolerance $2^{-52}$ and the fraction specifying his locally folded angle within its interval with absolute tolerance $2^{-53}$. Rounding the resulting posterior to a multiple of $2^{-48}$ gives the conservative error contract $\delta_{\rm num}=2^{-44}$. Classification into the 256 reliability cells then uses exact integer comparisons. These precisions concern classical records and computations, not optical state accuracy.

Assign a neutral soft input if Alice's public intensity cell or Bob's private intensity lies in a tail. This keeps these positions in the protocol while giving the decoder no evidence for either sign. The normalized reference intensity has marginal density
\begin{equation}
f_A(a)=e^{-a}\left(1-\frac{\Lambda_0}{2}+\frac{\Lambda_0a}{2}\right),
\end{equation}
and Bob's intensity has the same marginal. This density is at most one, so the two exact-record underflows cost at most $2^{-59}$. Its survival function is $\Pr(A\ge a)=e^{-a}(1+\Lambda_0a/2)$; hence the combined exact-record overflow probability is at most $2e^{-64}(1+32\Lambda_0)\le66e^{-64}$. We retain the conservative implementation allowance $\delta_{\rm tail}=2^{-59}+66e^{-63}$. It also leaves room for finite-precision threshold comparisons. Even if both compared intensities have relative error at most $\varepsilon_{\rm mach}=2^{-52}$, their combined tail probability is bounded by
\begin{equation}
\frac{2^{-59}}{1-\varepsilon_{\rm mach}}
 +66e^{-64/(1+\varepsilon_{\rm mach})}
 \le\delta_{\rm tail}.
\end{equation}
Thus the allowance covers the small displacement of both cutoffs, including the underflow cutoff; no tail class is discarded.

The signed-posterior density is bounded by
\begin{equation}
 B_{\Lambda_0,V_\phi}=\frac{\Lambda_0+2/\Lambda_0-2}{V_\phi}.
\end{equation}
To derive this density bound, condition on $T$. The signed variable $2p-1$ is uniform on $[-V_\phi a(T),V_\phi a(T)]$, where $a(T)$ was defined above. The conditional density of $p$ is therefore $1/[V_\phi a(T)]$ on its support. Averaging this upper bound gives
\begin{equation}
\begin{aligned}
&\frac{1}{V_\phi}\int_0^\infty\frac{f_T(T)}{a(T)}dT\\
 &=\frac{2}{\Lambda_0V_\phi}\int_0^\infty e^{-T}
   \left(1-\Lambda_0+\frac{\Lambda_0T}{2}\right)^2dT\\
 &=B_{\Lambda_0,V_\phi}.
\end{aligned}
\end{equation}
An input classification can change only if the exact posterior lies sufficiently near a cell boundary, or if a record lies in a tail. There are $2K-1$ signed-posterior boundaries. The density bound converts a bound on posterior displacement into a probability of crossing one of these boundaries. A displacement of magnitude at most $\delta=\delta_{\rm pub}+\delta_{\rm num}$ can cross a particular boundary only within an interval of length $2\delta$. Summing these lengths gives $(4K-2)\delta$. Multiplication by the density bound, addition of the tail allowance and a union bound over $n$ records give the session estimate. Coupling the exact reference inputs and implemented inputs on the same preparation records therefore bounds their session disagreement by
\begin{equation}
 \epsilon_{\rm in}\le n\left[(4K-2)B_{\Lambda_0,V_\phi}
 (\delta_{\rm pub}+\delta_{\rm num})+\delta_{\rm tail}\right].
 \label{pdimpl:frontend}
\end{equation}
The additional $n|\Lambda-\Lambda_0|$ allowance described above compares the declared source modulation with this reference channel. These are honest abort allowances; the security proof already conditions on Alice's actual public labels. This comparison does not assert that a cell-center posterior equals the exact posterior conditioned on a finite cell. Instead, it bounds the probability that the center approximation changes the finite input sent to the decoder. When the inputs agree, both procedures use the same bin likelihoods and the same decoding algorithm. The masks and completion allowances depend on Bob's local decoding inputs and Alice's public cells.

\subsection{A finite-precision decoder and its honest availability bound}
Bob decodes the polar-transformed coordinates in their prescribed order by successive cancellation. Each information-bit decision uses the channel likelihoods and the preceding decoded information bits; a disclosed coordinate is inserted as its true value when its turn is reached. Future disclosed coordinates are ignored until then. This choice keeps each decision aligned with the synthesized channel used in the polar construction: its conditioning includes the observation and preceding coordinates, but averages over future coordinates. Using future disclosed values could improve a decoder, but would require a different analysis. Ignoring them here supplies a fully specified algorithm whose finite error probability and arithmetic complexity can be bounded explicitly.

Very small likelihoods can occur even when their ratio remains informative. To avoid losing this ratio through numerical underflow, represent a positive likelihood as a binary64 mantissa in $[1,2)$ and a separate integer power of two. Products and sums then avoid probability underflow. Use round-to-nearest with $u=2^{-53}$, exclude fast-math transformations, and omit a smaller addend only when its exponent is more than $54$ below the larger one. Such an omission has relative error less than $u$. The positive, unnormalized pair updates are
\begin{align}
 f((a,b),(c,d))&=(ac+bd,ad+bc),\\
 g_0((a,b),(c,d))&=(ac,bd),\\
 g_1((a,b),(c,d))&=(bc,ad).
\end{align}
The first update predicts the parity of two binary variables by summing over their unknown second bit. Once that parity is known, the second-bit update retains the two products compatible with it, giving $g_0$ or $g_1$. Normalizing each pair is unnecessary: only its ratio affects a decision, and avoiding division makes the rounding analysis simpler.

Outward conversion of the initial high-precision likelihood enclosure gives binary64 endpoints $[l,h]$. This bracket must enclose the true bin likelihood and satisfy $h/l\le1+\epsilon_0$, where $\epsilon_0=2^{-50}$. Choose a representable initial value inside that interval; zeros are handled exactly. Here the exact initial likelihood means the true bin probability, not a bin-center substitute. If both it and its chosen approximation lie in $[l,h]$, their ratio lies between $(1+\epsilon_0)^{-1}$ and $1+\epsilon_0$. Repeated application of the positive sum/product updates makes every later likelihood a polynomial with nonnegative coefficients in these initial values. There is no cancellation that could amplify relative errors. For a degree-$d$ term, at most $d-1$ products and $d-1$ sums contribute rounding factors; this follows by counting the internal vertices of its binary computation tree. Each elementary relative factor lies between $1-u$ and $1+u$, including the allowed omission of a much smaller addend. A likelihood polynomial of degree $d$ therefore has at most $2(d-1)$ such factors on each monomial. In this paragraph, $d$ denotes the polynomial degree. Consequently its represented value $\widehat x$ and exact value $x$ satisfy
\begin{equation}
 (1-u)^{2(d-1)}(1+\epsilon_0)^{-d}
 \le \frac{\widehat x}{x}
 \le (1+u)^{2(d-1)}(1+\epsilon_0)^d.
\end{equation}
For $d\le b\le2^{24}$ both bounds lie within $1\pm\rho$, with $\rho=2^{-24}$. At the largest allowed degree, the absolute relative-error envelope is less than $1.864\times10^{-8}$, whereas $\rho\simeq5.961\times10^{-8}$. The same bound holds for sums of monomials because their coefficients are nonnegative. The decoder must decide only when the likelihood ordering survives these error bars. A zero decision is certified only if
\begin{equation}
 \widehat x_0(1-\rho)>\widehat x_1(1+\rho);
\end{equation}
the reversed inequality certifies a one. Binary significands allow these comparisons by exact integer products. Otherwise the decoder aborts. The reported bit, whenever certified, is the exact maximum-posterior bit. To quantify an ambiguous decision, put $\kappa=(1+\rho)/(1-\rho)$. If neither comparison is satisfied, the represented likelihood ratio lies between $\kappa^{-1}$ and $\kappa$. Accounting for the relative error of each likelihood places the exact ratio $x_0/x_1$ between $\kappa^{-2}$ and $\kappa^2$. Writing $p=x_0/(x_0+x_1)$ for this synthesized-bit posterior then gives
\begin{equation}
 |p-1/2|\le\frac{\rho}{1+\rho^2}\le\rho.
\end{equation}

To bound honest completion, consider a hypothetical decoder given every preceding true bit, and let $P_{e,i}$ be the mean error probability of its exact maximum-posterior decision on information coordinate $i$. Pointwise in its observation, an ambiguous decision has $\min(p,1-p)\ge1/2-\rho$. Let $q_i=\min(p,1-p)$ be the conditional error of the exact maximum-posterior decision. At an ambiguous observation the failure probability is one, but $1\le q_i/(1/2-\rho)$. At a certified observation the decision is exact and its conditional error is $q_i$, which obeys the same upper bound. Averaging over observations therefore bounds the probability of an error or numerical abort by $P_{e,i}/(1/2-\rho)$. The hypothetical decoder removes error propagation from the analysis. It need not be implemented: the actual decoder and this hypothetical decoder coincide until the first error or abort. A union bound, together with $P_{e,i}\le Z_i/2$, gives for $F$ honest independent frames
\begin{equation}
 \epsilon_{\rm dec}\le
 \frac{F\sum_{i\in\mathcal I}Z_i}{1-2\rho}.
\end{equation}
This includes numerical aborts of the specified arithmetic algorithm. The separate frontend coupling allowance accounts for finite public labels and posterior-bin classification. Channel-construction rounding is a physical degradation and requires no additional coupling penalty. The algorithm requires $O(n\log b)$ fixed-precision arithmetic operations and $O(b)$ working memory.

\subsection{Intensity-test thresholds, block budgets and rates}
For the parameter test, each party discloses an interval containing its normalized squared amplitude, $u_A=|\alpha|^2/q$ or $u_B=|\beta|^2/q$. These intervals are called radial cells because $|\alpha|$ and $|\beta|$ are radii in the complex preparation planes. They are calculated from the recorded modulation values and require no additional optical measurement. Both parties disclose them on the test positions, which do not contribute key bits. Their endpoints determine a lower-rounded test score with pointwise error at most $10^{-10}$, including overflow outcomes.

The honest radial variable $s=|a_0|^2+|b_0|^2$, where $a_0=\alpha^*/\sqrt q$ and $b_0=\beta^*/\sqrt q$, has distribution
\begin{equation}
 s\sim(1-\Lambda)\operatorname{Gamma}(2,1+\nu)
       +\Lambda\operatorname{Gamma}(3,1+\nu),
\end{equation}
with Gamma shape and scale in that order. In this subsection $s$ denotes the intensity sum used by the test, rather than the earlier decoder normalization scale. The density of $\operatorname{Gamma}(a,\sigma)$ is $s^{a-1}e^{-s/\sigma}/[\Gamma(a)\sigma^a]$ for $s\ge0$, where $\Gamma$ is Euler's Gamma function. The two mixture terms can be read directly from the accepted state. A thermal mode of mean $\nu$ gives a heterodyne intensity with exponential density of scale $1+\nu$, namely a Gamma law of shape one. Adding the two independent intensities gives shape two. The sum of the two diagonal photon-added contributions weights that joint density by its total intensity divided by $2(1+\nu)$, changing the normalized total-intensity law to shape three. The interference terms integrate to zero over the phase angles. This derives both mixture components and also shows why phase visibility does not enter the test law. In terms of the earlier normalized total $T=A+B$, the test variable is $s=(1+\nu)T$. These distributions give a direct way to evaluate the test's expected score and fluctuations under normal operation. To evaluate the bounded score $Q(s)=(3/2)^5e^{-s/2}[-s^2+14s/3-32/9]$, use
\begin{equation}
 \mathbb E[s^je^{-cs}]=\frac{(a)_j\sigma^j}{(1+c\sigma)^{a+j}}
\end{equation}
for Gamma shape $a$, scale $\sigma$, and rising factorial $(a)_j$, with $(a)_0=1$. Multiplying the Gamma density by $s^je^{-cs}$ changes its integral to $\Gamma(a+j)/(c+1/\sigma)^{a+j}$; division by the original normalization gives the stated identity. Here $j$ is a nonnegative integer and the applications use $c=1/2$ or $c=1$.

For an explicit moment calculation, let $M_j(c)$ denote the stated moment averaged over the shape-two and shape-three mixture, and put $\gamma_0=-32/9$, $\gamma_1=14/3$, $\gamma_2=-1$. Then
\begin{align}
\mu_Q&=(3/2)^5\sum_{j=0}^2\gamma_jM_j(1/2),\\
\mathbb E[Q^2]&=(3/2)^{10}\sum_{j,k=0}^2\gamma_j\gamma_kM_{j+k}(1),\\
V_h&=\mathbb E[Q^2]-\mu_Q^2.
\end{align}
Thus only finitely many Gamma moments are required; $V_h$ is the variance of the score under the honest model. If $m$ positions are tested, an honest test-abort allocation $\epsilon_h$ gives
\begin{equation}
 \delta_h=\frac{28\ln(1/\epsilon_h)}{3m}
 +\sqrt{\left[\frac{28\ln(1/\epsilon_h)}{3m}\right]^2
       +\frac{2V_h\ln(1/\epsilon_h)}m}.
\end{equation}
This margin is obtained by solving Bernstein's inequality, rather than by assuming a Gaussian distribution of the empirical mean. If $\overline Q$ is the mean of the $m$ independent honest test scores, the previously established bounds $Q\ge-27$ and $\mu_Q\le1$ give $\mu_Q-Q\le28$. Therefore
\begin{equation}
\Pr\{\overline Q<\mu_Q-\delta\}
 \le\exp\left[-\frac{m\delta^2}{2(V_h+28\delta/3)}\right].
\end{equation}
Setting this upper bound to $\epsilon_h$ and solving its quadratic inequality for the positive root gives the displayed $\delta_h$. The acceptance threshold is chosen below the honest mean, leaving a margin for downward statistical fluctuations and score rounding. Set $g_*\le\mu_Q-\delta_h-10^{-10}$, rounded down to a multiple of $10^{-12}$. Bernstein's lower-tail bound applies since $\mu_Q-Q\le28$; $10^{-10}$ covers the finite score's downward rounding. This honest variance never replaces the adversarial variance envelope in the composable proof.

The numbers of test and key positions are fixed before transmission. After the relay has committed its announcements, the allocation of accepted positions is chosen with fresh randomness independent of the preparation records. More tests tighten the secrecy estimate but leave fewer raw key bits; the design balances these competing effects.

For duration $T_{\rm acq}$, clock $f_0=1$ GHz and quantum duty $d_Q=0.90$, declare $M=\lfloor d_Qf_0T_{\rm acq}\rfloor$ attempted relay uses. Select $S_{\rm acc}=\lfloor0.99Mp_{\rm obs}\rfloor$ accepted positions, aborting on a shortfall. Choose $F$ full polar frames, so $n=Fb$, $m=S_{\rm acc}-n>0$, and $L_{\rm EC}=F(b-|\mathcal I|)$. Under the honest independent model, the total number of heralds is binomial with mean $Mp_{\rm obs}$. The declared accepted count is one percent below that mean. The multiplicative Chernoff inequality therefore bounds a shortfall by $e^{-0.01^2Mp_{\rm obs}/2}$, which is below $10^{-10}$ at every positive plotted point. Conditional accepted records have the same honest law regardless of where the heralds occurred; selecting test positions independently of their records permits the preceding test calculation. The security allowances are $\epsilon_r=\epsilon_s=2.5\times10^{-23}$, $\epsilon_{\rm PA}=\epsilon_{\rm auth}=2\times10^{-11}$ and verification length $v=36$. Let $k$ be the certified cap on virtual parity errors among the $n$ key positions from the preceding composable proof, and set $e_U=10^{-6}\lceil10^6k/n\rceil$. Define $h_2(x)=-x\log_2x-(1-x)\log_2(1-x)$, with $0\log_2 0=0$. Abort if $e_U\ge1/2$; otherwise the output length is\begin{equation}\ell=\left[\left\lfloor n[1-h_2(e_U)]-L_{\rm EC}-v-1-2\log_2(1/\epsilon_{\rm PA})\right\rfloor\right]_+.\label{pdimpl:length}\end{equation} Here $[x]_+=\max\{x,0\}$, and a zero output length also causes an abort. Rounding $k/n$ upwards is conservative because $h_2$ increases on $[0,1/2]$. The first term is the certified secret entropy supplied by the complementary-parity argument. The remaining terms charge the disclosed polar coordinates, verification tag, final accept/abort bit and privacy-amplification error, respectively. The finite decoder changes the honest probability of reaching this output; it does not change the general-attack entropy proof.

The protocol parameters are selected before transmission by optimizing over the six information sets in Table~\ref{pdimpl:masks}, admissible frame counts $F$, and $\epsilon_h\in\{10^{-6},10^{-5},10^{-4},10^{-3},0.003,0.01,0.03,0.05\}$. Directed high-precision arithmetic certifies the source modulation, herald counts, test threshold, adversarial confidence bound, integer error cap and output length. Let $p_{\rm pass}$ be the honest completion probability and $R_{\rm fin}=p_{\rm pass}\ell/M$ the expected rate. They obey
\begin{align}
 p_{\rm pass}^L&=\left[1-\epsilon_h-10^{-10}-\epsilon_{\rm in}
   -\epsilon_{\rm dec}-n|\Lambda-\Lambda_0|\right]_+,\\
 R_{\rm fin}&\ge p_{\rm pass}^L\frac\ell M,
 ~~ \text{throughput}=d_Qf_0R_{\rm fin}.
 \label{pdimpl:rate}
\end{align}
The completion bound adds the count-shortfall, test, finite-input, decoding and reference-channel mismatch allowances without requiring independence between these failures. To see this, consider the event on which the required number of records is obtained, the test passes, the reference and implemented decoder inputs agree, and every reconstructed frame is correct. The probability of its complement is at most the sum of the listed allowances. On this event verification succeeds for every seed because the strings agree. Thus no additional honest verification-failure cost is needed, although verification remains essential to correctness against an arbitrary attack. The separate allowance for $\Lambda-\Lambda_0$ transfers the reference-channel reconstruction statement to the actual declared honest modulation. The general-attack theorem determines the key length. The honest model determines the available successes, design parameters and completion probability.

The attempt count $M$ includes failed heralds and tests; reference allocation reduces the attempts available in a stated acquisition time. The benchmarks use physical fiber transmissivity $\tau^2$. Authentication-key consumption is subtracted separately for net expansion.

For the budget table, write $R_{\rm fin}^L=p_{\rm pass}^L\ell/M$ for the certified rate lower bound. Its benchmark ratio and corresponding throughput use this same lower bound.

\par\medskip\noindent
\begin{minipage}{\columnwidth}
\makeatletter\def\@captype{table}\makeatother
\centering\footnotesize
\caption{Certified finite budgets at $85$ dB using the polar information set with $b=2^{24}$ and $|\mathcal I|=3\,939\,906$. Both designs have $\Lambda_0=0.995$ and $N\simeq0.00541847$. The quantum duty is $0.9$ at a $1$ GHz nominal clock. Rate, benchmark-ratio and throughput entries report lower-bound calculations, rounded for display.}
\label{pdimpl:budget}
\begin{tabular}{lrr}\hline
Quantity & One day & Two days\\\hline
$M$ & $7.776\times10^{13}$ & $1.5552\times10^{14}$\\
$S_{\rm acc}$ & 37\,719\,403 & 75\,438\,806\\
$F$ & 1 & 3\\
$n$ & 16\,777\,216 & 50\,331\,648\\
$m$ & 20\,942\,187 & 25\,107\,158\\
$\epsilon_h$ & 0.05 & 0.05\\
$g_*$ & 0.976839718177 & 0.977011733706\\
$k$ & 544\,666 & 1\,570\,595\\
$e_U$ & 0.032465 & 0.031205\\
$L_{\rm EC}$ & 12\,837\,310 & 38\,511\,930\\
$\ell$ & 473\,508 & 1\,733\,181\\
$p_{\rm pass}^{L}$ & 0.948953 & 0.946860\\
$R_{\rm fin}^L$ & $5.77851\times10^{-9}$ & $1.05522\times10^{-8}$\\
$R_{\rm fin}^L/K_0$ & 1.2666 & 2.3130\\
Lower-bound bits/s & 5.2007 & 9.4970\\\hline\end{tabular}\end{minipage}
\par\medskip

Acquisition times exclude implementation-dependent decoding time. The model assumes exact photon-number classification, matched modes, independent phase noise and ideal trusted preparations. Detector misclassification, correlated reference backgrounds, source side channels and calibration drift require additional models for a deployed apparatus.


\section{Optimized phase disclosure under IID attacks}
\label{app:opd}

The optical preparation and heralding of CV-TF-QKD leave room for further optimization of the classical processing. The fixed CV-TF-PD map keeps reconciliation simple and supports the composable analysis. Here we describe a possible alternative, optimized phase disclosure (CV-TF-OPD). It retains the sign representation and reconstruction logic, while choosing the phase-disclosing party, phase-bin width, key owner, and retained classes from a declared family. For fixed optical parameters, these choices act on the same accepted optical state. The term optimized refers to this family, rather than to all possible protocols.

We first explain the processing through four steps and a common example, then derive its asymptotic IID rate and compare its performance with PD. The two selection rules act at different times: the disclosure policy is chosen from the intensities before either folded phase is announced, whereas the key owner is chosen after the phase bin is known. Neither rule gives a decoder access to the other user's undisclosed phase.

\subsection{Protocol and worked example}

As a recap, $\alpha$ and $\beta$ are the corrected preparation records, with $A=|\alpha|^2$ and $B=|\beta|^2$ as in Eq.~\eqref{main:correction}. Alice's sign $Z_A$ and folded phase $\theta_A$ are defined in Eq.~\eqref{proc:signmap}. A folded phase is the ordinary phase modulo $\pi$; its private sign distinguishes the two antipodal amplitudes. OPD uses this same representation, but discloses both intensities and allows either party to own the key.

\subsubsection*{Step 1: choose the disclosure policy from intensities}

Alice and Bob first announce $A$ and $B$. Let $o\in\mathcal O=\{A,B\}$ label the phase-disclosing party. The angular half-width $h$ is chosen from
\begin{equation}
 \mathcal H=\pi\{1/12,1/8,1/6,1/4,1/3,5/12\}.
 \label{proc:windows}
\end{equation}
Each candidate width defines two relative-phase bins,
\begin{equation}
 \begin{aligned}
 D_0(h)&=[-h,h), & \omega_0(h)&=2h/\pi,\\
 D_1(h)&=[h,\pi-h), & \omega_1(h)&=1-2h/\pi.
 \end{aligned}
 \label{opd:bins}
\end{equation}
The probabilities $\omega_d$ are the bin widths divided by $\pi$ in the phase-covariant accepted model. The half-open intervals fix the boundary convention.

Before acquisition, the parties fix a design $\vartheta=(N,t,\xi)$, using the optical parameters of Eq.~\eqref{asym:parameters}, and prepare the two selection rules. For every prospective public class and owner $j\in\{A,B\}$, the score $k_j^{(o)}(A,B,h,d;\vartheta)$ is the owner's conditional mutual information with the decoder minus Eve's conditional Holevo information. Both terms include the class's public disclosures, and the decoder uses only its own private record. A positive score identifies a useful class for that owner. The derivation of these scores is given after the protocol description. Phase covariance makes them independent of the absolute disclosed phase.

The first selection must account for both bins that could be announced. Its average score is
\begin{equation}
 \begin{split}
 &K_{o,h}(A,B;\vartheta)=\sum_{d=0}^1\omega_d(h)\\
 {}&\times\max\{0,k_A^{(o)}(A,B,h,d;\vartheta),
                  k_B^{(o)}(A,B,h,d;\vartheta)\}.
 \end{split}
 \label{proc:policyscore}
\end{equation}
The parties use the predetermined lookup
\begin{equation}
 f_\vartheta(A,B)=
 \underset{(o,h)\in\mathcal O\times\mathcal H}{\operatorname{arg\,max}}
 K_{o,h}(A,B;\vartheta).
 \label{proc:firstlookup}
\end{equation}
For ties, list Alice-first candidates in increasing $h$, followed by Bob-first candidates in increasing $h$, and take the first maximum. The lookup and this ordering are fixed before acquisition. Its input contains no phase or private sign.

As an example, reuse the three accepted plus-port records of Sec.~\ref{proc:pd}, all with $A_i=B_i=0.01$, and add a fourth plus-port round with
\begin{equation}
 \begin{aligned}
 \alpha_4&=\sqrt{0.11},\\
 \beta_4&=\sqrt{0.0275}\,e^{i\pi/6}.
 \end{aligned}
 \label{opd:fourth_record}
\end{equation}
Thus $Z_{A,4}=0$ and $\theta_{A,4}=0$. For $N=0.055$, total loss $90$ dB, and per-arm input-referred excess noise $\xi=10^{-5}$ SNU, the evaluated lookup gives
\begin{center}
\begin{tabular}{c|ccc}
rounds & $(A,B)$ & $f_\vartheta(A,B)$ & maximum $K$\\\hline
1--3 & $(0.01,0.01)$ & $(A,5\pi/12)$ & $0.355394$\\
4 & $(0.11,0.0275)$ & $(A,\pi/3)$ & $0.064201$
\end{tabular}
\end{center}
These are rounded numerical scores for the example design. The equal-intensity orientation tie is resolved in favor of Alice. The first three rounds receive the same initial policy because their announced intensities coincide, even though their private phases differ.

\subsubsection*{Step 2: disclose the selected phase and relative-phase bin}

The selected party announces its folded phase. For Alice-first disclosure, this is $\theta_A$ from Eq.~\eqref{proc:signmap}. Bob then writes his corrected record as
\begin{equation}
 \begin{gathered}
 \beta=(-1)^{Z_B}\sqrt B\,e^{i(\theta_A+\phi)},\\
 Z_B\in\{0,1\},\\
 -h\le\phi<\pi-h.
 \end{gathered}
 \label{opd:relative_fold}
\end{equation}
He obtains $\phi$ by reducing $\arg\beta-\theta_A$ modulo $\pi$ into the stated interval; removing an odd multiple of $\pi$ changes the sign. He announces only the bin $d$ containing $\phi$. For Bob-first disclosure, interchange the parties. At a zero amplitude, assign the corresponding sign and phase by a fixed convention; this event has zero probability in the continuous model.

The resulting public class is
\begin{equation}
 C=(s,A,B,o,h,\theta_o,d),
 \label{proc:opdclass}
\end{equation}
where $s$ is the relay-port announcement and $\theta_o$ is the disclosed folded phase. The other user's exact relative phase and both signs remain private.

For this example, Alice announces $(5\pi/6,0,0,0)$. Bob's first three folded relative phases are $\pi/6$, $\arccos(3/5)$, and $\arccos(-1/5)$. At $h=5\pi/12$, the first two lie in $D_0$ and the third in $D_1=[5\pi/12,7\pi/12)$. The fourth relative phase is $\pi/6\in D_0$ for $h=\pi/3$. Bob therefore announces the bin string $0010$ and keeps the corrected sign string $1010$ private. In the first round, reducing $-5\pi/6$ to $\pi/6$ removes an odd half-turn, which explains the sign change.

\subsubsection*{Step 3: select an owner or reject the public class}

The second lookup $g_\vartheta(C)$ compares the two owner scores for the announced class. Reject if both scores are nonpositive. Otherwise select Alice when her positive score is at least Bob's, and select Bob in the remaining case. A positive tie therefore selects Alice. These decisions depend only on public labels, not on either private sign. Each retained round supplies at most one raw key bit.

Continuing the example, the class scores and choices are
\begin{center}
\begin{tabular}{c|rrl}
rounds & $k_A$ & $k_B$ & $g_\vartheta(C)$\\\hline
1--2 & $ 0.426472$ & $ 0.366322$ & Alice\\
3 & $-0.065439$ & $-0.078821$ & reject\\
4 & $-0.121461$ & $ 0.096302$ & Bob
\end{tabular}
\end{center}
Alice's retained subblock is $10$ from rounds 1--2. Bob's retained subblock is the corrected sign $0$ from round 4. Round 3 contributes no key but remains in the relay-use denominator. The fourth round illustrates why disclosure and ownership are separate choices: Alice discloses the phase, but Bob owns the bit.

\subsubsection*{Step 4: reconstruct each owner subblock, verify, and extract}

For Alice-first disclosure and an Alice-owned bit, Bob uses his exact private relative phase. For a Bob-owned bit, Alice knows only the bin and must average over the phase she does not know. The relevant cosine averages are
\begin{equation}
 \begin{aligned}
 c_D&=\frac1{|D|}\int_D\cos\phi\,d\phi,\\
 c_{D_0}&=\frac{\sin h}{h},\\
 c_{D_1}&=0.
 \end{aligned}
 \label{proc:bincos}
\end{equation}
These geometric averages also hold with excess noise. The phase-independent coefficient of the cosine in Eqs.~\eqref{proc:noisyposterior00} and~\eqref{proc:noisyposterior} multiplies the realized cosine or its bin average. In particular, a decoder may use an exact phase only when that phase belongs to its own record.

Each owner sends a syndrome for its own retained subblock, and the other party reconstructs it using the available private information and public labels. The labels are side information to a code spanning many retained rounds; identical continuous records need not recur. Fresh universal-hash tags verify the two reconstructions, and a mismatch causes an abort. Privacy amplification is then applied to each owner subblock at a length accounting for its full public transcript. Bob-first disclosure uses the same rules with the users interchanged.

In the example, Bob's probabilities for Alice's first two bits are $p_1\simeq0.932796$ and $p_2\simeq0.200150$. Alice announces the parity $1\oplus0=1$. The compatible strings $10$ and $01$ have likelihoods $p_1(1-p_2)\simeq0.746097$ and $(1-p_1)p_2\simeq0.013451$, so Bob reconstructs $10$. For round 4, Alice averages over $D_0$ and obtains $\Pr(Z_B=0\mid Z_A=0,C)\simeq0.830773$, favoring Bob's actual bit zero. She cannot use his private value $\phi_4=\pi/6$ in place of that average. Fresh tags check both reconstructions. These short subblocks demonstrate decoding only and do not yield a certified secret key.

\begin{figure}[tbp]
\centering
\begin{tikzpicture}[
 bx/.style={draw=black!65,rounded corners=2pt,align=center,font=\footnotesize,inner xsep=4pt,inner ysep=5pt,line width=.45pt},
 wide/.style={bx,text width=7.35cm,fill=black!3},
 branch/.style={bx,text width=2.05cm,fill=blue!4},
 flow/.style={-{Latex[length=1.6mm]},line width=.5pt}]
\node[wide] (one) at (0,0) {\textbf{1. Announce intensities; apply $f_\vartheta(A,B)$}\\Choose orientation and width from the bin-averaged score.};
\node[wide,below=.35cm of one] (two) {\textbf{2. Selected party announces its folded phase}\\Other party folds, corrects its sign and announces only the bin.};
\node[wide,below=.35cm of two] (three) {\textbf{3. Apply $g_\vartheta(C)$ to the public class}\\Compare Alice-owner and Bob-owner scores.};
\node[branch,anchor=north,minimum height=1.1cm] (alice) at ([xshift=-2.6cm,yshift=-.45cm]three.south) {\textbf{Alice owns}\\Bob decodes.};
\node[branch,anchor=north,minimum height=1.1cm] (bob) at ([yshift=-.45cm]three.south) {\textbf{Bob owns}\\Alice decodes.};
\node[branch,fill=black!6,anchor=north,minimum height=1.1cm] (reject) at ([xshift=2.6cm,yshift=-.45cm]three.south) {\textbf{Reject}\\Both scores nonpositive.};
\node[wide,anchor=north] (four) at ([yshift=-.65cm]bob.south) {\textbf{4. Reconcile owner subblocks, verify and extract}\\Use an exact local phase only when the decoder holds it;\\otherwise average over its public bin.};
\draw[flow] (one) -- (two);
\draw[flow] (two) -- (three);
\draw[flow] (three.south) -- ++(0,-.2) -| (alice.north);
\draw[flow] (three.south) -- (bob.north);
\draw[flow] (three.south) -- ++(0,-.2) -| (reject.north);
\draw[flow] (alice.south) -- (alice.south |- four.north);
\draw[flow] (bob.south) -- (four.north);
\end{tikzpicture}
\caption{OPD processing. Disclosure is selected before phases are announced; ownership and rejection are selected after the bin is known. The two choices are distinct, and each retained owner subblock has its own reconstruction direction.}
\label{proc:opdflow}
\end{figure}
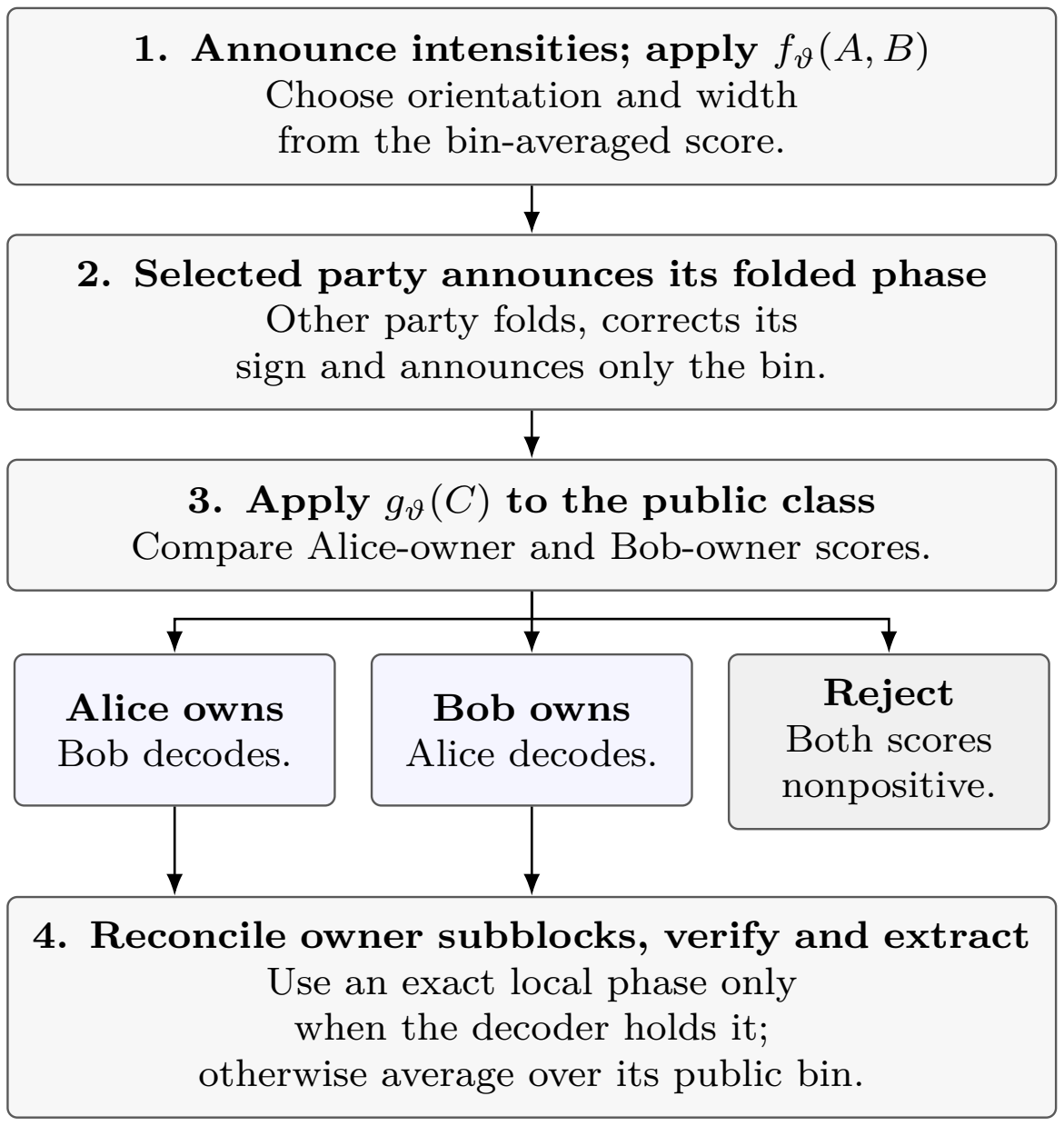

An approximate implementation must evaluate the achieved rate of its actual finite lookup table. It is not assigned the ideal maximum merely because it approximates the optimizing rule. Figure~\ref{proc:opdflow} summarizes the two selection stages and the ownership branches.

\subsection{Asymptotic IID security}

We now justify the scores used in the two lookups. We assume collective attacks and the complete asymptotic identification of the accepted preparation statistics described in Appendix~\ref{coll:appendix}. Eve may retain all her systems and measure them jointly after the public communication. The thermal-loss model specifies the identified state; excess noise or covariance data alone do not certify that non-Gaussian state. This appendix establishes IID asymptotic rates, without extending the composable proof of the fixed PD map to the additional OPD disclosures.

\subsubsection{Accepted state and the decoder's information}

As a recap, the accepted state is $\rho=\rho_{\lambda,\nu}$ in Eq.~\eqref{coll:rho}, with the optical parameters and acceptance probability of Eq.~\eqref{asym:parameters}. Eve holds a purification of this state, including the thermal noise. Ideal detection and phase alignment are assumed here. OPD changes the public measurement classes, while leaving the optical state unchanged at fixed $N,t,\xi$.

We use the intensity scale and normalized records of Eq.~\eqref{coll:scaled_intensities}: $s_0=q(1+\nu)$, $u=A/s_0$, and $v=B/s_0$, with $q=N/(1+N)$. Their accepted density $f_{uv}$ and contrast $V_\lambda(u,v)$ are given in Eq.~\eqref{coll:intensity_density}. Below, write $V=V_\lambda(u,v)$. The physical-intensity lookup of Step 1 is therefore evaluated as $f_\vartheta(s_0u,s_0v)$ in these normalized coordinates; this is a change of variables, not a second policy.

For Alice-first disclosure, substitute Alice's sign representation and Eq.~\eqref{opd:relative_fold} into the accepted preparation density. Conditional on the public class in Eq.~\eqref{proc:opdclass}, the joint sign and hidden-angle density is
\begin{equation}
 \begin{gathered}
 p(z_A,z_B,\phi\mid C)
 =\frac{1+(-1)^{z_A+z_B}V\cos\phi}{4|D_d|},\\
 \phi\in D_d.
 \end{gathered}
 \label{opd:joint}
\end{equation}
Summing over the four sign pairs cancels the interference term. The sign-averaged angle is therefore uniform, the bin probability is $\omega_d=|D_d|/\pi$, and each individual sign is uniform in every public class. This proves the weights used in Eq.~\eqref{proc:policyscore}. Conditioning on either one of the signs leaves the hidden angle uniform within its bin as well.

If Alice owns the bit, Bob knows both $Z_B$ and his exact private $\phi$. If Bob owns it, Alice knows $Z_A$ and the bin, but not $\phi$. Using the binary entropy $h_2$ defined in Eq.~\eqref{coll:entropies}, their conditional mutual informations are
\begin{equation}
 \begin{aligned}
 I_A^{(A)}(u,v,D)
 &=1-\frac1{|D|}\int_D h_2\!\left(\frac{1-V\cos\phi}{2}\right)d\phi,\\
 I_B^{(A)}(u,v,D)
 &=1-h_2\!\left(\frac{1-Vc_D}{2}\right).
 \end{aligned}
 \label{opd:information}
\end{equation}
The superscript names the phase-disclosing party and the subscript the key owner. Each expression is one bit of prior sign entropy minus the decoder's remaining uncertainty. In the first line, the decoder knows the realized angle, so its binary entropy is averaged over that angle. In the second line, the angle is unknown to the decoder, so the cosine is averaged before the binary entropy is taken. The bin averages are those in Eq.~\eqref{proc:bincos}. In particular, Alice has no information about Bob's sign in $D_1$. Bob-first disclosure exchanges the party labels and $u,v$.

\subsubsection{Eve's information after the actual announcements}

Use the conditional-spectrum map $\Gamma_\rho(F)$ from Eq.~\eqref{coll:Gamma}. Its spectrum is Eve's normalized conditional spectrum for the positive measurement effect $F$. Thus $S[\Gamma_\rho(F)]$ gives her entropy without constructing a particular purification. The derivation in Appendix~\ref{coll:appendix} also shows why undisclosed alternatives must be mixed before taking this entropy.

For Alice-first disclosure, a common phase rotation removes the announced $\theta_A$ from the spectral calculation. Reuse Alice's sign effects $A_z(u)$ from Eq.~\eqref{coll:effects}, and set $b_v=\sqrt{(1+\nu)v}$. Bob's disclosed intensity and phase bin now give the effects
\begin{equation}
 \begin{split}
 B_z(v,D)=\frac1{|D|}\int_D
 \ket{(-1)^zb_ve^{-i\phi}}\\
 {}\times\bra{(-1)^zb_ve^{-i\phi}}\,d\phi.
 \end{split}
 \label{opd:effects}
\end{equation}
The minus sign in the phase follows from the conjugation between physical preparation amplitudes and virtual heterodyne outcomes in Eq.~\eqref{coll:source_heterodyne}. Common positive heterodyne and Jacobian factors cancel on normalization. Write $A_\Sigma=A_0+A_1$ and $B_\Sigma=B_0+B_1$. The public class and the two possible owner-bit effects are
\begin{equation}
 \begin{aligned}
 F_C&=A_\Sigma\otimes B_\Sigma,\\
 F_{A,z,C}&=A_z\otimes B_\Sigma,\\
 F_{B,z,C}&=A_\Sigma\otimes B_z.
 \end{aligned}
 \label{opd:owner_effects}
\end{equation}
The sign sums describe distinct classical alternatives, not coherent superpositions. The phase integral in Eq.~\eqref{opd:effects} is likewise performed before evaluating entropy. Averaging the entropies of exact-phase states would give Eve the hidden angle and evaluate a different transcript. In contrast to PD, Bob's public effect is no longer the identity, because OPD discloses his intensity and a phase bin.

Total photon parity commutes with $\rho$ and $F_C$ and exchanges the effects for either owner's two signs. The sign-conditioned states therefore have equal spectra and equal probabilities. Eve's conditional Holevo information is
\begin{equation}
 \begin{gathered}
 \chi_j^{(A)}(u,v,D)=S[\Gamma_\rho(F_C)]
                  -S[\Gamma_\rho(F_{j,0,C})],\\
 j\in\{A,B\}.
 \end{gathered}
 \label{opd:holevo}
\end{equation}
This includes the information in both intensities, the announced folded phase, and the phase bin. Phase covariance removes the numerical dependence on the absolute folded phase, not the disclosure itself. For Bob-first disclosure, exchange the two retained modes.

\subsubsection{Achievable rate and the two selection rules}

The preceding expressions evaluate the score introduced in Step 1:
\begin{equation}
 \begin{split}
 k_j^{(o)}(A,B,h,d;\vartheta)
 &=I_j^{(o)}(u,v,D_d(h))\\
 &{}-\chi_j^{(o)}(u,v,D_d(h)),
 \end{split}
 \label{opd:score}
\end{equation}
where $u=A/s_0$ and $v=B/s_0$. For a uniform binary key, $H(Z_j\mid E,C)=1-\chi_j^{(o)}$, while ideal one-way reconciliation costs $H(Z_j\mid Y,C)=1-I_j^{(o)}$. Here $Y$ is the decoder's actual private information. Subtracting these quantities gives Eq.~\eqref{opd:score}, the conditional Devetak--Winter achievable rate~\cite{DW}.

For a fixed disclosure choice $(o,h)$, the bin occurs with probability $\omega_d(h)$. After learning that bin, the parties may select the better positive owner score or reject. Averaging over these possibilities gives exactly $K_{o,h}$ in Eq.~\eqref{proc:policyscore}. Since the disclosure choice must precede the bin announcement, the maximization over $(o,h)$ acts on this bin-averaged score. This justifies the first lookup in Eq.~\eqref{proc:firstlookup}; optimizing a window using the subsequently realized bin would describe a different protocol.

The owner and retention flag are deterministic functions of the public class, so they add no undisclosed conditioning variable. Under the IID assumption, conditioning on the per-round public classes preserves the product structure. Applying the conditional one-way rate to the two retained owner subblocks and averaging over the accepted intensity distribution gives
\begin{equation}
 \begin{split}
 R_{\rm OPD}^{\rm iid}
 =p_{\rm acc}\int_0^\infty\!\int_0^\infty f_{uv}(u,v)\\
 {}\times\max_{o,h}K_{o,h}(s_0u,s_0v;\vartheta)\,du\,dv.
 \end{split}
 \label{opd:rate}
\end{equation}
Rejected classes contribute zero without renormalizing the remaining probability. Every attempted relay use, including unsuccessful heralds, is counted in the denominator. Public disclosures are already included in the conditional information quantities; their raw message lengths are not subtracted again.

Equation~\eqref{opd:rate} is the optimum within the stated windows, disclosure orientations, binary owner choices, and class rejection at fixed optical parameters. Since the OPD transcript differs from the PD transcript, no general inequality between their rates at equal modulation follows. The two maps must be evaluated with their respective disclosures.

\subsection{Numerical evaluation and performance}

The phase-integrated effects can be evaluated directly in the photon-number basis. For an interval $D=[c-\Delta/2,c+\Delta/2)$, define $\operatorname{sinc}x=\sin(x)/x$, continuously extended at zero. Then
\begin{equation}
 \begin{split}
 &\langle m|B_z(v,D)|m'\rangle
 =\frac{e^{-b_v^2}b_v^{m+m'}}{\sqrt{m!m'!}}\\
 {}&\times(-1)^{z(m-m')}e^{-i(m-m')c}
 \operatorname{sinc}\!\left[\frac{(m-m')\Delta}{2}\right].
 \end{split}
 \label{opd:phase_kernel}
\end{equation}
The two intervals have $(c,\Delta)=(0,2h)$ and $(\pi/2,\pi-2h)$. This formula integrates the hidden phase exactly in Eve's effects. To avoid an explicit square root of the two-mode state, factor a positive effect as $F=LL^\dagger$. The nonzero eigenvalues of $\sqrt\rho F\sqrt\rho$ equal those of $L^\dagger\rho L$, because these are the products $XX^\dagger$ and $X^\dagger X$ with $X=\sqrt\rho L$. Dividing their common nonzero spectrum by its sum gives the entropy in Eq.~\eqref{coll:Gamma}. Any parity blocks of that spectrum must be concatenated before normalization.

For the intensity integral, put $T=u+v$ and $x=u/T$, with Jacobian $T$. Then
\begin{equation}
 \begin{gathered}
 f_{uv}(u,v)\,du\,dv=e^{-T}T W_\lambda(T)\,dT\,dx,\\
 0\le x\le1.
 \end{gathered}
 \label{opd:quadrature}
\end{equation}
This form permits generalized Gauss--Laguerre quadrature with weight $Te^{-T}$ and Gauss--Legendre quadrature in $x$. The classical phase integral in Eq.~\eqref{opd:information} is evaluated separately. Photon-number truncation is used only to approximate the entropy calculation, not to restrict Eve's allowed state or to discard preparation records.

\begin{figure}[t]
 \centering
 \includegraphics[width=0.9\columnwidth]{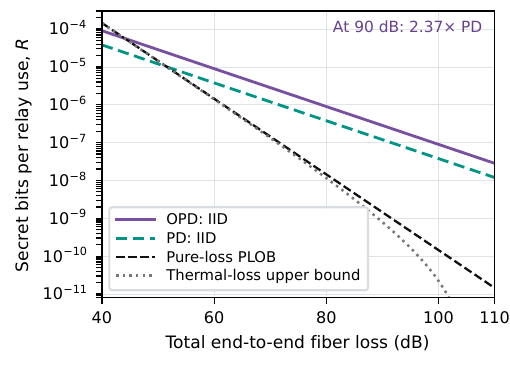}
 \caption{IID secret bits per attempted relay use. CV-TF-PD uses the map of the main text with $N=0.016196$; CV-TF-OPD uses Eq.~\eqref{opd:rate} with $N=0.055$ and the window set in Eq.~\eqref{proc:windows}. Both use $\xi=10^{-5}$ SNU per arm, ideal detection, and ideal phase alignment. Each modulation is held fixed over the loss sweep. The pure-loss PLOB and matched thermal-loss repeaterless bounds use the physical end-to-end fiber transmissivity. The gain concerns these specified IID designs.}
 \label{fig:opd_performance}
\end{figure}

Figure~\ref{fig:opd_performance} compares the PD design at fixed $N=0.016196$ with the OPD design at fixed $N=0.055$, using $\xi=10^{-5}$ SNU per arm. Thus the comparison includes the different useful modulation energies of the two processing maps. The extra disclosure is not assumed to be beneficial by itself: its information cost is included in Eq.~\eqref{opd:holevo}, while ownership and rejection are chosen using the resulting net score. The plotted parameters give a higher OPD rate than PD, and both exceed pure-loss PLOB at sufficiently high loss. At $90$ dB, the rates are $R_{\rm OPD}=2.84\times10^{-7}$ and $R_{\rm PD}=1.19\times10^{-7}$ bits per relay use, giving an OPD advantage of approximately $2.37$. The corresponding pure-loss bound is $1.44\times10^{-9}$, about $197$ times below the OPD rate.

The plotted OPD calculation uses $40$ nodes for each intensity variable in Eq.~\eqref{opd:quadrature}, $32$ nodes for the classical phase integral, and a total photon-number cutoff of $11$ for the entropy calculation. Increasing both intensity orders to $64$ and the cutoff to $13$ changes the rates by less than $0.02\%$ at $40$, $90$, and $110$ dB. An additional $96$-node calculation with $64$ classical phase nodes at $90$ dB differs from the plotted value by less than $0.04\%$. 

The direct-channel benchmarks are the pure-loss bound $K_0(\eta)$ and the matched thermal-loss bound in Eq.~\eqref{main:thermal_plob}. As in the main comparison, $\eta=10^{-L/10}$ at total loss $L$ dB, and the matched thermal channel is the cascade of the two arms. All plotted rates use the physical end-to-end fiber transmissivity; detector inefficiency and finite-data penalties are absent from this IID comparison.

For the high-loss limit at fixed $N$ and input-referred $\xi$, recall Eq.~\eqref{coll:highloss_parameters}: $\lambda\to\lambda_0=N/(N+\xi/2)$ and $\nu\to N$ as $t\to0$. Let $\mathcal K_{\rm OPD}(\lambda,\nu)$ denote the integral in Eq.~\eqref{opd:rate}, without $p_{\rm acc}$. Each retained class score is between zero and one. For every public class of nonzero probability, the binary classical--quantum conditional information varies continuously with the state parameters. Finite maximization preserves continuity. The unit bound on retained scores and the integrable exponential intensity weight then permit dominated convergence, giving
\begin{equation}
 \begin{split}
 R_{\rm OPD}^{\rm iid}
 =2(N+\xi/2)\mathcal K_{\rm OPD}(\lambda_0,N)\sqrt\eta\\
 {}+o(\sqrt\eta).
 \end{split}
 \label{opd:scaling}
\end{equation}
For the plotted OPD parameters, the limiting coefficient in Eq.~\eqref{opd:scaling} is approximately $8.97\times10^{-3}$. OPD can therefore improve the coefficient of the twin-field scaling while preserving its $\sqrt\eta$ dependence whenever the limiting secret fraction is positive. A finite implementation would use finite intensity and phase cells and predetermined finite lookup tables, with its rate evaluated for those actual disclosures. Extending that implementation to coherent attacks and finite-key composable security requires an additional proof.

\end{document}